\documentclass[12pt,a4paper,twoside,prd,showpacs,nofootinbib,preprintnumbers,superscriptaddress]{revtex4-1}

\usepackage{hyperref}
\newcommand{\beq}{\begin{equation}}
\newcommand{\eeq}{\end{equation}}
\usepackage[pdftex]{graphicx}
\usepackage{xcolor}
\usepackage{amssymb,amsmath,bm,natbib}
\usepackage{appendix}

\usepackage{units}
\usepackage{subfig}
\usepackage{cleveref}
\usepackage{booktabs}
\usepackage{xspace}
\AtBeginDocument{
\heavyrulewidth=.08em
\lightrulewidth=.05em
\cmidrulewidth=.03em
\belowrulesep=.65ex
\belowbottomsep=0pt
\aboverulesep=.4ex
\abovetopsep=0pt
\cmidrulesep=\doublerulesep
\cmidrulekern=.5em
\defaultaddspace=.5em
}
\usepackage{todonotes}
\presetkeys{todonotes}{line,color=blue!30, size=\tiny}{}
\newcommand{\nexp}{PIONEER}

\newcommand{\eps}{\varepsilon}

\usepackage{amsmath}

\usepackage{siunitx}

\usepackage[ampersand]{easylist}
\usepackage{gensymb}
\usepackage[T1]{fontenc}

\newcommand{\be}{\begin{equation}}
\newcommand{\ee}{\end{equation}}
\newcommand{\bl}{\vspace{1mm}\begin{easylist}}
\newcommand{\el}{\end{easylist}\vspace{1mm}}

\newcommand{\pdar}{\mbox{$\pi$DAR}}

\newcommand{\pdif}{\mbox{$\pi$DIF}}
\newcommand{\mdif}{\mbox{$\mu$DIF}}
\newcommand{\pie}{$\pi\to e \nu$}
\newcommand{\pme}{$\pi\to\mu\to e$}
\newcommand{\app}{\mbox{$\approx$}}
\newcommand{\remu}{\mbox{$R_{e/\mu}$}}
\newcommand{\tmu}{\mbox{4.1\,MeV}}
\newcommand{\gbl}{\mbox{G4Beamline}}
\newcommand{\atar}{\mbox{ATAR}}
\newcommand{\calo}{\mbox{CALO}}
\newcommand{\track}{\mbox{TRACKER}}
\newcommand{\beam}{\mbox{BEAM}}

\newcommand{\ttrack}{\mbox{\fontfamily{lmss}\selectfont TRACK}}
\newcommand{\tcaloh}{\mbox{\fontfamily{lmss}\selectfont CaloH}}
\newcommand{\tpi}{\mbox{\fontfamily{lmss}\selectfont PI}}
\newcommand{\tprompt}{\mbox{\fontfamily{lmss}\selectfont PROMPT}}
\newcommand{\thcal}{\mbox{58\,MeV}}

\newcommand{\vud}{\ensuremath{\left|V_{ud}\right|}\xspace}
\newcommand{\vus}{\ensuremath{\left|V_{us}\right|}\xspace}

\newcommand{\dedx}{\ensuremath{\mathrm{d}\hspace{-0.1em}E/\mathrm{d}x}\xspace}

\usepackage[margin=0pt,font=normalsize,labelfont=bf,labelsep=endash,justification=RaggedRight]{caption}

\begin{document}
\presetkeys{todonotes}{disable}
\newpage

\title{PSI Ring Cyclotron Proposal R-22-01.1\\ PIONEER: Studies of Rare Pion Decays}



\author{W.~Altmannshofer}\affiliation {Santa Cruz Institute for Particle Physics (SCIPP), University of California Santa Cruz, 1156 High street, Santa Cruz (CA) 95064 USA}
\author{H.~Binney}\affiliation{Department of Physics, University of Washington, Box 351560, Seattle, Washington 98195 USA }
\author{E.~Blucher}\affiliation{Enrico Fermi Institute and Department of Physics, University of Chicago,
5720 South Ellis Avenue,
Chicago, IL 60637 USA}
\author{D.~Bryman}\affiliation{Department of Physics \& Astronomy, University of British Columbia 6224 Agricultural Road, Vancouver V6T 1Z1 Canada}\affiliation{TRIUMF, 4004 Wesbrook Mall, Vancouver V6T 2A3 Canada}
\author{L.~Caminada}\affiliation{Paul Scherrer Institute, 5232 Villigen PSI Switzerland}
\author{S.~Chen}\affiliation { Department of Engineering Physics, Tsinghua University, 30 Shuangqing Road, Haidian District, Beijing, 100084 P. R. China   }
\author{V.~Cirigliano}\affiliation{Institute for Nuclear Theory, University of Washington, Seattle WA 98195-1550 USA}
\author{S.~Corrodi}\affiliation{Argonne National Laboratory, High Energy Physics Division, 9700 S Cass Ave, Lemont, IL 60439 USA}
\author{A. Crivellin}\affiliation{Paul Scherrer Institute, 5232 Villigen PSI Switzerland}\affiliation{Physik-Institut
University of Zurich 
Winterthurerstrasse  190
CH-8057 Zurich
Switzerland}\affiliation{Division of Theoretical Physics, CERN,Espl. des Particules 1, 1211 Meyrin  Switzerland}
\author{S.~Cuen-Rochin}\affiliation{Tecnol\'{o}gico de Monterrey, School of Engineering and Sciences, Blvd. Pedro Infante 3773 Pte, Culiacan 80100 Mexico}
\author{A.~DiCanto}\affiliation{Physics Department, Brookhaven National Laboratory, Upton, NY, 11973 USA}
\author{L.~Doria}\affiliation{PRISMA$^+$ Cluster of Excellence and Johannes Gutenberg Universit\"at Mainz, Institut für Kernphysik, J.-J.-Becher-Weg 45, 55128 Mainz Germany}
\author{A.~Gaponenko}\affiliation{Fermi National Accelerator Laboratory (FNAL), P.O. Box 500, Batavia IL 60510-5011 USA}
\author{A.~Garcia}\affiliation{Department of Physics, University of Washington, Box 351560, Seattle, Washington 98195 USA }
\author{L.~Gibbons}\affiliation{Department of Physics, Cornell University, 109 Clark Hall, Ithaca, New York 14853 USA}
\author{C.~Glaser}\affiliation{Department of Physics,  University of Virginia,  P.O. Box 400714, 382 McCormick Road, Charlottesville, VA 22904-4714 USA}
\author{M.~Escobar~Godoy}\affiliation {Santa Cruz Institute for Particle Physics (SCIPP), University of California Santa Cruz, 1156 High street, Santa Cruz (CA) 95064 USA}
\author{D.~G\"oldi}\affiliation{ETH Zurich, Main building, Rämistrasse 101, 8092 Zurich Switzerland}
\author{S.~Gori}\affiliation {Santa Cruz Institute for Particle Physics (SCIPP), University of California Santa Cruz, 1156 High street, Santa Cruz (CA) 95064 USA}
\author{T.~Gorringe}\affiliation{Department of Physics and Astronomy, University of Kentucky, 505 Rose Street Lexington, Kentucky 40506-0055 USA}
\author{D.~Hertzog}\affiliation{Department of Physics, University of Washington, Box 351560, Seattle, Washington 98195 USA }
\author{Z.~Hodge}\affiliation{Department of Physics, University of Washington, Box 351560, Seattle, Washington 98195 USA }
\author{M.~Hoferichter}\affiliation{Albert Einstein Center for Fundamental Physics, Institute for Theoretical Physics, University of Bern, Sidlerstrasse 5, 3012 Bern  Switzerland}
\author{S.~Ito}\affiliation{KEK, High Energy Accelerator Research Organization, 1-1, Oho, Tsukuba-city, Ibaraki 305-0801 Japan}
\author{T.~Iwamoto}\affiliation{International Center for Elementary Particle Physics (ICEPP), The University of Tokyo, 7-3-1 Hongo, Bunkyo-ku, Tokyo 113-0033 Japan}
\author{P.~Kammel}\affiliation{Department of Physics, University of Washington, Box 351560, Seattle, Washington 98195 USA }
\author{B.~Kiburg}\affiliation{Fermi National Accelerator Laboratory (FNAL), P.O. Box 500, Batavia IL 60510-5011 USA}
\author{K.~Labe}\affiliation{Department of Physics, Cornell University, 109 Clark Hall, Ithaca, New York 14853 USA}
\author{J.~LaBounty}\affiliation{Department of Physics, University of Washington, Box 351560, Seattle, Washington 98195  USA }
\author{U.~Langenegger}\affiliation{Paul Scherrer Institute, 5232 Villigen PSI Switzerland}
\author {C.~Malbrunot}\affiliation{TRIUMF, 4004 Wesbrook Mall, Vancouver V6T 2A3 Canada}
\author{S.M.~Mazza}\affiliation {Santa Cruz Institute for Particle Physics (SCIPP), University of California Santa Cruz, 1156 High street, Santa Cruz (CA) 95064 USA}
\author{S.~Mihara}\affiliation{KEK, High Energy Accelerator Research Organization, 1-1, Oho, Tsukuba-city, Ibaraki 305-0801 Japan}
\author{R.~Mischke}\affiliation{TRIUMF, 4004 Wesbrook Mall, Vancouver V6T 2A3 Canada}
\author{T.~Mori}\affiliation{International Center for Elementary Particle Physics (ICEPP), The University of Tokyo, 7-3-1 Hongo, Bunkyo-ku, Tokyo 113-0033 Japan}
\author{J.~Mott}\affiliation{Fermi National Accelerator Laboratory (FNAL), P.O. Box 500, Batavia IL 60510-5011 USA}
\author{T.~Numao}\affiliation{TRIUMF, 4004 Wesbrook Mall, Vancouver V6T 2A3 Canada}
\author{W.~Ootani}\affiliation{International Center for Elementary Particle Physics (ICEPP), The University of Tokyo, 7-3-1 Hongo, Bunkyo-ku, Tokyo 113-0033 Japan}
\author{J.~Ott}\affiliation {Santa Cruz Institute for Particle Physics (SCIPP), University of California Santa Cruz, 1156 High street, Santa Cruz (CA) 95064 USA}
\author{K.~Pachal}\affiliation{TRIUMF, 4004 Wesbrook Mall, Vancouver V6T 2A3 Canada}
\author{C.~Polly}\affiliation{Fermi National Accelerator Laboratory (FNAL), P.O. Box 500, Batavia IL 60510-5011 USA}
\author{D.~Po\v{c}ani\'c}\affiliation{Department of Physics, University of Virginia, P.O. Box 400714, 382 McCormick Road, Charlottesville, VA 22904-4714 USA}
\author{X.~Qian}\affiliation{Physics Department, Brookhaven National Laboratory, Upton, NY, 11973 USA}
\author{D.~Ries}\affiliation{Department of Chemistry – TRIGA site, Johannes Gutenberg University Mainz, Fritz-Strassmann-Weg 2, 55128 Mainz Germany}
\author{R.~Roehnelt}\affiliation{Department of Physics, University of Washington, Box 351560, Seattle, Washington 98195 USA }
\author{B.~Schumm}\affiliation {Santa Cruz Institute for Particle Physics (SCIPP), University of California Santa Cruz, 1156 High street, Santa Cruz (CA) 95064 USA}
\author{P.~Schwendimann}\affiliation{Department of Physics, University of Washington, Box 351560, Seattle, Washington 98195 USA }
\author{A.~Seiden}\affiliation {Santa Cruz Institute for Particle Physics (SCIPP), University of California Santa Cruz, 1156 High street, Santa Cruz (CA) 95064 USA}
\author{A.~Sher}\affiliation{TRIUMF, 4004 Wesbrook Mall, Vancouver V6T 2A3 Canada}
\author{R.~Shrock}\affiliation {C. N. Yang Institute for Theoretical Physics and Department of Physics and Astronomy, Stony Brook University, Stony Brook, NY 11794 USA}
\author{A.~Soter}\affiliation{ETH Zurich, Main building, Rämistrasse 101, 8092 Zurich Switzerland}
\author{T.~Sullivan}\affiliation{Department of Physics and Astronomy, Elliott Building, University of Victoria, Victoria, BC V8P 5C2 Canada}
\author{M.~Tarka}\affiliation {Santa Cruz Institute for Particle Physics (SCIPP), University of California Santa Cruz, 1156 High street, Santa Cruz (CA) 95064 USA}
\author{V.~Tischenko}\affiliation{Physics Department, Brookhaven National Laboratory, Upton, NY, 11973 USA}
\author{A.~Tricoli}\affiliation{Physics Department, Brookhaven National Laboratory, Upton, NY, 11973 USA}
\author{B.~Velghe}\affiliation{TRIUMF, 4004 Wesbrook Mall, Vancouver V6T 2A3 Canada}

\author{V.~Wong}\affiliation{TRIUMF, 4004 Wesbrook Mall, Vancouver V6T 2A3 Canada}
\author{E.~Worcester}\affiliation{Physics Department, Brookhaven National Laboratory, Upton, NY, 11973 USA}
\author{M.~Worcester}\affiliation{Instrumentation Division, Brookhaven National Laboratory, Upton, NY, 11973 USA}
\author{C.~Zhang}\affiliation{Physics Department, Brookhaven National Laboratory, Upton, NY, 11973 USA}

\newpage

\begin{abstract}
ABSTRACT: A next-generation rare pion decay experiment, PIONEER, is strongly motivated by several inconsistencies between Standard Model (SM) predictions and data pointing towards the potential violation of lepton flavor universality. It will probe non-SM explanations of these anomalies through sensitivity to quantum effects of new particles even if their masses are at very high scales.
Measurement of the charged-pion branching ratio to electrons vs.\ muons $R_{e/\mu}$ is extremely sensitive to a wide variety of new physics effects. At present, the SM prediction for $R_{e/\mu}$ is known to 1 part in $10^4$, which is 15 times more precise than the current experimental result.  An experiment reaching the theoretical accuracy will test lepton flavor universality at an unprecedented level, probing mass scales up to the PeV range. Measurement of  the rare process of pion beta decay, $\pi^+\to \pi^0 e^+ \nu (\gamma)$, with  3 to 10-fold improvement in  sensitivity, will determine \vud in a theoretically pristine manner and test CKM unitarity, which is very important in light of the recently emerged tensions. In addition, various exotic rare decays involving  sterile neutrinos and  axions will be searched for with unprecedented sensitivity.  The experiment design benefits from experience with the recent PIENU and PEN experiments at TRIUMF and the Paul Scherrer Institut (PSI).  Excellent energy and time resolutions, greatly increased calorimeter depth, high-speed detector and electronics response, large solid angle coverage, and complete event reconstruction   are all critical aspects of the approach. The PIONEER experiment design includes  a 3$\pi$ sr 25 radiation length  calorimeter, a segmented low gain avalanche detector stopping target, a positron tracker, and other detectors. Using intense pion beams, and state-of-the-art instrumentation and computational resources, the experiments can be performed at the PSI ring cyclotron.
\end{abstract}
\date{January 10, 2022}
\maketitle
\newpage

\newpage
\tableofcontents
\newpage

\section{Introduction}

Precise low-energy measurements of observables that can be very accurately calculated in the Standard Model (SM)  offer highly sensitive tests of new physics (NP). In light of the existing intriguing hints for lepton flavor universality (LFU) violating NP~\cite{Crivellin:2021sff,Fischer:2021sqw,Bryman:2021teu}, the ratio $R_{e/\mu} = \Gamma(\pi^+\rightarrow e^+\nu(\gamma))/\Gamma(\pi^+\rightarrow \mu^+\nu(\gamma))$ for pion decays to positrons relative to muons is especially promising: 
it is one of the most precisely known observables involving quarks within the SM and NP can even have (chirally) enhanced effects, making it an extremely sensitive probe of NP.
However, while the uncertainty of the SM calculation for $R_{e/\mu}$ is 
very small (with relative precision $1.2\times 10^{-4}$~\cite{Cirigliano:2007xi}), the current experimental world average is about a factor 15 less precise, limiting the NP reach.

A new experiment, PIONEER, is proposed at Paul Scherrer Institute (PSI), where high-intensity pion beams can be delivered. In Phase~I, it  will bridge the gap of a factor 15  between theoretical and experimental precision for $R_{e/\mu}$. With measurements at the $0.01\%$ level in precision, NP up to the PeV scale~\cite{Bryman:2011zz} may be revealed. Such precision would  contribute to  stringent tests of LFU in a context where several intriguing hints of LFU violation (LFUV) have emerged. In addition, it will allow extended searches for exotics such as heavy neutral leptons and dark sector processes. In later Phases (II,III),  PIONEER  will  also study  pion beta decay $\pi^+\to \pi^0 e^+ \nu (\gamma)$ ultimately aiming at  an order of magnitude improvement in precision to determine \vud in a theoretically pristine manner and test CKM unitarity, for which there is presently a $ 3\sigma$ tension~\cite{ParticleDataGroup:2020ssz}. PIONEER is an ambitious program that will span more than a decade of research activity. 

While we focus on the measurement of the  \pie~ branching ratio $R_{e/\mu}$, the following sections discuss the theoretical motivation for pursuing the full rare pion decay program.  
Discussions of the \nexp\ detector concepts, simulations, estimated sensitivities, and planning for realization  follow. In the final section, we discuss aspects related to training, equity, diversity and inclusion.
Appendices contain greater detail on some of the topics.

\section {Theory}

While no particles or interactions beyond those of the SM have been observed so far, intriguing hints for LFUV have been accumulated in recent years~\cite{Crivellin:2021sff,Fischer:2021sqw,Bryman:2021teu}. In particular, the measurements of the ratios of branching ratios (Br)  $R(D^{(*)})=Br[ B\to D^{(*)}\tau\nu_\tau$]/Br[$B\to D^{(*)}\ell\nu_\ell$]~\cite{Lees:2012xj,Aaij:2017deq,Abdesselam:2019dgh} , where $\ell=\mu, e$, and  $R(K^{(*)})=Br[B\to K^{(*)} \mu^+ \mu^-$]$/$Br[$B\to K^{(*)} e^+ e^-$]~\cite{Aaij:2017vbb,LHCb:2019hip,LHCb:2021trn} deviate from the SM expectation by more than $3\sigma$~\cite{Amhis:2019ckw,Murgui:2019czp,Shi:2019gxi,Blanke:2019qrx,Kumbhakar:2019avh} and $4\sigma$~\cite{Alguero:2019ptt,Aebischer:2019mlg,Ciuchini:2019usw,Arbey:2019duh}, respectively. In addition, anomalous magnetic moments $(g-2)_\ell$ ($\ell=e,\mu,\tau$) of charged leptons are intrinsically related to LFUV, as they are chirality flipping quantities. Here, the longstanding discrepancy in $(g-2)_\mu$, just reaffirmed at the level of $4.2\sigma$~\cite{Bennett:2006fi,Muong-2:2021ojo,Aoyama:2020ynm}, can be considered as another hint of LFUV, since, if compared to $(g-2)_e$, the NP contribution scales with a power of the lepton mass~\cite{Davoudiasl:2018fbb,Crivellin:2018qmi}.  In addition, there is a hint for LFUV in the difference of the forward-backward asymmetries ($\Delta A_{\rm FB}$) in $B\to D^*\mu\nu$ vs $B\to D^*e\nu$~\cite{Bobeth:2021lya,Carvunis:2021dss}. As another possible indication of LFUV, CMS observed an excess in non-resonant di-electron pairs with respect to di-muons~\cite{Sirunyan:2021khd}. Furthermore, the possible deficit in first-row unitarity of the Cabibbo-Kobayashi-Maskawa (CKM) matrix, known as the Cabibbo angle anomaly (CAA) (see Fig.~\ref{fig:CKM} (left)), can also be viewed as a sign of LFUV~\cite{Coutinho:2019aiy,Crivellin:2020lzu}. For these reasons, there is very strong motivation for an upgraded $R_{e/\mu}$ experiment whose precision matches that of the SM prediction (see Sec.~\ref{sec:LFUV}). Moreover, the significance of the CAA depends crucially on experimental input quantities used for the extraction of CKM matrix elements as well as a number of theory corrections. Here, an improved measurement of pion beta decay would allow one to extract \vud in a theoretically pristine manner, requiring a gain of a factor $3$ in experimental precision when combined with $K_{\ell 3}$ decays~\cite{Czarnecki:2019iwz} and an order of magnitude for a stand-alone extraction (see Sec.~\ref{sec:CKM}). Finally, an experiment capable of addressing these physics goals would at the same time be able to improve sensitivity to a host of exotic decays, as described in Sec.~\ref{sec:BSM}.  

\subsection{Lepton flavor universality tests and $R_{e/\mu}$}
\label{sec:LFUV}

The branching ratio 
$R_{e/\mu} = \frac{\Gamma\left(\pi^+ \rightarrow e^+ \nu (\gamma) \right)}{\Gamma\left(\pi^+ \rightarrow \mu^+ \nu (\gamma)\right)}$
for pion decays to electrons over muons provides the best test of electron--muon universality in charged-current weak interactions. In the SM, $R_{e/\mu}$ has been calculated with extraordinary precision at the $10^{-4}$ level as \cite{Cirigliano:2007xi,Cirigliano:2007ga,Marciano:1993sh}
\begin{equation}
\label{Remu_SM}
    R_{e/\mu} \hspace{0.1cm}\text{(SM)} = 1.23524(15)\times10^{-4},
\end{equation}
perhaps the most precisely calculated weak interaction observable involving quarks.\footnote{Reference~\cite{Bryman:2011zz} estimates the uncertainty due to the unknown non-leading-logarithmic contributions of $O(\alpha^2 \log (m_\mu/m_e)$ in a different way compared to Ref.~\cite{Cirigliano:2007xi}. This leads to a larger total uncertainty, i.e.,  $R_{e/\mu} \hspace{0.1cm}\text{(SM)} = 1.23524(19)\times10^{-4}$.} Because the
uncertainty of the SM calculation for $R_{e/\mu}$ is very small and the decay $\pi^+ \rightarrow e^+ \nu$ is helicity-suppressed by the $V-A$ structure of charged currents, a measurement of $R_{e/\mu}$ is extremely sensitive to the presence of pseudoscalar (and scalar) couplings absent from the SM; a disagreement with the theoretical expectation would unambiguously imply the existence of NP. With measurements of 0.01\% experimental precision, NP at the PeV scale can be probed~\cite{Bryman:2011zz}, even up to several PeV in specific models such as leptoquarks. 

The uncertainty of the SM prediction~\eqref{Remu_SM} for $R_{e/\mu}$ arises from low-energy constants in chiral perturbation theory, which absorb the divergences in the two-loop calculation of Refs.~\cite{Cirigliano:2007xi,Cirigliano:2007ga}, but whose finite parts need to be determined by other means. Fortunately, in the case of  $R_{e/\mu}$ these non-perturbative uncertainties only affect the SM prediction at the relative precision of $10^{-4}$, more than an order of magnitude beyond the current experimental precision~\cite{ParticleDataGroup:2020ssz,PiENu:2015seu,Bryman:1985bv,Britton:1993cj,Czapek:1993kc}
\begin{equation}
   R_{e/\mu} \hspace{0.1cm}\text{(exp)} =1.2327(23)\times10^{-4}.
\end{equation}
$R_{e/\mu}$ thus provides a unique opportunity for a pristine test of LFU in the quark sector. 

The comparison between theory and experiment  provides a stringent test of the $e$--$\mu$ universality of the weak interaction. We  express the results in terms of the effective couplings $A_\ell$ multiplying the low-energy charged current  contact interaction 
\begin{equation}
{\cal L}_{CC} = A_\ell [\bar u \gamma^\mu P_L d] [ \bar \nu_\ell \gamma_\mu P_L \ell ],
\end{equation}
where $P_L \equiv (1-\gamma_5)/2$. In the SM at tree level the couplings are given by  $A_\ell = - 2 \sqrt{2} G_F V_{ud}$ and thus satisfy LFU, i.e., $A_\ell / A_{\ell^\prime} = 1$.  
The measurement of $R_{e/\mu}$ results in 
 \begin{equation}
     \left( \frac{A_\mu}{A_e}  \right)_{ R_{e/ \mu} }= 1.0010(9)\,,\label{gpi}
 \end{equation}
which is in agreement with the SM expectation and provides the best available test of LFU. 
A deviation from $A_\ell / A_{\ell^\prime} = 1$ can originate from various mechanisms. In the literature it is common to interpret deviations from $A_\ell/A_{\ell^\prime} = 1$ in terms of flavor-dependent couplings $g_\ell$ of the $W$-boson to 
the leptonic current, in which case $A_\ell \propto g_\ell$.  We note that in the context of modified $W$-boson couplings LFU tested with $R_{e/\mu} $  probes the couplings of a longitudinally polarized $W$-boson, whereas tests using purely leptonic reactions such as $\tau\to \ell \nu_\tau \nu_{\ell}$ ($\ell=e,\mu$) test the couplings of transversely polarized $W$-boson and are thus complementary. 

\begin{figure}[t!]
\centering
\includegraphics[width=0.9\textwidth]{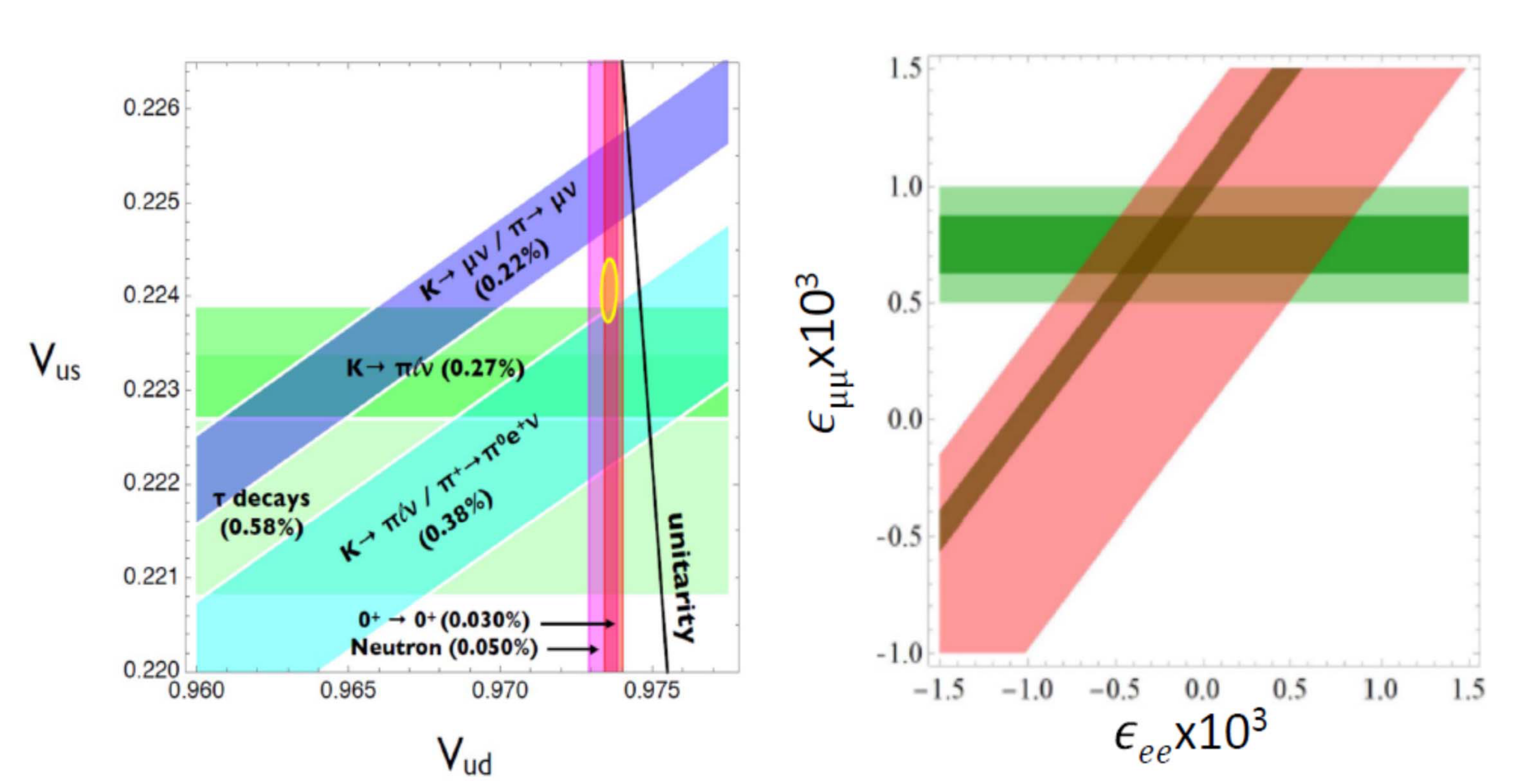}
%
%
%
%
\caption{Left: Tensions in the first-row CKM unitarity test (see text)~\cite{Bryman:2021teu}. Right: Constraints (1$\,\sigma$) on modified $W\ell\nu$ couplings from CKM unitarity (green) and LFUV (red) (adapted from Ref.~\cite{Crivellin:2020lzu}). The light bands  show the current status and the dark bands include the expected PIONEER  sensitivity. The SM values of $A_\ell$ are assumed to be modified by $1+\eps_{\ell\ell}$ where $\ell=e,\mu$.}
\label{fig:CKM}
\end{figure}

Assuming that LFUV originates from modified $W\ell\nu$ couplings, the determination of CKM elements will also be affected. Importantly, beta decays have an enhanced sensitivity to a modified $W\mu\nu$ coupling, due to a CKM enhancement by $\left|V_{ud} / V_{us}\right|^2 \sim 20$. Such a modification of the $W\ell\nu$ couplings would also affect $R_{e/\mu}$, albeit for a different flavor combination (see Fig.~\ref{fig:CKM} (right)). This connection provides further motivation for an improved $R_{e/\mu}$ measurement, especially because the sensitivity to LFUV would be comparable to future improved constraints from beta decays. Moreover, recent global fits to electroweak observables and tests of LFU show a preference for $R_{e/\mu}$ larger than its SM expectation \cite{Coutinho:2019aiy,Crivellin:2020ebi}. Furthermore, $R(K^{*})$ can be correlated to $R_{e/\mu}$~\cite{Capdevila:2020rrl} and a combined explanation of the deficit in the first-row CKM unitarity and the CMS excess in di-electrons even predicts that $R_{e/\mu}$ should be larger than its SM value~\cite{Crivellin:2021rbf}.

\subsection{CKM unitarity and pion beta decay}
\label{sec:CKM}

The detector optimized for a next-generation $R_{e/\mu}$ experiment will also be ideally suited for a high-precision measurement of pion beta decay. Precision measurements of beta decays of neutrons,
nuclei, and mesons provide very accurate determinations of the elements \vud and \vus of the
CKM quark-mixing matrix \cite{Cabibbo:1963yz,Kobayashi:1973fv}. Recent theoretical
developments on radiative corrections and form factors have led to a $ 3\sigma$ tension with CKM
unitarity illustrated in 
Fig.~\ref{fig:CKM} (left)~\cite{Bryman:2021teu,Cirigliano:2021yto}, and with specific assumptions on theory corrections and processes considered, higher significances have been obtained~\cite{Seng:2021nar,Aoki:2021kgd}. However, in these determinations hadronic~\cite{Marciano:2005ec,Seng:2018yzq,Seng:2018qru,Czarnecki:2019mwq,Seng:2020wjq,Hayen:2020cxh,Shiells:2020fqp} and nuclear~\cite{Miller:2008my,Miller:2009cg,Gorchtein:2018fxl,Hardy:2020qwl} corrections play an important role, to the extent that they dominate the systematic uncertainty of the \vud extraction from superallowed beta decays~\cite{Hardy:2020qwl}.  A determination from pion beta decay would be much cleaner, provided it could be measured to sufficient precision.

The branching ratio for pion beta decay was most accurately measured by the PiBeta experiment at PSI
\cite{Pocanic5,Frlez:2003vg,Pocanic:2003pf,Frlez:2003pe,Bychkov:2008ws} to be
\begin{equation}\frac{\Gamma(\pi^+ \to  \pi^0 e^+ \nu)}{\Gamma(\text{Total})}= 1.036 \pm 0.004 (\text{stat}) \pm 0.004(\text{syst}) \pm 0.003(\pi\to e\nu) \times 10^{-8},
\end{equation}
where the first uncertainty is statistical, the second systematic, and the third is the $\pi\to e\nu$ branching ratio uncertainty.
Pion beta decay, $\pi^+ \rightarrow \pi^0 e^+ \nu (\gamma)$, provides the theoretically cleanest determination of the magnitude of the CKM matrix element \vud. With current input one obtains $\vud = 0.9739(28)_{\textrm{exp}}(1)_{\textrm{th}}$, where the experimental uncertainty comes almost entirely from the  $\pi^+ \rightarrow \pi^0 e^+ \nu (\gamma)$ branching ratio (BRPB) \cite{Pocanic:2003pf} (the pion lifetime contributes ${\delta}V_{ud} = 0.0001$), and the theory uncertainty has been reduced from $({\delta}V_{ud})_{\textrm{th}} = 0.0005$ \cite{Sirlin:1977sv,Cirigliano:2002ng,Passera:2011ae} to $({\delta}V_{ud})_{\textrm{th}} = 0.0001$ via a lattice QCD calculation of the radiative corrections \cite{Feng:2020zdc}. The current precision of 0.3\% on \vud makes $\pi^+ \rightarrow \pi^0 e^+ \nu (\gamma)$ not presently relevant for the CKM unitarity tests because superallowed nuclear beta decays provide a nominal precision of 
0.03\%. 

In order to make $\pi^+ \rightarrow \pi^0 e^+ \nu (\gamma)$ important for CKM unitarity tests, two precision experimental stages can be identified:
(1) As advocated in Ref.~\cite{Czarnecki:2019iwz}, a three-fold improvement in BRPB precision compared to Ref.~\cite{Pocanic:2014jka} would allow for a 0.2\% determination of $\left|V_{us}/V_{ud}\right|$, via improving the measurement of the ratio 
   \begin{equation}
   \label{rv}
        R_V = \frac{\Gamma\left(K \rightarrow \pi l \nu (\gamma) \right)}{\Gamma\left(\pi^+ \rightarrow \pi^0 e^+ 
\nu (\gamma)\right)}=1.9884(115)_{\pi}(93)_K\times 10^7,    \end{equation} 
    where the uncertainties are due to the pion partial width and the $K_L$ lifetime and branching ratio. Equation~\eqref{rv} is
    independent of the Fermi constant and SM short-distance corrections, but subject to structure-dependent long-distance radiative corrections. This
    would match the precision of the current extraction of $\left|V_{us} / V_{ud}\right|$ from the axial channels~\cite{Marciano:2004uf}, which proceeds via 
  \begin{equation}
        R_A = \frac{\Gamma\left(K \rightarrow \mu \nu (\gamma) \right)}{\Gamma\left(\pi \rightarrow \mu \nu (\gamma)\right)}=1.3367(25),
   \end{equation}
    (see Fig. \ref{fig:CKM}), thus providing a new competitive constraint on the \vus--\vud plane and probing NP that might affect vector and axial-vector channels in different ways.
    The theoretical case for this approach was recently strengthened by improved analysis of radiative corrections in $K \to \pi e \nu $ decays \cite{Seng:2021wcf}.  
(2)  In the second phase, an order of magnitude improvement  in the
BRPB precision will be sought. This would provide the theoretically cleanest extraction of \vud at the 0.02\% level, comparable to the current value from superallowed beta decays~\cite{Hardy:2020qwl}. 

\subsection{Constraints on New Physics}

There are several ways that (heavy) NP can affect $R_{e/\mu}$ and the extraction of \vud from beta decays, relevant for this proposal. Concerning the former, there are several possibilities: a modified $W\ell\nu$ coupling and a contribution to an $\ell\nu u d$ operator ($\ell=\mu,e$), while only modified $W\mu\nu$ couplings and an $e\nu u d$ operator affect the latter. However, \vud is also sensitive to a direct NP contribution to muon decays via an $e\mu\nu \nu$ operator entering through the Fermi constant, necessary to extract \vud from beta decays~\cite{Crivellin:2021njn}.

The following NP models (see Ref.~\cite{deBlas:2017xtg} for a complete categorization) give potential observable tree-level effects in $R_{e/\mu}$ and/or in the determination of \vud from beta decays (direct, or indirect via muon decay): 
\begin{itemize}
    \item A $W^\prime$ bosons can generate tree-level effects in beta decays and in $R_{e/\mu}$ as well as modified $W\ell\nu$ couplings via mixing with the SM $W$.
    \item Vector-like leptons (VLL) affect $W\ell\nu$ couplings via their mixing with SM leptons after EW symmetry breaking (see Ref.~\cite{Crivellin:2020ebi} for a recent analysis).
    \item A singly charged $SU(2)_L$ singlet scalar can give a necessarily constructive tree-level effect in muon decay such that the CAA can be solved~\cite{Crivellin:2020klg,Marzocca:2021azj}.
    \item An $SU(2)_L$ triplet scalar can give a necessarily destructive effect in muon decay.
    \item A neutral vector boson ($Z^\prime$) can give a constructive effect in muon decay if it has flavor-violating couplings.
    \item A leptoquark can generate a left-handed vector current and/or a scalar current affecting beta decays and $R_{e/\mu}$.
    \item A charged Higgs gives rise to pseudoscalar operators such that chirally enhanced effects in $R_{e/\mu}$ can be generated.
\end{itemize}
Therefore, the PIONEER results will be important for a very wide range of SM extensions containing one or more of these particles.

\subsection{Heavy Neutrinos and other Dark Sector Physics}
\label{sec:BSM}

The PIONEER experiment will also achieve improved sensitivity in probing for effects of heavy neutrinos. If a heavy neutrino has a sufficiently low
mass that it can be emitted, the signature will be a monochromatic peak
in the energy of the outgoing charged lepton at an anomalously low value.
Even if the heavy neutrino has a mass greater than the kinematic limit
for emission, there will, in general, still be an apparent violation of
$e$--$\mu$ universality, i.e., a deviation in the measured ratio $R_{e/\mu} \equiv BR(\pi^+ \to e^+ \nu_e)/BR(\pi^+ \to \mu^+ \nu_\mu)$ from its SM value 
\cite{Shrock:1980vy,Shrock:1980ct,Shrock:1981wq}. Constraints from early experiments on $\pi^+_{\ell 2}$ decays were reported in 
Refs.~\cite{Abela:1981nf,Minehart:1981fv,Bryman:1983cja,Azuelos:1986eg,Britton:1992pg}.
Recent results on $\pi^+_{\ell 2}$ decays include
\cite{PiENu:2015seu,PIENU:2017wbj,PIENU:2019usb}.
Constraints from the non-observation of neutrinoless double beta
decays imply that the heavy neutrinos of relevance here are Dirac fermions.  This
can be arranged in various NP scenarios.  There are also constraints
from primordial nucleosynthesis.  Heavy neutrinos with masses of 
relevance here are consistent with these primordial nucleosynthesis 
constraints in several NP theories \cite{Arguelles:2021dqn}.
Recent limits include \cite{Bryman:2019ssi,Bryman:2019bjg}.

Looking beyond sterile neutrinos, dark sectors that consist of particles that interact very feebly with the SM are highly motivated extensions of the SM. There is an ever-growing interest in exploring the parameter space of such scenarios. The PIONEER experiment has unique capabilities to search for pion decays to various light NP states.
 One example is the three-body decay $\pi^+ \to \ell^+ \nu_\ell X$, where $X$ could, for example, be an axion-like-particle (ALP) that mixes with the neutral pion~\cite{Altmannshofer:2019yji}, or a light gauge boson coupling to differences of lepton numbers~\cite{Dror:2020fbh}. Another interesting process is the three-body decay $\pi^+ \to \ell^+ X Y$, where both $X$ and $Y$ are light dark sector particles~\cite{Batell:2017cmf}. Existing limits from PIENU~\cite{PIENU:2021clt} already probe some of these models. Moreover, PIONEER can also look for lepton-flavor-violating decays of the muon into light NP particles $\mu \to e X$.

\section{PIONEER Experiment}
\subsection{Experiment Overview and Strategy}

\begin{figure}[h!]
\centering
\includegraphics[width=0.4\textwidth]{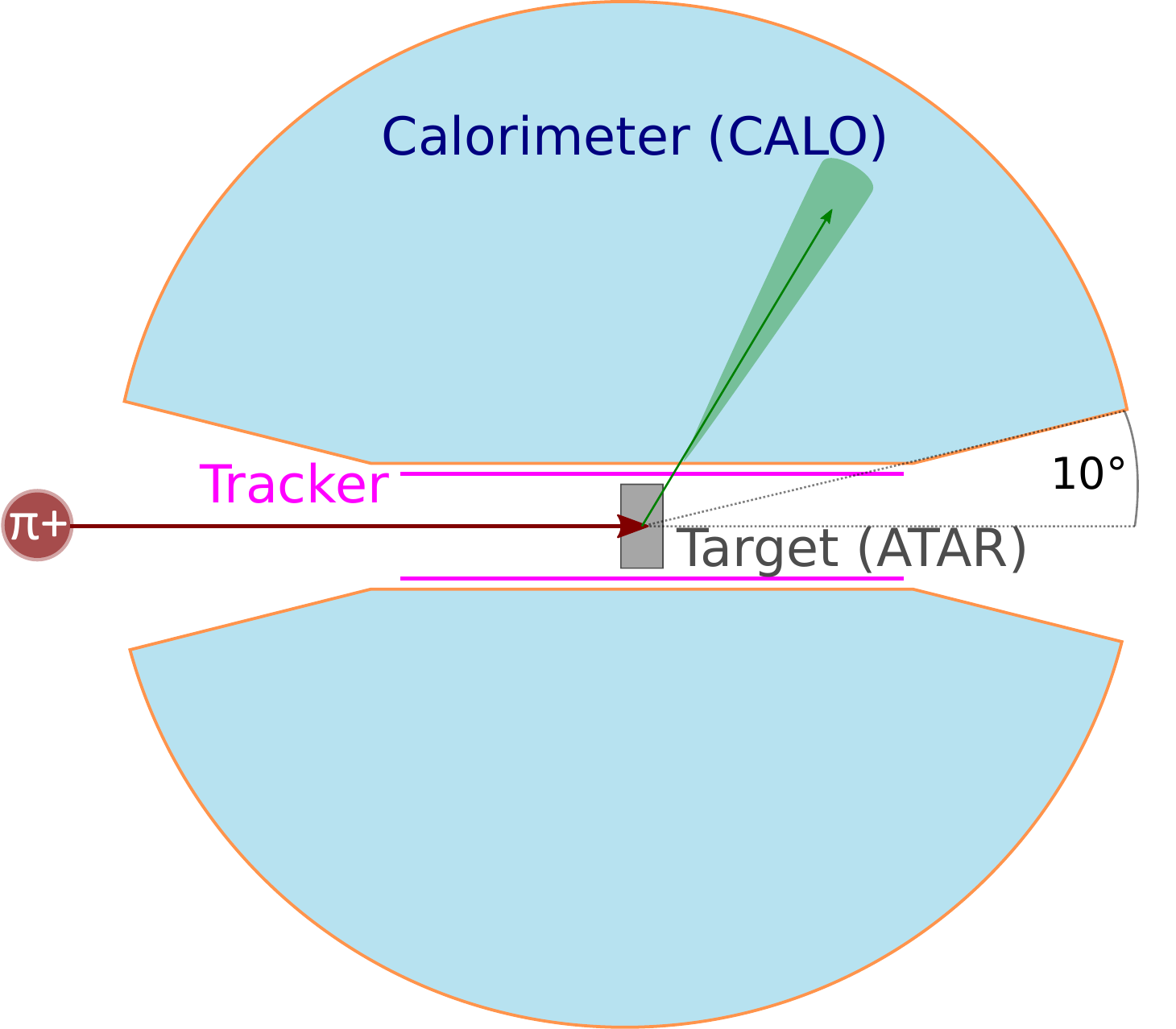}
\caption{Layout of the PIONEER rare pion decay experiment.  The intense positive pion beam enters from the left and   is brought to rest in a highly segmented active target (ATAR).   Decay positron trajectories are measured from the ATAR to an outer electromagnetic calorimeter (CALO) through a tracker.  The CALO records the positron energy, time and location.
}
\label{fig:GenericDetector}
\end{figure}

The main challenge in developing a next generation experiment for a high precision measurement of rare pion  decays is  accurately assessing the performance of the chosen detector technology in suppressing sources of systematic uncertainties and handling increased rates. The PIONEER detector design concept, described in the next sections, is based on
the experience gathered with the PIENU \cite{PiENu:2015seu} and PEN/PiBeta \cite{Pocanic1,PEN:2018kgj, Pocanic:2014jka} experiments, which are reviewed in the Appendix.
Generically, the detector will have the features sketched out in Fig.~\ref{fig:GenericDetector}.  An intense pion beam is brought to rest in an instrumented (active) target (ATAR) and an electromagnetic calorimeter (CALO) surrounds the stopping target. A cylindrical tracker surrounding the ATAR is used to link the locations of pions stopping in the target to showers in the calorimeter.
Features of the PIONEER approach will include improved time and energy resolutions, greatly increased calorimeter depth, high-speed detector and electronic response, large solid angle coverage, and complete event reconstruction. The proposed detector will include a 3$\pi$ sr, 25 radiation length ($X_0$) electromagnetic calorimeter, an advanced design segmented stopping target, and  beam and  positron trackers.

Phase I of \nexp\ aims to measure $R_{e/\mu}$ with precision of 0.01\%, where the uncertainty budget is equally allocated to statistics and systematics; $2\times 10^8$ $\pi^+\to e^+ \nu$ events are required.
This Proposal is specifically focused on the $R_{e/\mu}$ measurement. 
However, we envision extension of PIONEER in Phase II (III), the details of which will be proposed  in  future. These future phases will focus on a 3-fold (10-fold) improvement in the measurement of the ultra-rare pion beta decay  process,
 $\pi^+ \to \pi^0 e^+ \nu$.
The $\pi^+ \to \pi^0 e^+ \nu$ branching ratio is $ 10^4$ times smaller than the \pie~ channel and will require running with a 100x more intense pion flux.  The event identification is more straightforward owing to the characteristic signature of the $\pi^0 \rightarrow \gamma\gamma$ decay in the calorimeter.
While the optimization of the beam properties, instrumentation, and stopping target details for the pion beta decay experiment may require replacements of some systems,  our aim is that the core calorimeter, mechanics, tracker, and DAQ systems will be designed to meet the needs of both experiments with limited modifications.
 We emphasize that the requirements for measuring $R_{e/\mu}$ drive the \nexp~ design.
In subsequent sections we describe the beam,
target, positron detectors, calorimeter, DAQ, and electronics aspects of the experiment. The Appendices include additional technical information and priority R\&D plans.

 \subsubsection{Requirements for measuring $R_{e/\mu}$ }
 At rest, the pion lifetime is 26\,ns and the muon lifetime is 2197\,ns.  The monoenergetic positron from $\pi \rightarrow e\nu$ has an energy of 69.3\,MeV.  Positrons from ordinary muon decay form the Michel spectrum from 0 to an endpoint of 52.3\,MeV.  In principle, the monoenergetic $e^+$ from \pie~ is well isolated above the Michel endpoint and can be easily identified using a high-resolution, hermetic calorimeter. To determine $R_{e/\mu}$ we measure the ratio of positrons emitted from \pie~ and \pme~ decays for which many systematic effects such as solid angle acceptance cancel to first order. However, counting all $\pi \rightarrow e$ events with a precision of one part in $10^4$ requires determining the low-energy tail of the electromagnetic shower and radiative decays that hide under the Michel spectrum from the $\pi \rightarrow  \mu \rightarrow e$ chain, which has  four  orders  of magnitude higher rate.

\begin{figure}[h!]
\centering
\includegraphics[width=0.7\textwidth]{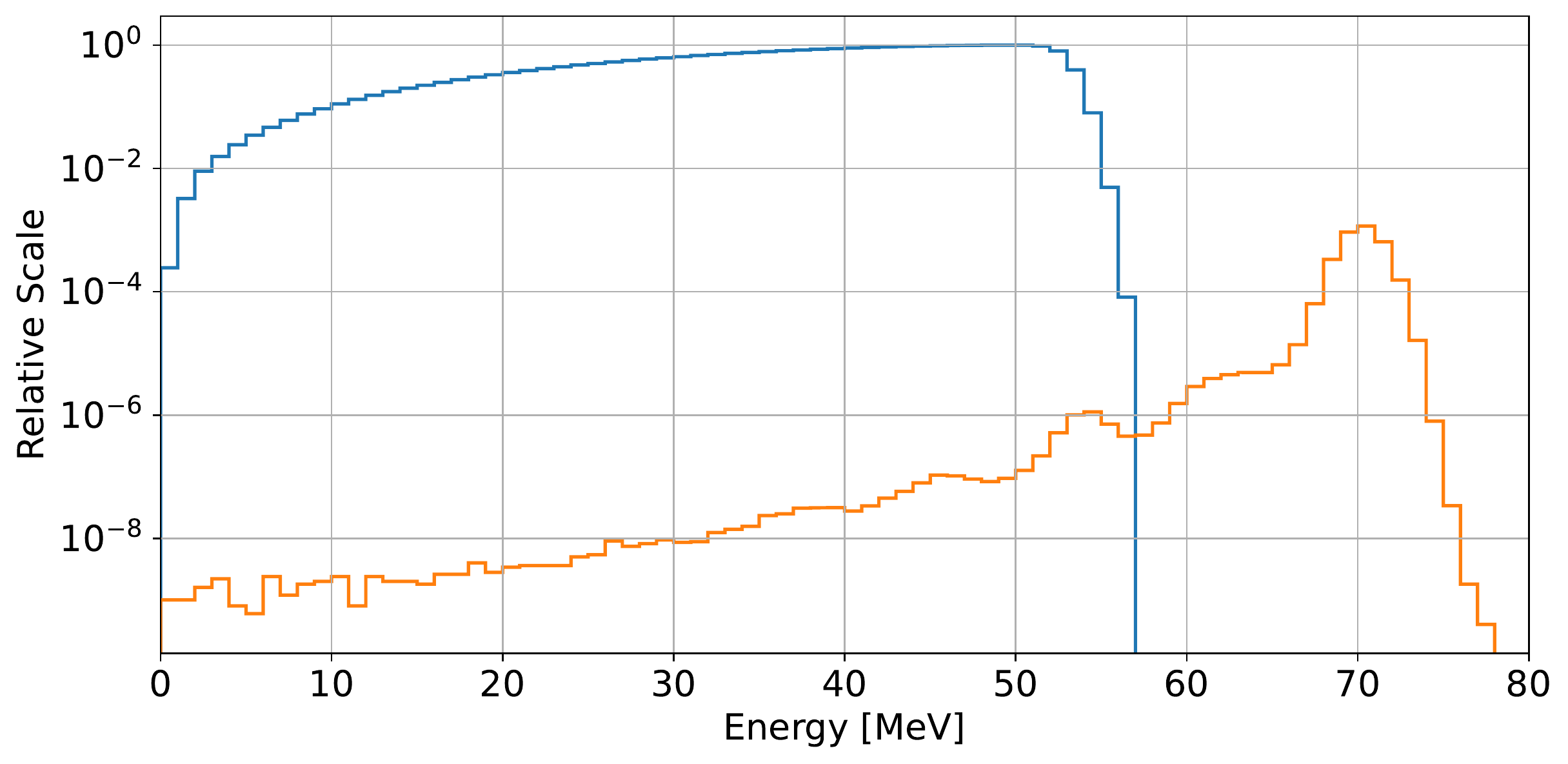}
\caption{The positron energy spectra from  muon decays (blue) and from \pie~ decays (orange) for a calorimeter 
resolution of 1.5\% and a 
depth of $25\,X_0$. 
The simulation includes energy losses owing to photonuclear interactions. 
}
\label{fig:Signal-Tail}
\end{figure}

 Figure~\ref{fig:Signal-Tail} illustrates the relationship between the two channels and their respective positron energy spectra.  Here, we have modeled the spectrum from both channels assuming a high resolution, $25\,X_0$ calorimeter.  There remains an unavoidable tail fraction below 53\,MeV that must be determined accurately in order to obtain the branching ratio.
That challenge was critical to previous generations of  experiments and was responsible for the leading systematic uncertainty in the 
PIENU experiment at TRIUMF.
\nexp~ will minimize the intrinsic tail fraction 
through the use of a $25\,X_0$ LXe calorimeter, the design of which is based on the considerable experience of the MEG Collaboration.

The \pie~ branching ratio $R_{e/\mu}$ will be obtained by first separating events into high- and low-energy regions at an energy cut value ($E_{cut}$) as discussed in the Appendices  for the PIENU experiment. The time spectra will be fit in each region with the $\pi^+ \rightarrow e^+ \nu$ and $\pi^+ \rightarrow \mu^+ \rightarrow e^+$ timing distribution shapes, along with  backgrounds originating from different sources including event pile-up effects,  pion decays in flight, and effects from old muon decays.

\begin{figure}[h!]
\centering
\includegraphics[width = \textwidth]{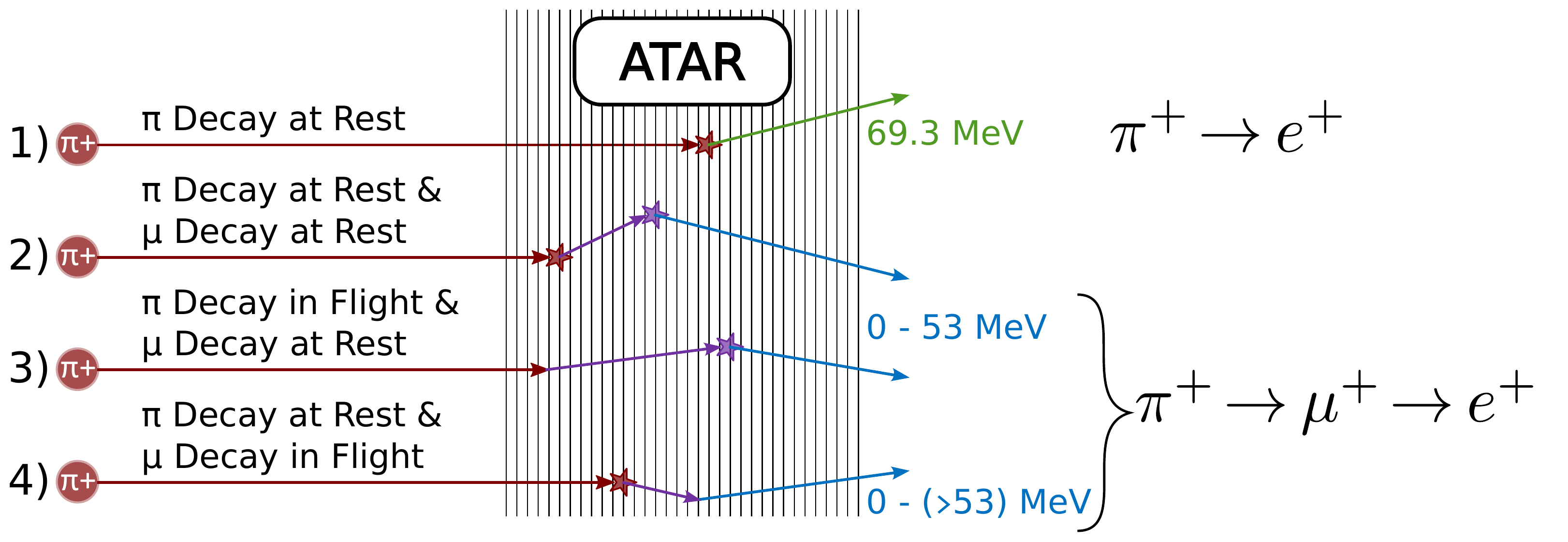}
\caption{Illustration of event types in the $\pi \rightarrow e$ and $\pi \rightarrow \mu \rightarrow e$ chains. The stars are indicative of the energy deposited at the Bragg peak for pions (red) or muons (purple) that stop in the segmented ATAR.  The main channels of interest include 1) the $\pi \rightarrow e$ ``signal channel'' decay that emits a monoenergetic 69.3\,MeV positron; 2) the dominant $\pi \rightarrow \mu$ decay, where the 4.1\,MeV muon travels up to 0.8\,mm and also stops in the ATAR before emitting a  positron.  Events 3) and 4) represent situations that can confuse the classification of events into categories 1) or 2). In 3) the pion decays within the ATAR  prior to stopping; the muon stop can then appear as a pion stop.  In 4), the pion stops, and the decay muon (very rarely, but importantly) decays in the short time prior to stopping. Because it is a decay in flight, the Lorentz boost can push the positron  energy beyond the 52.3\,MeV endpoint.  The ATAR is being designed to distinguish these event patterns.
}
\label{fig:EventTypes}
\end{figure}

\nexp~ will incorporate improvements to the previous techniques including a deeper calorimeter.
However, it is also important to be able to create triggers that can isolate $\pi \rightarrow e$ from $\pi \rightarrow \mu \rightarrow e$ chains within the stopping target, identify pion and muon decays in flight, as well as identify pileup from long-lived muons remaining in the target from earlier pion stops.
Figure~\ref{fig:EventTypes} illustrates several key processes that will occur at different rates in the stopping target.  The signal event (1) has a pion decay at rest; its Bragg peak energy deposition as well as its depth within the target identifies it as a pion stop. The much higher background (and normalization) channel (2) is illustrated by that same pion stop followed by a decay to a 4.1\,MeV muon, which travels through a number of planes and then stops, leaving behind an image of its own short trajectory and Bragg peak. If the muon decays in the prompt window being observed for the pion decays (typically 3 -- 50\,ns after the pion stop), its Michel positron will be recorded in the CALO.  More rare but subtle processes 
may also occur.
A  pion can also decay in flight within the target to a muon before it stops (3).  
In very rare cases, the emitted 4.1 MeV muon can decay in flight before stopping (4) providing a Lorentz boost to the emitted Michel positron, which  contaminates the signal spectrum above the ordinary stopped muon endpoint.  

To distinguish event types, we will use an active target that can provide 4D tracking (at the level of 150\,$\mu$m in space  and $<$1\,ns in time) and energy measurements from the O(30)~keV  signals for positrons to the 4000\,keV Bragg peaks of stopping pions and muons.  As discussed in Sec.~\ref{sec:ATAR}, our collaboration is focusing on the new low gain avalanche detector (LGAD) sensors as a centerpiece of the experiment.  Simulations using optimized LGAD parameters provide confidence that triggers can be constructed to isolate and measure all event types. 

The calorimeter tail fraction for \pie~ events  will be measured {\it in situ} by suppressing the \pme~ decays using information provided by the active target. \pme~ events can be identified and suppressed  by the presence of the 4.1\,MeV pulse from $\pi \to \mu \nu$ decay, use of a narrow time window, $\pi-\mu$ particle identification, and tracking information to identify pion decay-in-flight (\pdif), and muon decay-in-flight (\mdif) following pion decay-at-rest (\pdar).  As discussed  in Sec.~\Ref{Sims}, we anticipate that $\mu-e$ backgrounds in the tail region can be suppressed to a level that will allow the uncertainty in the tail fraction to contribute $<0.01\%$ to the error in $R_{e/\mu}$.

The experiment will require a continuous wave low-momentum pion beam that can be focused to a small spot size and  stop within the ATAR dimensions.  Ideal characteristics include a relatively low momentum of $55\,$\,MeV/$c$ ($\pm2\%$) and a flux of 300\,kHz.  At this low momentum, a separator is very effective to reduce background from beamline muons and positrons.   The $\pi$E5 beam can  provide the needed flux.  We are also investigating use of the $\pi$E1 line.
Because of the high data rate, state-of-the-art triggering, fast digitizing electronics, and high bandwidth data acquisition systems are required.

 \subsubsection{Requirements for measuring pion beta decay }

In Phase II (III), pion beta decay $\pi^+\to\pi^0 e^+ \nu$ will be measured by observing the characteristic (nearly) back-to-back gammas from $\pi^0$ decay normalized to \pie~ decay as in \cite{Pocanic5,Frlez:2003vg,Pocanic:2003pf,Frlez:2003pe,Bychkov:2008ws}. In \nexp\ we also expect to observe the low-energy positron absorbed in the ATAR in coincidence with the gammas in the calorimeter.
The Phase II (III) pion beta decay experiment will require $7\times 10^5$ ($7\times 10^6$) events at an intrinsic branching ratio of $ 10^{-8}$.  This will require running at a significantly higher pion flux of $\geq 10$\,MHz.  The beam momentum and emittance may  be higher than for the \pie~ measurement to achieve the higher flux.  
The higher rate can be handled because of the  gamma ray coincidence identification and nearly fixed energy sum. The  $\pi$E5 beamline appears to have the necessary properties for this measurement. 

\subsubsection{Simulations Guiding the Design}
The discussion of the beamline and detector that follows is based on simulations reported in Section~\ref{Sims}.  Briefly, there are three distinct efforts: beamline and upstream detectors, simulation of the ATAR and event topologies, and simulation of the calorimeter response.
A G4beamline~\cite{Roberts:2008zzc} model of $\pi$E5 is being used to extrapolate from the existing surface muon tunes used in the MEG program to the $55-70$\,MeV/$c$ range of interest for  positive pions required for PIONEER.  
Descriptions of the ATAR and CALO are included in the GEANT-4~\cite{GEANT4:2002zbu} models.
The models include inactive materials such as walls, cables, and windows and a
Tracker detector that surrounds the ATAR.
\subsection{Beam}
\subsubsection{Beam requirements and target setup}

The basic beam requirements for PIONEER are summarized in Table~\ref{tab:beam}. The values given are initial estimates, and some of the requirements,
e.g. in the longitudinal and transverse beam extent, might be relaxed pending further studies.
In the following we will only discuss PIONEER Phase I beam requirements, but the parameters for the full experimental program are also considered.


\begin{table}[htb]
\centering
\begin{tabular}{ccccccc}
\toprule
Phase & p  & $\Delta$p/p & $\Delta$Z 	& $\Delta$X x $\Delta$Y& $\Delta$X',$\Delta$Y'  & R$_\pi$ \\ 
      & (MeV/c) & (\%) & (mm) 				& (mm$^2)$ 				& 						 & (10$^6$/s) \\ 
\hline 
I & 55-70 & 2 & 1 & 10x10 &  $\pm$10\degree & 0.3 \\ 
II,III & \app\ 85 & $\le$ 5 & 3 & 15x15 & $\pm$10\degree  & 20 \\ 
\bottomrule 
\end{tabular} 
\caption{Required beam properties. $\Delta$Z and $\Delta$X $\times$ $\Delta$Y are longitudinal (FWHM) range width and transverse (FWHM) beam sizes at target location, respectively.}
\label{tab:beam}
\vspace{-0mm}
\end{table}

The 0.01\% precision goal of the \remu\ measurement places stringent requirements on the  beam and target as discussed below. 
\bl 
\ListProperties(Margin=1cm, Hang1=true,Hide2=0,Numbers1=r,Numbers2=r,Progressive*=.5 cm,Start1=1)
& Pion stops: The efficiency for
detecting the $\pi-\mu$ sequence should be the same as for a $\pi$ stop; thus all decay muons must be  fully contained in  \atar. 
This drives the requirement for the small longitudinal and lateral size of the stopped $\pi$ distribution. Upstream  detectors (if necessary) will  be minimized, as they
increase the beam spot by scattering. 

& Positron detection: The \atar\ should  be small to reduce the positron energy deposited in it, though largely this loss can be reconstructed. The constraint on the beam entrance divergence is necessary to keep the  opening angle of the beam entrance cone
to 10\degree\ relative to its axis. (A larger cone will decrease the solid angle for positron detection and, in particular,  increase the surface for shower leakage.) This requirement limits the phase space acceptance to 175 $\pi\,$cm$\,$mrad.
\el

\begin{figure}[htb]
\centering
\includegraphics[width=1\textwidth]{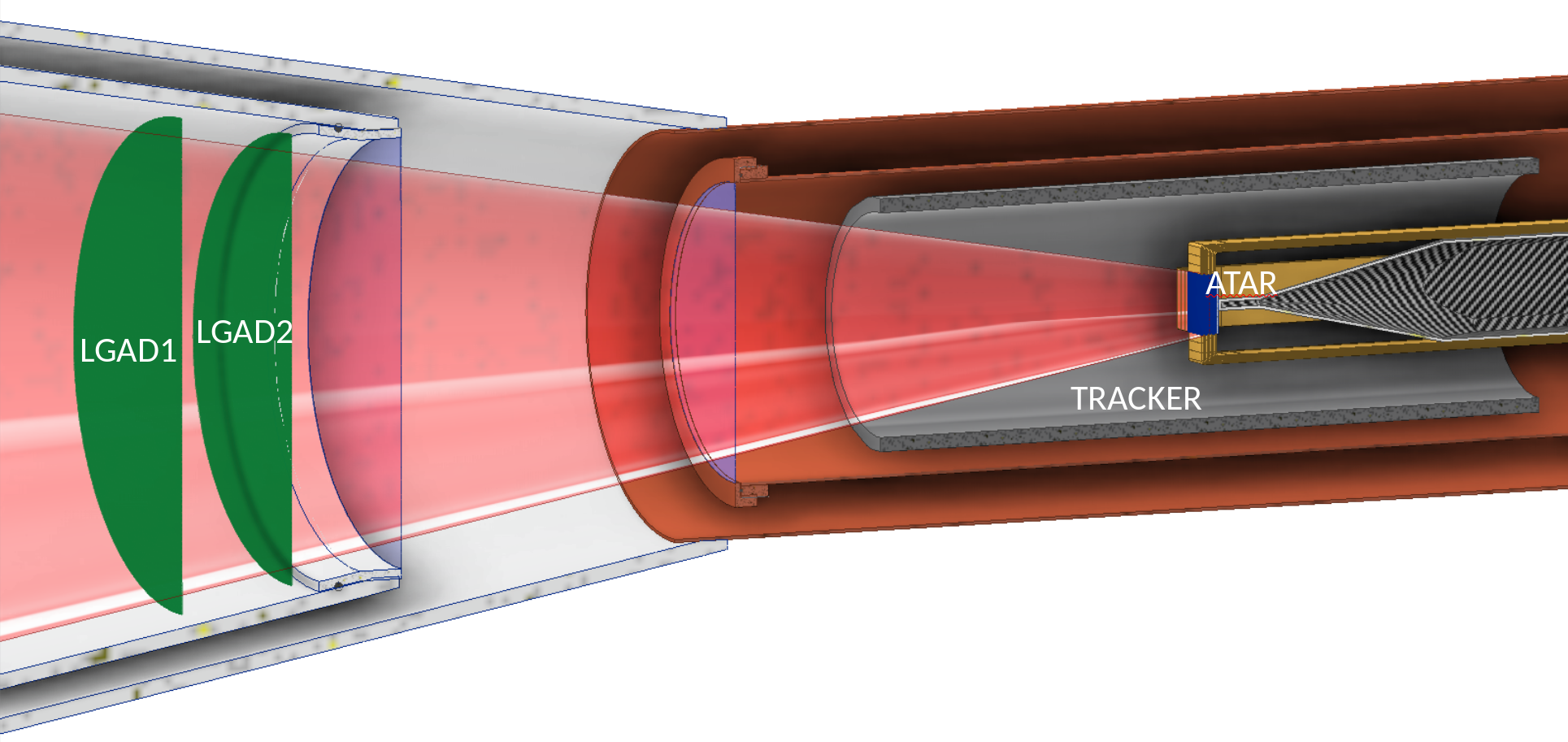}
\caption{Beam counters and \atar. An active degrader disk (not shown) is optional, and only required for p$>$55 MeV/c. The beam detectors systems LGAD1 and LGAD2 provide an event by event trajectory for entering particles.  It is conceivable that they can be rotated out of the beam for production running at the lowest operating momenta.}
\label{fig:atar}
\end{figure}

\begin{figure}[htb]
\centering
\includegraphics[width=0.9\textwidth]{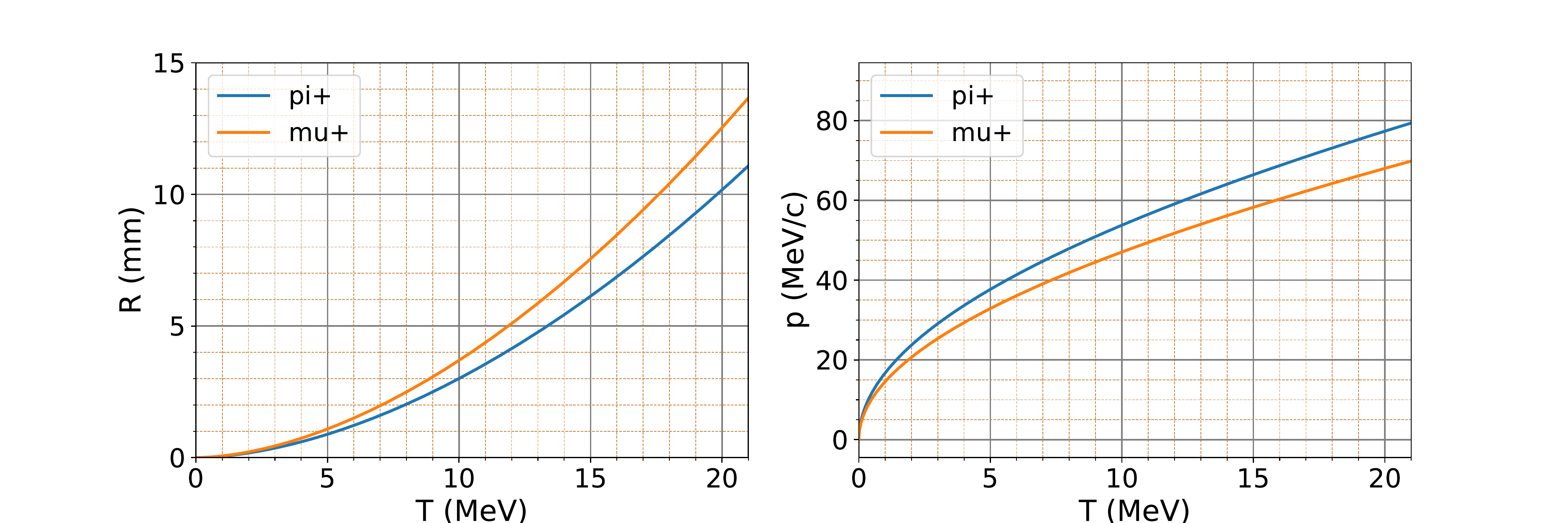}
\caption{ Range-energy (left) and  momentum-energy (right) relations for pions and muons. Muons from \pdar\ have 4.12 MeV kinetic energy and travel 
0.8 mm in Si.}
\label{fig:r-p}
\end{figure}

Figure~\ref{fig:atar} sketches the baseline set-up for the beam-target section of the experiment. For a more complete description of the overall \atar-\calo\ design we refer
to Sec.~\ref{sect:Mech-Eng}.
Pions selected by an E$\times$B separator pass through a beam window of $70\,\mu$m Mylar, two LGAD detector planes (two crossed layers of $50\,\mu$m thick AC LGADS, each, denoted as BEAM in this proposal)
to stop in the ATAR. 
No degrader or upsteam LGAD detectors may be necessary if a
suitable beam at p=55\,MeV/$c$ can be established. 
As sufficient beam intensity is expected, the beam spot may be improved by
collimation just upstream of the detector entrance. Collimators would also serve to suppress muons and electrons displaced by the angular deflection in the separator but
their design has to be carefully optimized between the competing requirements of collimating close to the target focus while absorbing positrons and gammas before they reach the
\calo. Some important aspects of the stopping beam can be understood from the pion range-energy relations shown in Fig.~\ref{fig:r-p}.\footnote{These were
calculated by SRIM (http://www.srim.org/) to describe the  slowing down of particles at low energies.}. 

A momentum of 55\,MeV/$c$ has several advantages including that the separator will work more efficiently
and a range width $\Delta Z$ of less than 0.4\,mm is expected for a 2\% $\Delta$p/p beam momentum acceptance. (The stopping range width would increase to 0.8\,mm at 70\,MeV/$c$.)
The range-energy curve also shows how the separation between pions and muons of the same energy improves with the energy deposited in \atar.

\subsubsection{Beam line at $\pi$E5}
The PSI High Intensity Proton Accelerator (HIPA) chain consists of a Cockcroft-Walton pre-accelerator, delivering 870\,keV protons to a set of two isochronous cyclotrons which further accelerate to the maximum 590\,MeV. 
In the Ring cyclotron, the protons are accelerated by four copper resonator cavities operated in continuous wave (CW) mode at a frequency of 50.6 MHz, giving the periodic 19.75\,ns beam microstructure and their final energy of 590\,MeV.
The $\pi$E5 channel is a high-intensity low-energy muon and pion beamline, with a maximum momentum of 120\,MeV/$c$, that views the second graphite production target, Target~E, at 165$^{\circ}$ with respect to the proton beam axis.
The $\pi$E5 channel is used exclusively for particle physics experiments, and is currently home to the MEG~II and Mu3e experiments.
The channel length from Target~E to end of the QSB43 quadrupole magnet and the entrance to the experimental area is 13\,m, giving approximately 7\,m of free space between the concrete wall and current MEG~II infrastructure.
In this area the SEP41 E$\times$B Wien Filter for particle separation, QSK quadrupole triplet for focusing the beam on the \atar\, and the PIONEER installation could be placed.
\begin{figure}[h]
	\centering
	\includegraphics[width=1\linewidth]{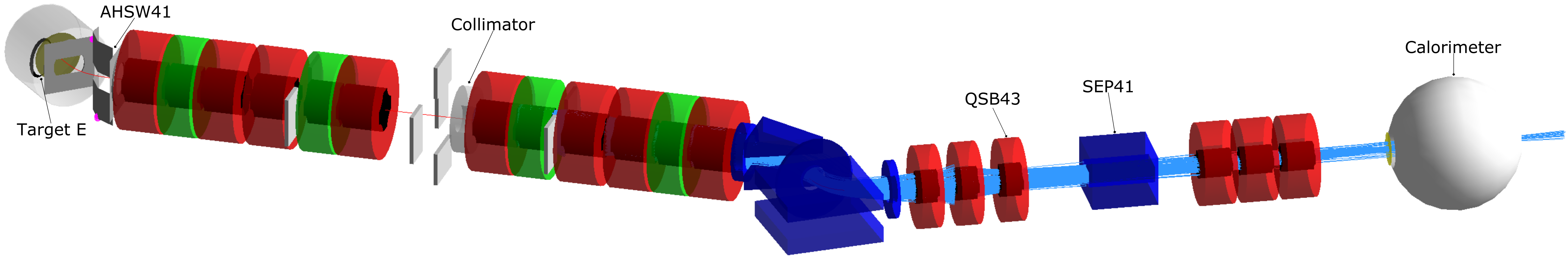}
	\caption{The \gbl\ model of the $\pi$E5 beamline including a stand-in calorimeter.}
	\label{fig:g4blsim_pie5_labeled}
\end{figure}
\subsubsection{Beam optimization and studies} \label{sec:beam_studies}

Simulations for both pion production at Target~E and transport of particles down the $\pi$E5 channel were performed using the \gbl\ toolkit ~\cite{Roberts:2008zzc}.
A figure of the $\pi$E5 beamline from the \gbl\ simulation, beginning at Target E and ending at the PIONEER calorimeter, is shown in Fig.~\ref{fig:g4blsim_pie5_labeled}.
The pion production simulation utilizes the parameterized cross sections developed for the High Intensity Muon Beamline (HiMB) project \cite{PhysRevAccelBeams.19.024701}.
All simulation results are scaled to the nominal 2.2~mA proton beam current at PSI.

\begin{figure}[htb]
	\centering
	\includegraphics[width=.8\linewidth]{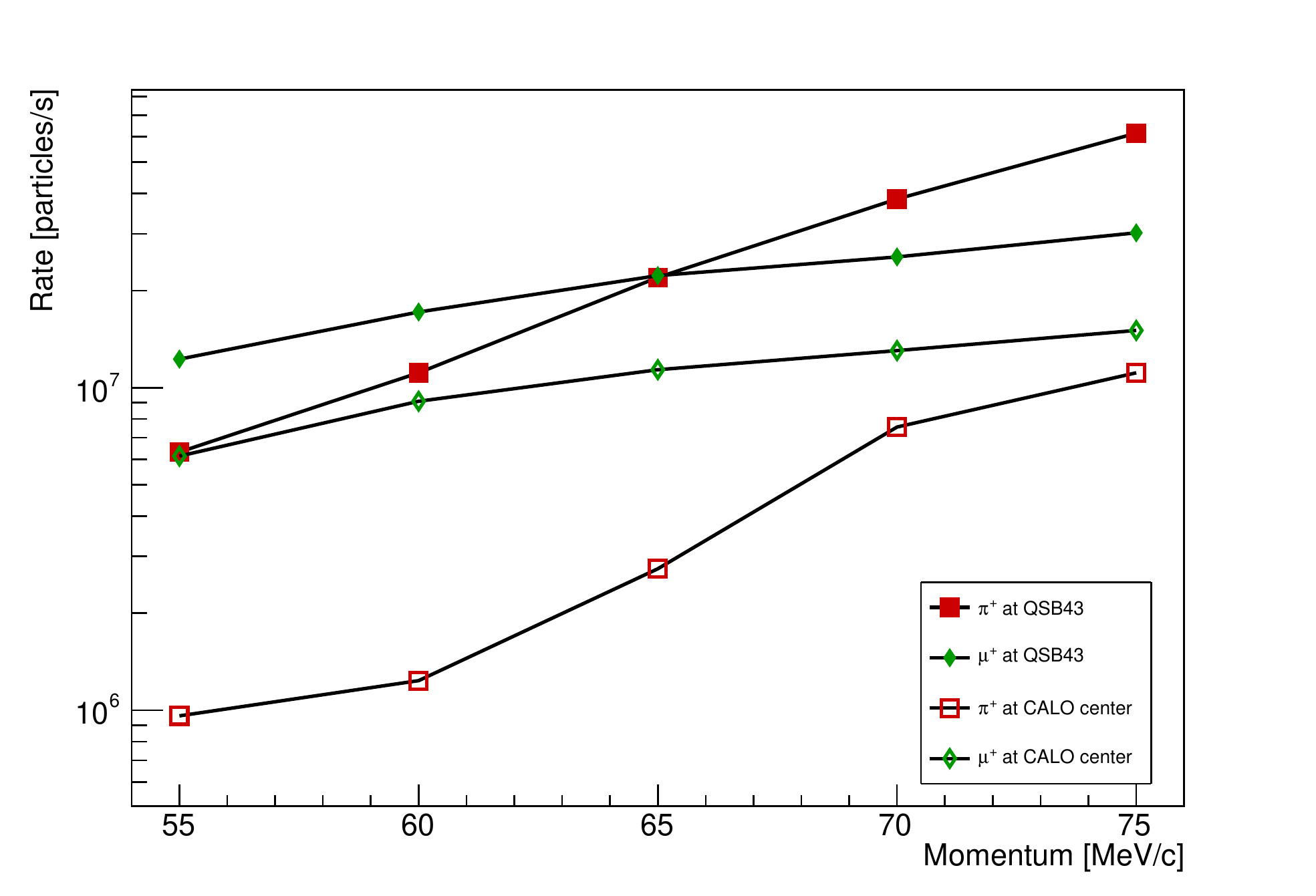}
	\caption{The rates for pions and muons at the entrance to the experimental area (QSB43) and reaching the center of the calorimeter, calculated in the $\pi$E5 \gbl\ simulation.}
	\label{fig:piE5rates}
\end{figure}

The \gbl\ simulation was originally tuned for 28\,MeV/$c$ muons and has been rescaled for the respective pion momenta.
Some additional tuning was performed for the last two dipoles and quadrupole triplets to optimize transmission through the SEP41 and on to the \atar.
The physical volume of the SEP41 is included but its electric and magnetic fields are not enabled for these simulations.
Rates for pions and muons reaching the entrance to the experimental area (QSB43) and 4.8\,m downstream at the center of the calorimeter volume (\calo\ center) for beamline central momenta ranging from 55 to 75\,MeV/$c$ are shown in Fig.~\ref{fig:piE5rates}.
The rates for those particles at the lower and upper end of the possible momentum ranges are listed in Table~\ref{tbl:rates}.
The pion rates calculated in the simulation at the center of the calorimeter for a 2\% momentum-bite 
are sufficient for the \remu\ measurement planned in Phase I of PIONEER, even when
further losses in the SEP41 or upstream collimation, necessary for reducing background events in the \atar\ and calorimeter, are included.
For PIONEER Phase II, III increasing the momentum and the momentum bite will be required.

\begin{figure}[htb]
	\centering
	\subfloat{
    \includegraphics[width=0.4\textwidth,height=0.4\textwidth]{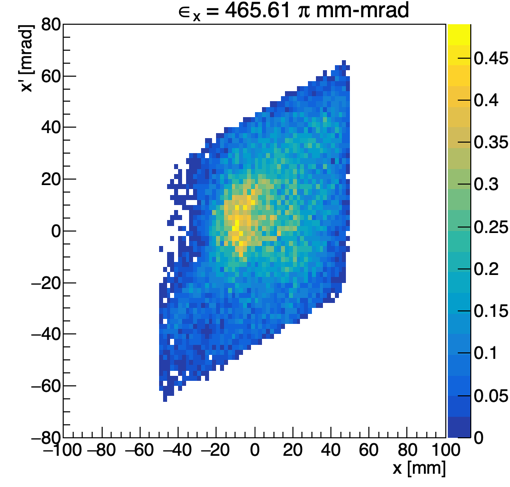}}
    \qquad
    \subfloat{
    \includegraphics[width=0.4\textwidth,height=0.4\textwidth]{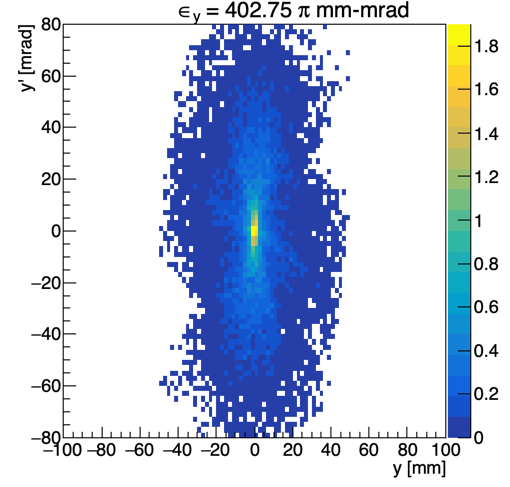}}
	\caption{Preliminary pion phase space distributions at the \calo\ center position for 65~MeV/$c$ pions with decay disabled from the \gbl\ simulation of $\pi$E5. Left: Horizontal. Right: Vertical. The calculated emitance $\epsilon$ for each is included in the plot as text. The non-Gaussian structure of these phase spaces present a challenge in correctly modeling the beam transport.}
	\label{fig:phasespace}
\end{figure}

\begin{center}
\begin{table}
\begin{tabular}{p{4cm}|p{2cm}<{\centering}|p{2cm}<{\centering}|
                p{2cm}<{\centering}}
    \toprule
	Beamline Position& $p_{\pi}$ (MeV/$c$) & $\pi^{+}$ Rate&$\mu^{+}$ Rate \\
	\hline
	QSB43           & 55 & 6.3 & 12.3 \\
	\calo\ Center   & 55 & 1.0 & 6.1 \\
	\hline
	QSB43           & 75 & 61.5 & 30.2 \\
	\calo\ Center   & 75 & 11.1 & 15.1 \\
	\bottomrule
\end{tabular}
\caption{Particle Rates ($\times 10^{6}$ $s^{-1}$) in $\pi$E5 for a 2\% momentum-bite calculated in the \gbl\ simulation. 
}\vspace{-3mm}
\label{tbl:rates}
\end{table}
\end{center}
A very preliminary study to focus the beam onto the calorimeter center positions by simulating 65~MeV/c pions with their decay disabled is displayed in Fig.~\ref{fig:phasespace}. The horizontal ($x-x^{\prime}$) phase space indicates a pronounced non-Gaussian nature, while the vertical ($y-y^{\prime}$) phase space demonstrates a well defined focus. The complex structure of of the horizontal phase space emphasizes that full \gbl\
simulations are required to tune the beam and, in particular, experimental verification is indispensable to achieve the desired small beam spot 
on \atar. At the moment, there is no readily available experimental data that could verify our preliminary simulations for $\pi^+$ beams. The above used simulation code was verified so far using surface muon beams of 28 MeV/c momentum only. The last experiment using pion beams at $\pi$E5 was the PiHe experiment that needed high $\pi^-$ rates for pionic helium spectroscopy~\cite{Hori:2020iom}.

Key experimental parameters, including the design of the beam conduit into the calorimeter and the geometry of the \atar\, rely on a detailed understanding of the $\pi^+$ beam properties at $\pi$E5 at the targeted momentum range (55-75\,MeV/$c$).  
The most important parameters to be determined are the $\pi^+ / \mu^+ / e^+$ rates as a function of beam momentum, the momentum bite $\Delta p/p$, the final focus and the capability to select a clean $\pi^+$ beam by applying a Wien (E$\times$B) filter.
 Measurements of available $\pi^{+}$ rates and comparison to simulation will be critical in determining achievable pion rates at the \atar.
For this purpose we propose to carry out detailed measurement program of the $\pi$E5 pion beam at PSI, adapting SiPM-based beamline instrumentation that was used to characterize pion and muon rates in the past. Because
of their urgency in informing the detector design we request a first study in 2022 (beam request submitted separately).



\subsection{Active target (ATAR)}
\label{sec:ATAR}

\begin{figure}[htbp]
\centering
\vspace{-20mm}
\includegraphics[width=0.9\textwidth]{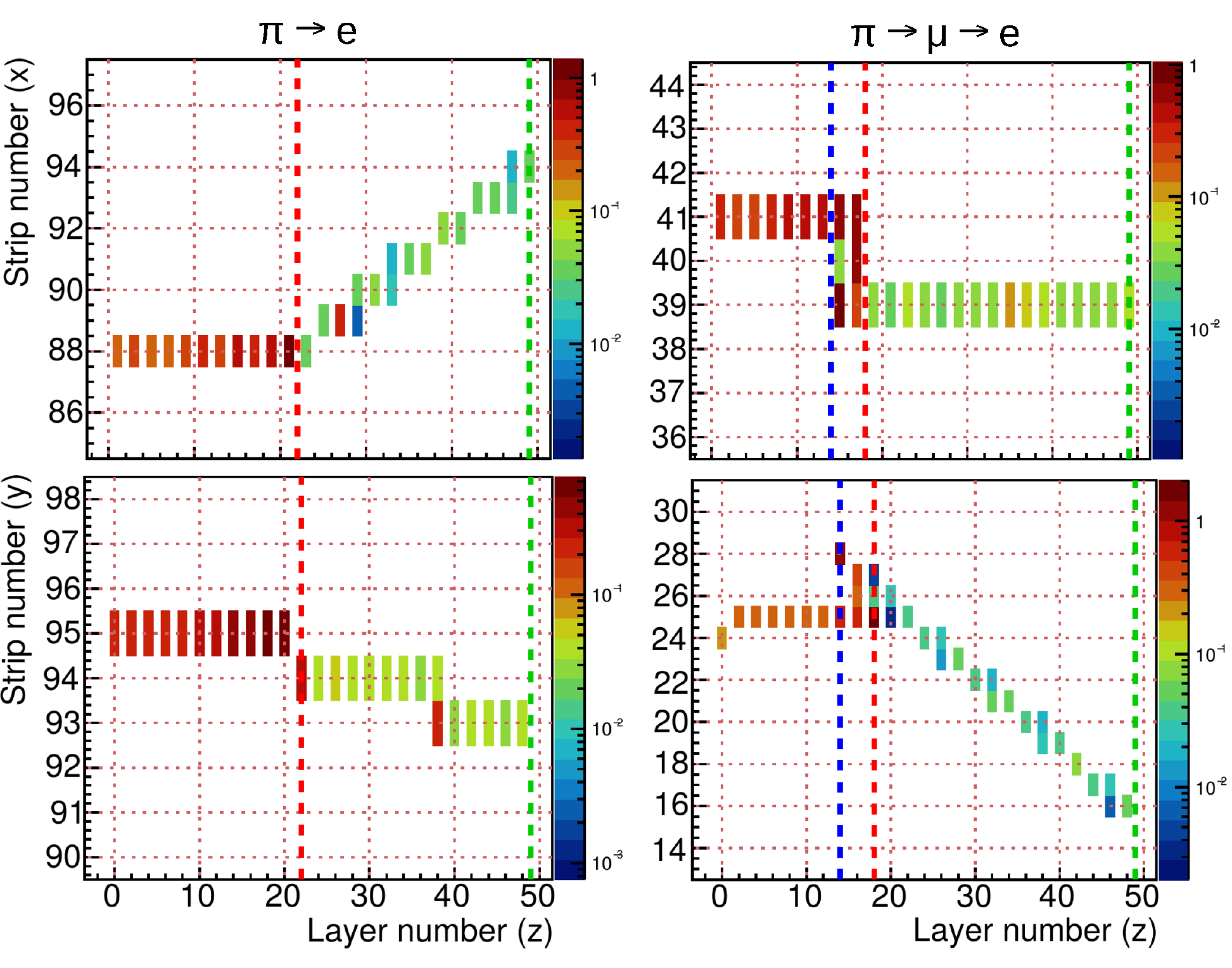}
\caption{Simulated example displays of pion decay events in the \atar. Pions enter horizontally from the left; the  red dotted lines show the positions of the pion stops.
The color of the bars indicates the deposited energy. Left column: X-Z (top) and Y-Z (bottom) strip  views of a \pie\ event.  Right column: Same views of a \pme\ event. The  blue dotted line shows the position of the decay muon stop.}
\label{fig:Event_display}
\end{figure}

A highly segmented active target (\atar)~\cite{Mazza:2021adt} is a key new feature of the proposed PIONEER experiment;  technical details and alternative designs of the sub-detector can be found in Appendix~\ref{sec:ATAR_appendix} with a detailed timeline described in Appendix~\ref{sec:ATAR_timeline_app}. 
The \atar\ will define the fiducial pion stop region, provide high resolution timing information,
and will furnish selective event triggers. Examples of event displays for \pie\ and \pme\ events are shown in Fig.~\ref{fig:Event_display}.

For the measurement of the calorimeter response to \pie\ decays,  the \atar\  will suppress the $\pi \rightarrow \mu \rightarrow e$ decay at rest (\pdar), and pion and muon decay-in-flight (DIF)  low energy  backgrounds.
DAR events are identified by a good pulse pair resolution observing the \tmu\ $\pi\to\mu$ decay, and a tight positron observation window of about one pion lifetime.  \pdif\ events are  identified  by kinks in the trajectories, \dedx measurements along the track, and observed range in the target. 
Furthermore, the \atar\ will suppress accidental muon stops that precede the trigger signal; these were a significant source of pile-up background in the previous experiments. 
Confirmation that  the observed positron belongs to the pion stopping vertex will also be obtained using tracking in the \atar\ and in additional tracking detectors (discussed in Sec.~\ref{sec:positron_detectors}).

To fulfill these goals the \atar\ must be able to detect both the exiting $e^+$, a minimum ionizing particle (MIP), and larger ($\thicksim$100 MIPs) energy deposits from $\pi^+$ and $\mu^+$.
The large dynamic range O(2000) of the signals is a significant challenge for the readout electronics  in the amplification and digitization stages.
The position of the energy deposits needs to be identified with sufficient granularity along the beam direction and in the transverse plane.
Furthermore, to identify single components of the decay processes a \SI{\sim 1.5}{\ns} pulse pair resolution is needed.

\begin{figure}[htbp]
\centering
\includegraphics[width=0.4\textwidth]{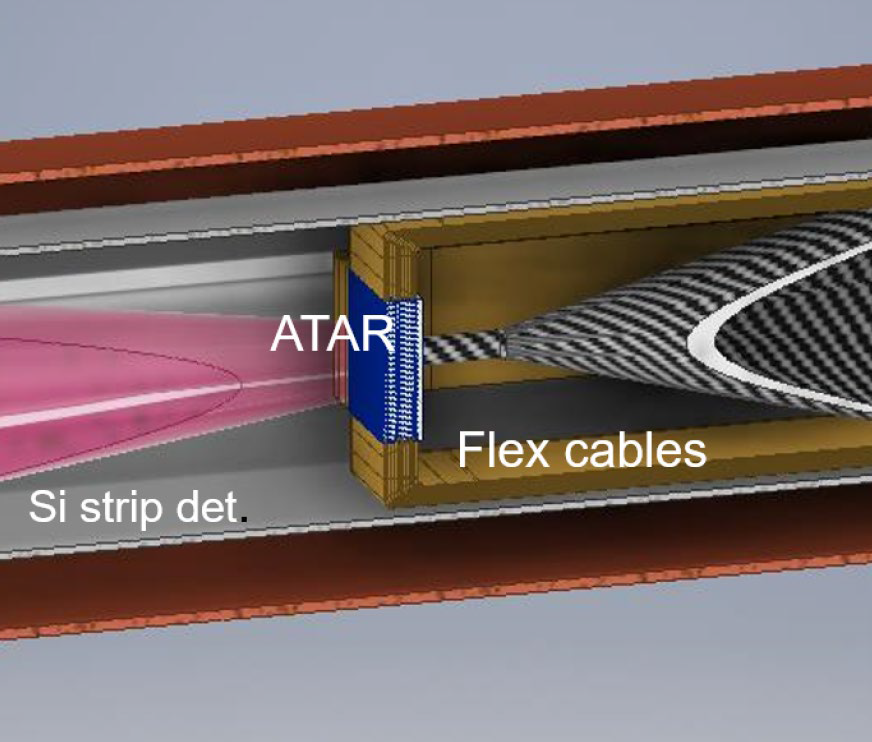}
\includegraphics[width=0.55\textwidth]{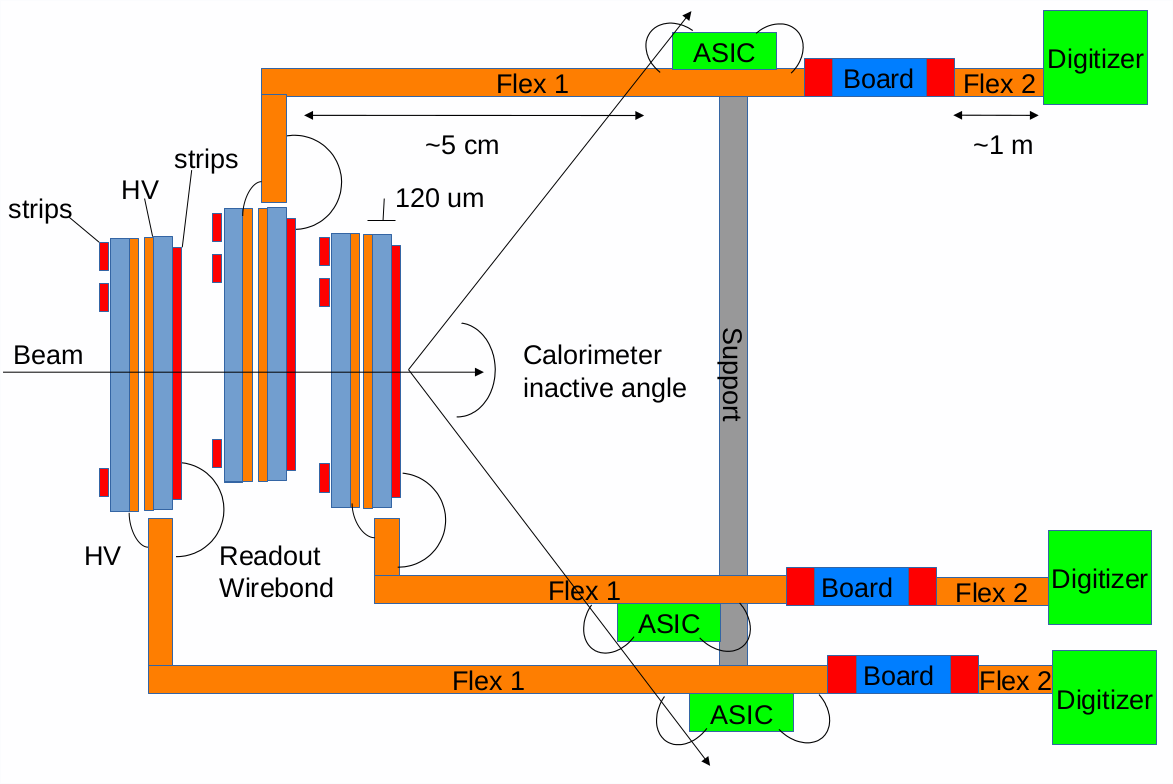}
\caption{Left: \atar\ position in the beam line. Right: Concept schematic design of the \atar. The flex from the first, third and fifth sensors is directed in and out of the page. The modules are attached on the HV side and with a few  \si{\micro\metre} of separation on the strip side. 48 sensors are coupled in 24 pairs. }
\label{fig:ATAR_scheme}
\end{figure}

The \atar\ tentative design dimensions are 2$\times$2\,\si{\cm\squared} transverse to the beam. In the beam direction individual silicon sensors are tightly stacked with a total thickness of \SI{\sim 6}{\mm}.
A strip geometry with the electronic readout connected on the side of the active region via wire bonding is foreseen.
The strips are oriented at \ang{90} to each other in subsequent staggered planes to provide measurement of both coordinates of interest and to allow space for the readout and wire bonds.
The preliminary sensor geometry has strips with a pitch of \SI{200}{\micro\metre}, so that a sensor would have 100 strips mated to a chip with 100 channels and \SI{2}{\cm} width, a standard dimension for microchips. 
A best estimate at present for the sensor thickness is around \SI{120}{\micro\metre} to avoid support structures for the sensor, which would introduce dead area. 
To reach a total thickness of about \SI{6}{\mm}, $\thicksim$50 planes are needed, so in total the \atar\  would comprise $\thicksim$5000 channels.
The detectors are paired with the high-voltage facing each other in a pair to avoid ground and high voltage in proximity. 
In the preliminary design, shown in Fig.~\ref{fig:ATAR_scheme}, the strips are wire bonded, with a connection alternating on the four sides, to a flex that brings the signal to a readout chip positioned a few cm away from the active volume.
This removes the chip from the path of the exiting positrons, reducing its degradation of their energy resolution.
The flex will be produced with aluminium traces so that the cable material per flex would be \SI{\sim 0.025}{g/\cm\squared}. 
In the figures in this proposal, all cables are shown exiting downstream.
A readout from both ends is being investigated to reduce the average material traversed by exiting positrons. 
At present, the maximum material in the path of the positrons occurs when 12 flexes are traversed. 
The readout ASIC sits on the first flex that tapers out to accommodate the additional traces. 
Then the flex is connected, via connector, to a PCB connected to a second flex that brings the amplified signal to the digitizers in the back end.

The chosen technology for the \atar\ is Low Gain Avalanche Detector (LGAD)~\cite{bib:LGAD} (detailed description in Appendix~\ref{sec:LGAD_app}), thin silicon detectors with moderate internal gain.
A \SI{120}{\micro\metre} thick LGAD sensor, coupled to fast electronics, has a time resolution of less than \SI{100}{\ps} on the rising edge and can separate a single hit from two overlapping hits if they arrive more than \SI{1.5}{\ns} apart.
Preliminary studies made with X-rays coming from the Stanford light source (SSRL)~\cite{GALLOWAY20195} and PSI~\cite{Andra:ay5534} show that LGADs have an energy resolution of around \SI{10}{\percent}; more detailed results are reported in Appendix~\ref{sec:LGAD_app}.
The effect of gain saturation that was reported in the community in the past year~\cite{gainsuppr} also needs to be studied.
The LGAD technology was chosen over standard silicon technology because of the intrinsic gain and thin bulk. 
Normal silicon detectors without gain need 300--500\,\si{\micro\metre} of active silicon to have enough collected charge for high MIP detection efficiency.  This would degrade granularity in the beam direction, time resolution and pulse separation discrimination. 
A brief discussion on alternative technologies is presented in Appendix~\ref{sec:LGAD_app}.

Current standard LGADs have a limitation in terms of granularity and active area. To achieve a \SI{\sim 100}{\percent}  active area several technologies still at prototype level are being evaluated for \nexp, such as AC-LGADs~\cite{Apresyan:2020ipp}.
Studies were made on strip AC-LGAD prototypes from BNL as shown in Appendix~\ref{sec:ACLGAD_app}.
Results on sensor simulation  are also reported in Appendix~\ref{sec:ACLGAD_app} and
additional LGAD technologies are being evaluated as reported in Appendix~\ref{sec:TILGAD_app}.

To read out the \atar\ sensors two crucial electronic components need to be identified: an amplifier chip and a digitization board.
Preliminary tests on  available solutions are reported in Appendix~\ref{sec:ATAR_electronics_app}. 
Since the amplification chip has to be positioned away from active region, the effect of placing a short (\SI{5}{\cm}) flex cable between the sensor and the amplification stage will be studied.
To successfully reconstruct the decay chains the \atar\ is expected to be fully digitized at each triggered event. 
The fast charge collection time of thin LGADs will allow separation of subsequent charge depositions as mentioned before.  
To achieve this goal a high bandwidth digitizer with sufficient sampling rate has to be used. 
A discussion on  available amplifier chips and digitizers is reported in Appendix~\ref{sec:ATAR_electronics_app}. 

Relative to previous experiments, the use of a highly segmented active target allows for discrimination against backgrounds by looking for deposition patterns internal to the target that are associated with the various signals' decay topologies.

\subsection{Cylindrical Tracker}
\label{sec:positron_detectors}
A dual layer cylindrical silicon strip tracker is situated between the \atar\ and the calorimeter to measure the  positron position in two dimensions (along the beam direction, $z$, and azimuthal angle, $\phi$), and time, as indicated in Fig.~\ref{fig:atar}.
The detector has  an inner diameter of \SI{5}{\cm} and a length of \SI{25}{\cm}. 
The readout ASIC will be wire bonded at a location outside  the calorimeter active solid angle.
Overlapping lengths of long strips (about \SI{10}{\cm}) are needed to cover the entire region.

Two layers of strips with a small stereo angle  between them will provide 
O(mm) $z$ resolution and \SI{300}{\micro\metre} resolution in the direction perpendicular to the strips.
An alternative under consideration is to connect  two or three strip sensors in a line reading out both ends to obtain O(mm) position information along the strip position using charge attenuation information.

The silicon strip sensors may be constructed with  either regular silicon or LGADs. 
Regular silicon would have \SI{\sim 300}{\micro\metre} of thickness for positron detection with a time resolution \SI{< 1}{\ns}, while LGADs can be \SI{50}{\micro\metre} or less with a time resolution of \SI{< 50}{\ps}.
However, LGAD silicon strips with length of 10\,cm present an additional challenge
while for the ATLAS detector standard silicon strips with \SI{\sim 10}{\cm} of length have been successfully fabricated on 8-inch wafers \cite{CERN-LHCC-2017-005}.
\subsection{LXe Calorimeter} 
\subsubsection {Overview}

\label{cal}

Due to its fast timing properties, high light yield with excellent energy resolution and highly uniform response, liquid xenon (LXe) 
read out by UV sensitive phototubes and state-of-the-art VUV SiPMs is considered for the calorimeter. Here, experience is drawn from the  MEG \cite{MEG:2016leq} and  MEG-II \cite{MEG:2018vi} experiments which use a large scale,  high rate  LXe detector to search for the lepton flavor violating muon decay, $\mu^+\rightarrow e^+ \gamma$.
 Experiments searching for the elusive dark matter (e.g. XENON, LUX-ZEPLIN) and (neutrinoless) double beta decay events (KamLAND-Zen, (n)EXO) also use detectors with similar scale liquid xenon cryostats. Unlike those latter experiments, MEG, like PIONEER, only detects scintillation light (other experiments rely on both scintillation and charge collection) and is a high rate experiment. In MEG, the rate of muon stops in the target was $\unit[3\times10^7]{\mu^+/s}$ and the LXe calorimeter rate due to photons was $\mathcal{O}$(\unit[0.3]{MHz}). The anticipated PIONEER calorimeter rate (due to  pion stop rate of 300\,kHz  in the ATAR for the $R_{e/\mu}$ measurement) is of the order handled by the  MEG LXe calorimeter. The main difference comes from the nature of the particles entering the calorimeter (gammas in the case of MEG and positrons for PIONEER) and the geometry  of the calorimeter leading to different event ``signatures''. The PIONEER LXe detector is foreseen to be a 25 radiation length, 3$\pi$~sr sphere surrounding the ATAR. A discussion on the capability of a large single volume to handle pileup events is given in Sec.~\ref{sec:calo_simu}.

 The homogeneity of the LXe detector is an advantage in achieving the high energy resolution which is important for determining accurately the low energy ``tail'' fraction of $\pi\rightarrow e \nu$ events. MEG  currently reports  an energy resolution of $\sigma=1.8\%$  for 50\,MeV gammas and they continue to study possible improvements. The baseline energy resolution goal for PIONEER at 70\,MeV is 1.5\%.
 
 The  energy resolution is impacted by the efficiency of collection of scintillation light which is itself influenced by design parameters (like photo-sensor coverage) and physical or technical parameters (like the light attenuation due to impurities in LXe, reflection of VUV light on surfaces, photo sensor refraction index, the level of dark current which impact the photo-electron threshold for summing the energies of the different photo-sensors, etc). In addition to the finite energy resolution of the calorimeter, photo-nuclear interactions, shower leakage and geometrical acceptance contribute to the low energy tail. The impact of photo-nuclear effects in LXe, for which little literature exists, will be determined by simulation and bench-marked against available data and new measurements. 
 
\begin{figure}[h!]
\centering
\includegraphics[width=0.9\textwidth]{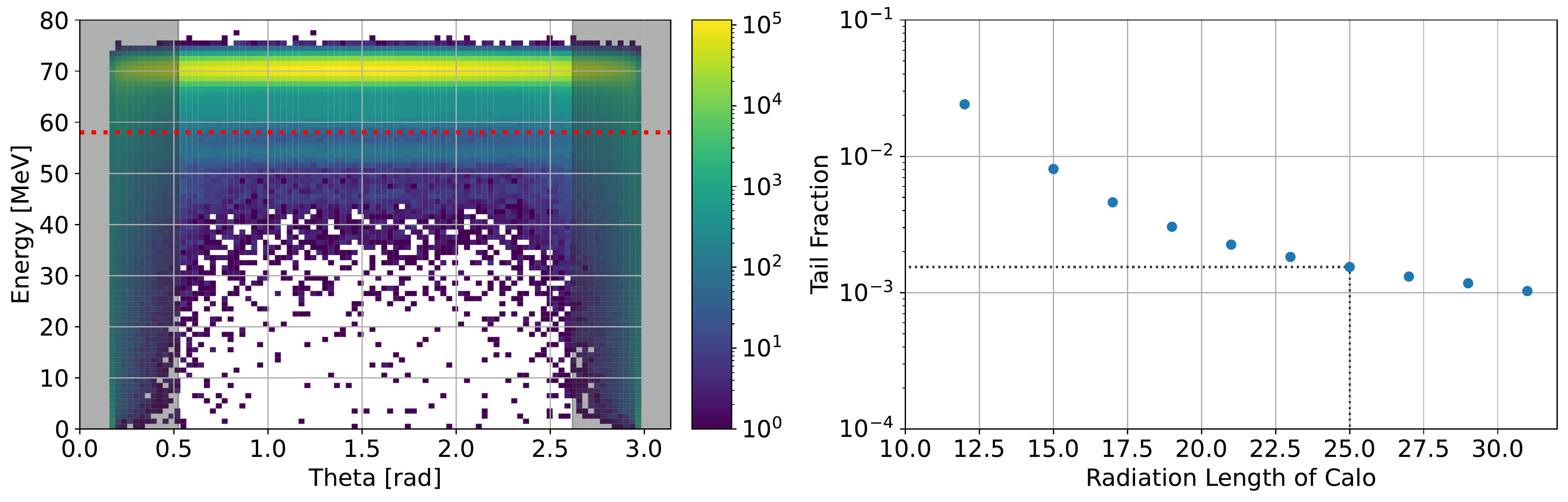}
\caption{Left:  The energy deposited by monoenergetic 69.3\,MeV $e^{+}$ from \pie~ decays in a 25\,$X_0$ calorimeter with an energy resolution of 1.5\% vs. the angle Theta with respect to the beam axis.  The grey bands indicate the boundaries of the fiducial volume region (here, $30^{\circ}$).  Right: The shower tail fraction below 58\,MeV vs. the calorimeter depth in radiation lengths for the 69.3\,MeV $\pi \rightarrow e \nu$ events. 
}
\label{fig:tail_fraction}
\end{figure}

A GEANT-4 simulation of a bare LXe calorimeter geometry was used to determine the residual tail fraction below the Michel end point versus calorimeter depth for \pie~ events.  Figure~\ref{fig:tail_fraction}, Left, shows the energy deposited in the spherical calorimeter vs. the angle Theta with respect to the beam axis for a 25\,$X_0$ calorimeter depth.  Figure~\ref{fig:tail_fraction}, Right, shows the fraction of the energy deposited that is below  58\,MeV  vs. depth.  The volume of LXe required scales as the radius cubed and the required photo-sensor coverage scales as the radius squared.  These practical factors are optimized for smaller depth.  The containment of the shower slowly improves with increased depth.
The depth was chosen to be 25\,$X_0$, which we use throughout this proposal.   
This choice, along with the spherical shape and $>\unit[3\pi]{sr}$ coverage, limits the effect of secondaries being missed by the calorimeter (e.g.~Bhabha scattered events) and will significantly reduce the contribution of shower leakage. We will continue to carefully model the calorimeter geometrical parameters and assess the expected detector sensitivity; see discussion in Sec.~\ref{sec:calo_simu}. 

\begin{figure}[h!]
\centering
\includegraphics[width=\textwidth]{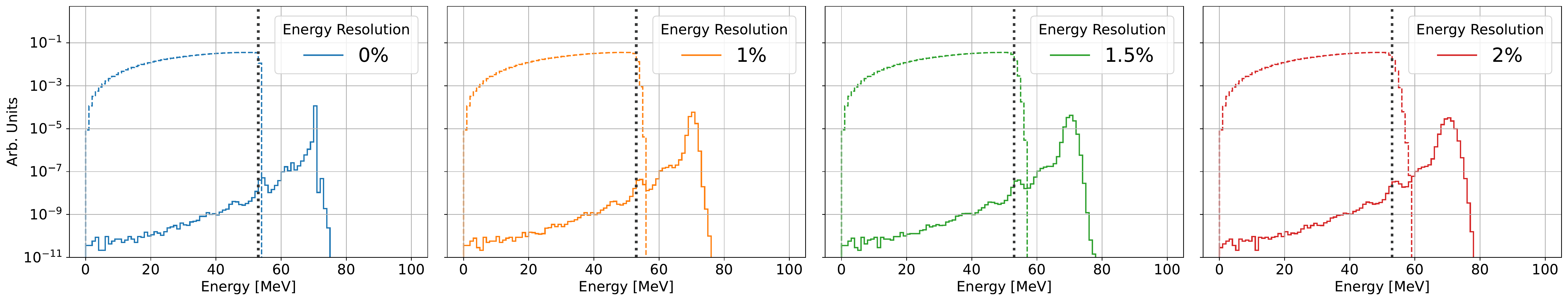}
\caption{The effect of energy resolution on a 25\,$X_0$ calorimeter. The events indicated with a solid line are monoenergetic 69.3\,MeV $e^+$ from $\pi \rightarrow e \nu$ decay. The events indicated with a dashed line are the Michel spectrum from muon decay. The dashed grey line at 53 MeV represents the nominal endpoint of the Michel spectrum.
}
\label{fig:resolution_scan}
\end{figure}

Figure \ref{fig:resolution_scan} shows the energy spectrum for a 25\,$X_0$ calorimeter of varying intrinsic resolution; these results indicate that the tail correction is relatively insensitive to resolution. 
We are aiming for  resolution $\sigma=1.5\%$, based on the performance of the  MEG calorimeter. 

 The xenon scintillation light absorption length has been  measured for MEG to be more than \unit[1]{m} \cite{Baldini:2004ph}.
 However, the light absorption length can be significantly reduced through absorption by impurities such as H$_2$O and O$_2$.
A large scale detector requires purification at the ppb level  which was achieved by MEG.
The purification system and the cryostat design (needed to maintain the xenon at \unit[165]{K} \cite{Mihara:2011zza}) of MEG will be considered for scaling up the design for PIONEER (see Sec.~\ref{sec:LXe_Cryogenics-purification}).

 MEG-II performed extensive R\&D in collaboration with Hamamatsu to develop large MPPC (Multi-Pixel Photon Counter) detectors ($12 \times$\SI{12}{\milli \meter \squared}) while maintaining optimal performance characteristics of  a small decay constant and good gain uniformity.
Efficient detection of VUV light requires careful choice of the silicon layer assembly and coatings. For their chosen sensor, the MEG-II collaboration reported a photon detection efficiency (PDE) of $\sim20\%$ \cite{MEG:2018vi} but observed later degradation of the PDE in LXe \cite{meg2sipm} which is under investigation. Below we discuss details of  photosensor options considered for PIONEER.

\subsubsection {Photosensors}\label{sec: LXe_photosensors}

In order to readout the liquid xenon scintillation light, special photo sensors are necessary due to the short scintillation wavelength of 175\,nm and the low operation temperature of 165\,K. Two types of photo sensors were developed for this purpose by Hamamatsu K.K. and the MEG/MEG II collaboration: 2-inch photomultipliers (PMT) and multi-pixel photo sensors (MPPC).
A round-shaped 2-inch PMT sensitive to VUV light (R9869) has been developed using
Bialkali  (K-Cs-Sb) photocathode and a synthetic silica window. It achieved a quantum efficiency of \SI{15}{\percent}. Aluminum strips were added in the photocathode to prevent an increase of the sheet resistance at low temperature. 

In order to improve the granularity for the inner surface in the MEG-II experiment, a silicon photomultiplier, called MPPC (S10943-4372) with an active area of 12$\times$12\,mm$^{2}$, was developed. The main developments consisted in removing the protective layer, matching the optical properties of the silicon surface to that of the LXe, and reducing the thickness of the insensitive layer. Besides achieving higher granularity, there are several advantages to using SiPM with respect to photomultiplier tubes: they are insensitive to magnetic fields, the single photoelectron peak can be used for calibration of the sensor and the required supply voltage is relatively low (less than \SI{100}{\volt}).

However, as mentioned above, the PDE of the SiPMs were found to drastically decrease under exposure to the muon beam. Investigations are ongoing to understand the causes. It was however shown that the PDE could be recovered through an ``annealing'' procedure performed during the accelerator shutdown period. 
The gain of the PMTs was also found to decrease due to damage to the dynode induced by photoelectrons. The PMT lifetime is however sufficiently long and thus should not be of concern for PIONEER.

Considering the above, the baseline design for PIONEER is the use of PMTs on the outer surface targeting a coverage of \SI{20}{\percent} (1000~phototubes) of the surface. The choice of the photosensor technology might evolve depending on the developments regarding  SiPM performance degradation.

Simulations will be performed to study the impact on track identification capabilities and energy resolution  of adding SiPMs on the inner surface regions of the calorimeter. The ATAR is small (\SI{130}{\milli\meter}) 
and surrounded by a positron tracker which can measure the positron's incident position right before it enters the LXe calorimeter. In these conditions the addition of reflective material coatings on the inner surface may be sufficient for the PIONEER needs. However, it is notably difficult to accurately simulate UV optical properties on large scales. Agreement between data and photon transport simulations are sometimes poor, mainly because of the large number of optical interfaces, i.e., surfaces with a certain shape, absorbance and reflectance, that are not well known for the UV light from LXe (optical properties are indeed most often measured in vacuum). Apart from differences in refraction indices between vacuum and LXe, LXe may also alter the optical properties of some materials (e.g. \cite{Silva:2009ip, LZ:2015kxe, Neves:2016tcw}). It is thus particularly important to perform in-situ measurements in LXe to evaluate both the performance of SiPM in the LXe environment and deduce the appropriateness of surfaces for coatings (as well as possibly segmentation for pile-up rejection, see discussion in Sec.~\ref{sec:calo_simu}). For this purpose we have planned measurements at facilities where  LXe setups are  available.



\subsubsection{Cryogenics and Purification}\label{sec:LXe_Cryogenics-purification}
\label{cryo-puri}

A system consisting of the following equipment is required in order to stabilize a large volume of liquid xenon in the detector:

\begin{enumerate}
\item High-pressure gas tanks for xenon storage,
\item Gas xenon purification system,
\item Liquid xenon purification system,
\item Refrigeration,
\item Temperature and pressure control system.
\end{enumerate}

Piping for transporting xenon either in the gas or liquid phase is required as appropriate. In addition to these, a cryostat that can temporarily hold xenon in a liquid state during emergencies or maintenance ensures efficient operation of the detector. \\

\ListProperties(Start1=1)
\textbf{1. High-pressure gas tanks for xenon storage}:
Xenon is stored in high-pressure gas tanks at room temperature during  detector maintenance periods. A storage tank that can withstand a pressure of 5.8 MPa at 17$^\mathrm{o}$C (1.1g/cm$^3$, the critical point of xenon) is required. After detector operation, the xenon is recovered in the storage tank by cooling the tank with liquid nitrogen. This requires a liquid nitrogen bath to cool the high-pressure gas tank.

\textbf{2. Gas xenon purification system}:
When operating the LXe detector, it is necessary to reduce the xenon gas pressure provided from the high-pressure tanks and then to remove impurities in the gas before filling in the detector cryostat. Metal-heated getter purifier technology has been used in previous experiments and will be employed also in the PIONEER experiment.

\textbf{3. Liquid Xenon Purification System}:
Based on the experience of the MEG experiment \cite{MEG:2016leq}, it is known that even if the xenon gas is liquefied in the detector after impurity removal by the gaseous-phase purification system, impurities remaining on the detector surface may dissolve into the liquid, causing absorption of scintillation light and degradation of detector performance. Among these impurities, water is the one most likely to affect the propagation of liquid xenon scintillation as its absorption cross section overlaps with the xenon scintillation light spectrum. In order to efficiently remove the water dissolved into the liquid, the xenon is circulated in the liquid phase by a low-temperature liquid pump through a filter containing molecular sieves with a pore size 13A, which is suitable to remove water molecules. In the MEG experiment, 70L/hour circulation is achieved with this system, resulting in preparing sufficiently pure liquid in a reasonable time.

\textbf{4. Small refrigerator}:
As used in the MEG experiment, a small refrigerator installed on the top the detector is used to liquefy xenon and re-liquefy the constantly evaporating xenon gas. The use of a pulse-tube refrigerator is suitable from the viewpoint of low vibration noise, but to obtain a high refrigeration power, a GM-type refrigerator can be installed at a distance from the detector with a vacuum insulated pipe to transfer the liquid to the detector. Both types of refrigerators have been used in MEG, and there are no technical difficulties.

\textbf{5. Temperature and pressure control system}:
The cryostat is always subject to heat by conduction through the support structure and cables as well as heat generation from electronics circuits and photo sensors placed in the cryostat. After liquefaction completes, the liquid xenon evaporates continuously due to the heat generated by these, and the pressure rises. In order to control the temperature and pressure of the liquid xenon, xenon is re-liquefied by the refrigerator. In order to maintain the liquid state at a temperature and pressure suitable for operation, the refrigerator cooling power is controlled by changing the heater output on the cold head while the refrigerator continues to operate at maximum capacity. This method has already been used in MEG and many other experiments using liquid xenon, and there is no problem in maintaining the stability of the liquid.
\subsubsection{Calorimeter Mechanical Engineering}\label{sec:Calorimeter_Mech_Eng}

\label{sect:Mech-Eng}

\subsubsection*{Requirements}

Mechanical engineering considerations for the LXe calorimeter are discussed below.

\bl [itemize]
\ListProperties(Start1=1)
& The \calo\ mechanical design has to support a load of about 9 tons of LXe as well as
the required cryogenic envelope, consisting of a cold vessel and an insulation vacuum. Ideally, the heat load
of the new system should be compatible with existing MEG cooling and purification infrastructure. 
& The beam region consist of an entrance cone to focus a  beam with 10\degree\ divergence onto a beam spot of
about 1 cm$^2$. The entrance side will support the \beam\ detectors. The exit side should be in air, allow the
insertion of \atar\ and \track\ and their associated electronics, and provide sufficient heat removal.
& The \atar\ target region should be surrounded by thin beam pipes. The goal is to
minimize material in the \calo\ acceptance of close to degree=140\degree\ in polar angle, while covering
the full azimuth. 
& The \atar\ and \track\ region is very space limited, while increasing the inner pipe radii is problematic, both in terms
of solid angle and window strengths. A compromise between these conflicting requirements needs to be optimized.
& The PIONEER detector is designed for a measurement program extending over a decade. Both routine services as well as major
repairs and upgrades must be supported by the design.
\el

\subsubsection*{Conceptual design}

\begin{figure}[htbp]
	\centering
	\includegraphics[width=0.8\textwidth]{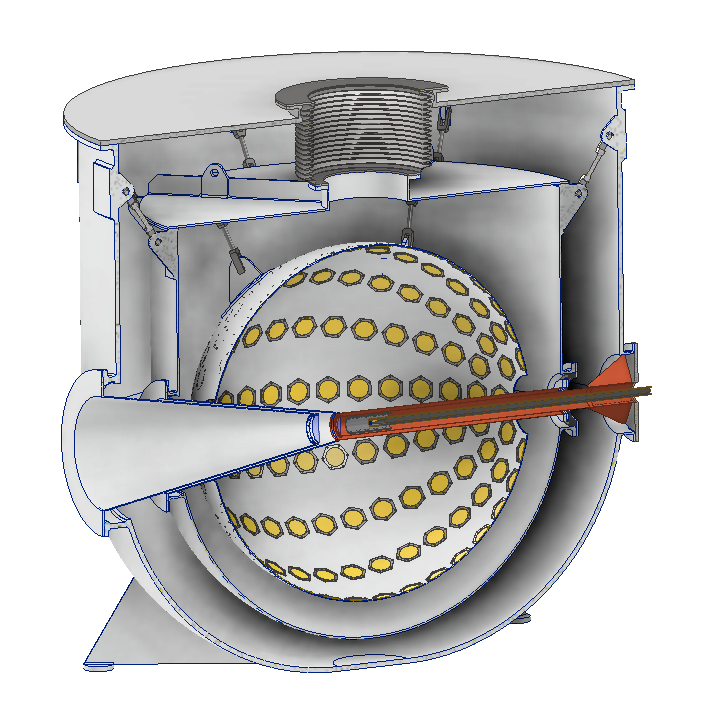} 
	\caption{Concept design of the liquid xenon calorimeter.  For scale, the lid is 3.05 m diameter. The yellow circles are merely representative of the photosensors; they are not placed accurately.}
	\label{fig:LXe_Calo_XSec}
\end{figure}

The calorimeter is built of interactive functional blocks incorporating a top load concept with the intent to be modular and to allow for ease of maintenance.  The functional blocks are described below and a drawing is shown in Figure~\ref{fig:LXe_Calo_XSec}.

%

\ListProperties(Start1=1)
\textbf{1. Liquid xenon vessel}:
This vessel has a spherical bottom to follow the contour of the detector mounting surface. This both reduces bending stress in the walls and reduces the volume of xenon required to cover the detectors.  Running through this vessel horizontally is a cone-tube pathway.  The tube portion, which surrounds the target, is envisioned as beryllium. The cone may be of a variety of metals.  The connection between the two components is designed to be a braze or weld.

\textbf{2. Detector mounting surface}:
Beyond the basic requirement of being spherical and holding detectors, the inner radius is set at 0.9\,m to achieve the required number of radiation lengths of xenon.

\textbf{3. Thermal insulation}:
To achieve the best insulation, the calorimeter has been designed to be a vacuum dewar.  The support of the liquid xenon vessel is achieved by hanging it on turnbuckles made of material which does not conduct heat well.  Stainless steel is seen as the most likely option.  If additional reduction in heat load is needed the link could be changed to titanium, allowing for a reduced cross section.  The more complicated portion of the thermal insulation is maintaining vacuum around the incoming beam and the target space.  This is made easier by having a break in this tube.  The beam tube loads from one side, the target tube from the other.  Both terminate in thin vacuum barrier windows which allow beam passage.  To accommodate both temperature induced dimensional changes and deflection from weight loading, there are two bellows present.  The tube running through the xenon vessel has bellows on the target loading side.  There is a large bellows on the top of the  xenon vessel to allow cabling and piping to pass through the vacuum space.

\textbf{4. Target air space}:
The tube the target resides in is envisioned as beryllium.  This tube is mounted on a flange on the vacuum vessel and protrudes past the centerline by approximately 175\,mm.  The inside end of the tube is capped with a window that allows passage of the beam and also acts as a vacuum barrier.  This tube also contains the tracking detector and the necessary cabling.  

\textbf{5. Beam path to the target}:
The beam is shown focusing with a center line to the beam edge angle of 10 degrees. 
The inner cone is between the beam (at atmospheric pressure) and the vacuum (insulation).  The outer cone is the barrier  between the vacuum and the xenon. Further studies might lead to a modification of the downstream exit pipe to mirror the entrance side, and the optimization of the ATAR holder,
to allow pass through of \pdif\ muons. 
\subsubsection{Other Calorimeter Options}

A LYSO crystal calorimeter is also being investigated as an alternative to LXe; it would provide natural segmentation as in  the PEN/PiBETA experiments. We discuss this further in the Appendix.

\subsection{Trigger and Backend electronics}
All triggers will start with a PI signal, which is a loose condition for an incident beam particle. PI can be defined as a coincidence
of the BEAM detectors (LGAD telescope layers). The key point is that this trigger must not introduce any bias between \pie\ and \pme\ events. For the estimates we assume that a separator is installed, so that the rate R$_{PI}$=300\,kHz corresponds to the design beam rate and is dominated by pions.

\begin{figure}[htb]
\centering
\includegraphics[width=0.7\textwidth]{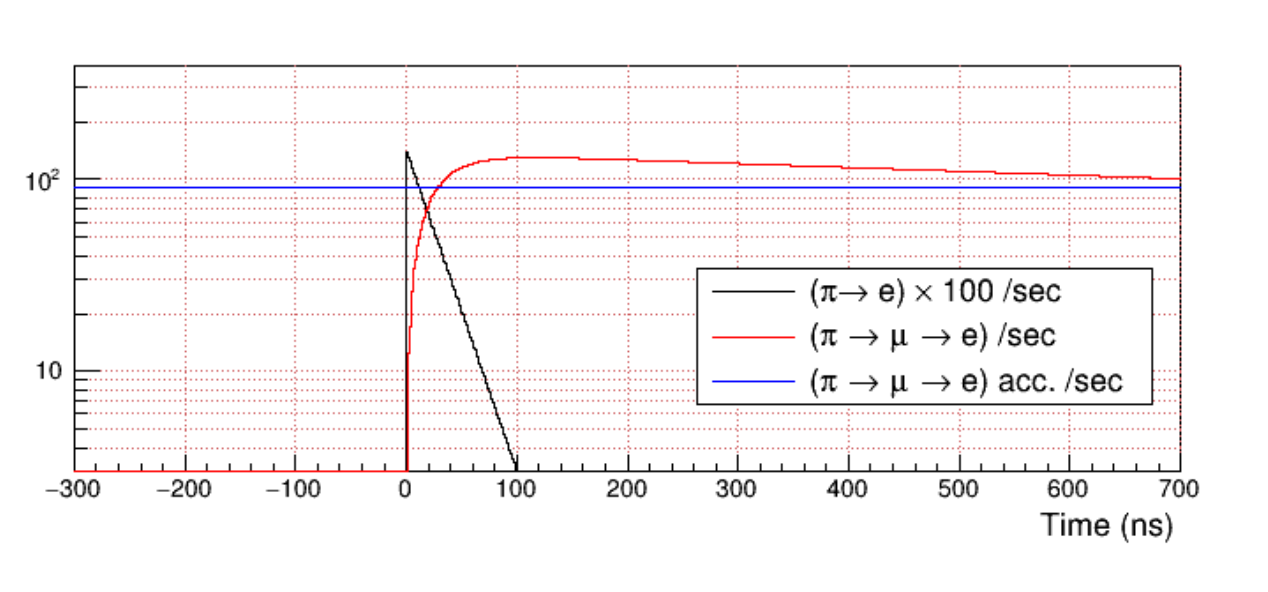}
\caption{Positron rates after PI per second in 1\,ns time bins. \pie\ positron rates are multiplied by a factor 100.  \pme\ rates generated by PI shown in red and positron rates from old muons, i.e. from accidental pions, shown in blue. }
\label{fig:time}
\end{figure}

The main time distributions in the vicinity of the PI signal are sketched in Fig.~\ref{fig:time}. After an initial build-up with the pion lifetime,  positron rates from \pme\ reach their maximum before decreasing with the muon lifetime. The constant accidental rate from muon decays of other pions stopped in ATAR is high.

After requiring the PI signal, the following main triggers are formed:
 \bl [enumerate]
\ListProperties(Start1=1)
& \tpi: This is a minimum bias trigger, with the PI signal prescaled by about k=1000.
& \tcaloh: Selection of high energy (E$_{tot} \gtrsim$ \thcal) events detected by the \calo\ within a time range TR=[-300,700] ns relative to PI.
& \ttrack: All events with \track\ hit within time range TR relative to PI, prescaled by about k=50. We note that the probability to observe a \pme\ positron in TR is 0.19, while the probably for detecting an old muon positron is 0.3, thus accidentals are a significant part of this trigger.
& \tprompt: Selected prompt events with a \track\ hit in time range [2,32] ns relative to PI, potentially prescaling required.
\el

For all triggers a full event readout will be initiated. 
The properties and required bandwidth for these trigger classes are summarized in Table~\ref{tab:triggers}. Additional data flow from the \track\ and \beam\ counters is small, and not included.

\begin{table}[htbp]
  \centering
  \small
    \begin{tabular}{cccc|ccc|ccc|cc}
    \toprule
    triggers & prescale &  range & rate & \multicolumn{3}{c|}{\calo} & \multicolumn{3}{c|}{\atar\ digitizer} & \multicolumn{2}{c}{\atar\ high thres} \\
          &       &  TR(ns)     &   (kHz)     & $\Delta$T(ns) & chan  & MB/s  &$\Delta$T(ns) & chan  & MB/s  & chan  & MB/s \\
       \hline
    \tpi  		& 1000  & -300,700 	& 0.3   & 200  	& 1000 	& 120 	& 30    & 66    & 2.4 & 20    & 0.012 \\
    \tcaloh 	& 1     & -300,700 	& 0.1  	& 200 	& 1000 	& 40 	& 30    & 66    & 0.8 & 20    & 0.004 \\
    \ttrack 	& 50  	& -300,700 	& 3.4  	& 200  	& 1000 	& 1360  & 30 	& 66    & 27 & 20    & 0.014 \\
    \tprompt 	& 1     & 2,32  	& 5     & 200  	& 1000 	& 2000 	& 30    & 66    & 40  & 20    & 0.2 \\
    \bottomrule
    \end{tabular}%
  \caption{Main triggers: time range TR and  trigger rates. For detector systems readout island length $\Delta$T, average number of channels and required readout bandwidth are given.  }
  \label{tab:triggers}%
\end{table}%

The main purpose of \tcaloh\ and \ttrack\ is the selection of true \pie\ and \pme\ events, respectively. The $E_{tot}$ \calo\ energy sum will be calculated in the trigger processor with a latency of 200\,ns. Events above the Michel continuum with $E_{tot} >E_H \approx$ \thcal\ will be assigned to \tcaloh.

The full event readout includes the \calo\ waveforms, the \atar\ and the \beam\ and \track\ detectors.

The \calo\ data  will comprise islands of contiguously digitized ADC samples from the SiPMs / PMTs readout of the calorimeter. We assume 2-bytes per sample, 200 samples per island, and 1000 islands channels per trigger. Thus a trigger rate of 5\,kHz generates 2\,GB/s.

The zero suppressed \atar\ waveforms of typical length of $\Delta$T=30 ns will be stored,  
with the readout window extended if the signal remains above threshold at the end of the island.
The dominant \pme\ event will fire about 66 strips. Depending on strip cross talk, which differs for LGAD technology choices, i.e. AC-LGAD vs. TI-LGADS, this strip count could be 3-times higher.  With a 1 GSa/s digitization and 4 Bytes required per hit, 66 channels of 30 ns islands correspond to 30*66*4=8 kB. Thus a trigger rate of 5\,kHz generates 40 MB/s.

In addition (see \atar\ high thres in Table~\ref{tab:triggers}) the trigger processor will also receive high threshold logic signals from the \atar\ low and high threshold discriminators located at the frontend of the ATAR digitizers and will time stamp those with coarse resolution (few ns). Those hits will always be read out for a time period of 10\,$\mu$s preceding PI. This data can determine whether  a previous pion stopped at the stop location of PI, to potentially apply a local pile-up for those events. 

By far the most challenging is the \tprompt\ trigger which is of critical importance for the experiment. 
\bl
\ListProperties(Margin=2cm, Hang1=true,Hide2=0,Numbers1=r,Numbers2=r,Progressive*=.5 cm,Start1=1)
& It will trigger on all prompt \ttrack\ events, where the \pme\ events are suppressed by a factor 6.4$\times 10^{-3}$ in [2,32] ns. This should verify that the \tcaloh\ trigger does not miss events by its more complicated E$_{tot}$ trigger configuration. If required this part of the trigger can be prescaled.
& \tprompt\ is critical to measure \pie\ events which are lost because the CALO response falls below $E_H$. The uncertainty in those losses were the limiting systematics in the last generation of experiments. We consider additional \atar\ requirements at trigger level, like hits in the most upstream layers and no hit in the most down stream layers within the \atar\ time resolution of 100\,ps. We will design the trigger processor and DAQ bandwidth such that all \tprompt\ events can be transferred in streaming mode to a front end computer farm. Thus the stored data volume will be reduced in a combination of trigger processor and software decisions, by applying \atar\ cuts on track topology and energies to suppress \pme\ events. 
Ideally, this should be done in an unbiased way. However, as the main purpose
is a measurement of the loss term of \pie\ events, small biases can be tolerated and even prescaling can be considered if unavoidable.

&
In addition, the \tprompt\ trigger will select a sample of \pdar-\mdif\ events. Positrons from immediate \mdif\ decay are indistinguishable from the
\pie\ radiative tail events. Measurements of the spectrum of positrons after PI, where the muon deposits only a fraction
of its 4\,MeV energy, will constrain this background class.
\el

\subsubsection{Trigger design}
\label{Trigger}
\begin{figure}[tb]
\begin{center}
\includegraphics[width=0.9\textwidth]{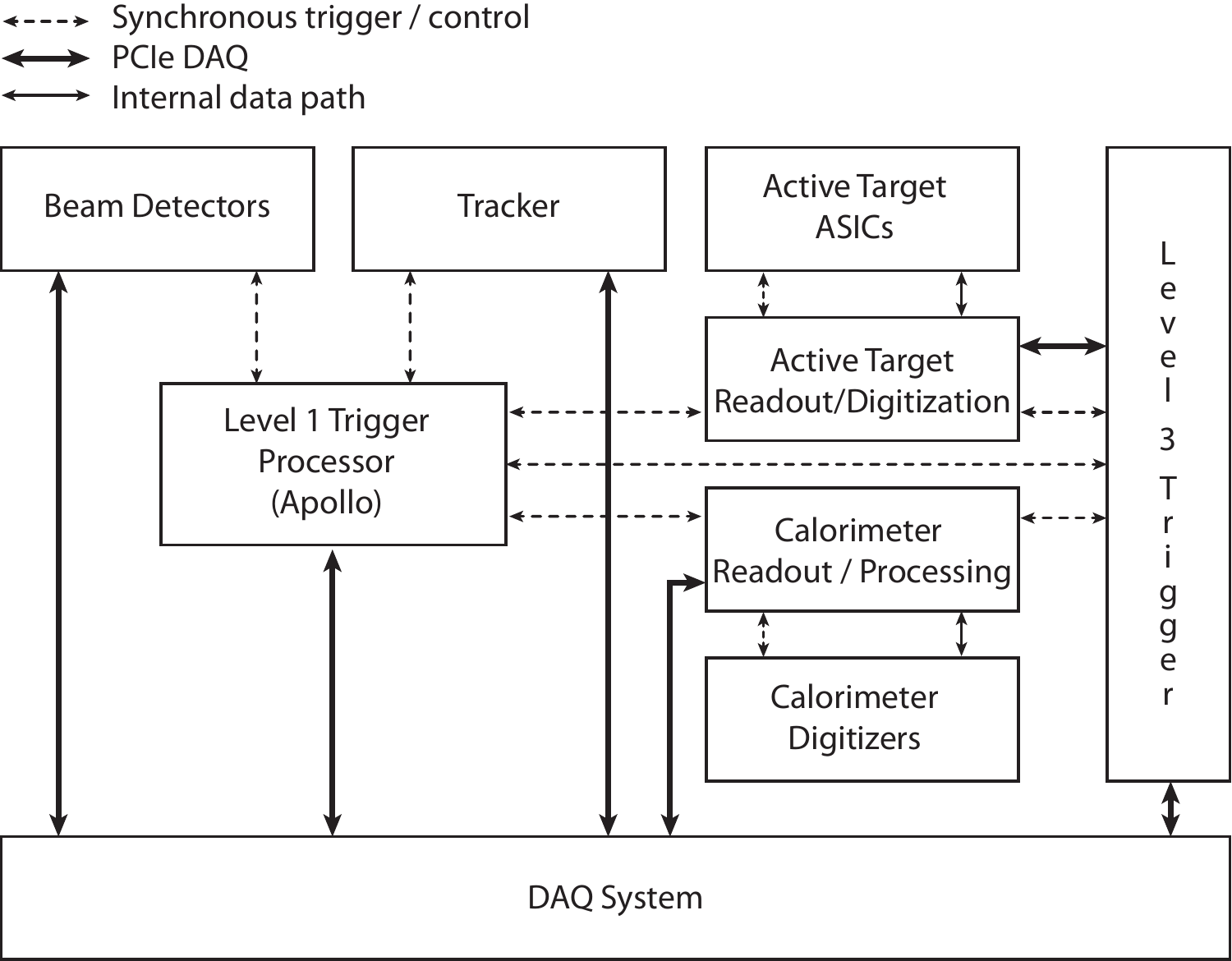}
\caption{Proposed topology of the trigger and subdetector readout systems.}
\label{fig:readout}
\end{center}
\end{figure}

Figure~\ref{fig:readout} shows the planned topology for readout and triggering of PIONEER.  The design takes advantage of
the APOLLO~\cite{ApolloTWEPP21} platform designed for the trigger track finder and pixel readout in the CMS experiment at
the LHC.  The Command Module (CM) component of the APOLLO board provides a highly configurable environment supporting a dense
high speed IO environment via
25 Gbps Samtec Firefly~\cite{firefly} optical inputs.  The platform already
supports the flexible high speed synchronous command and triggering TCDS2 (Trigger and Timing Control and Distribution
System)~\cite{CMSP2TDR} for the CMS phase II upgrade, as well as options for direct IO for receiving timing signals or
generating trigger signals for devices expecting analog triggers.  We anticipate using a single FPGA version of the board (utilizing the Xilinx VU13P) with an interposer in the second FPGA slot to bring all the serial links to the VU13P.
While by some measures this board likely provides more power than needed for PIONEER, it has already undergone significant prototyping with full production anticipated in 2024, and can satisfy several of the needs of the system with a single platform.  The engineering group that designed the CM will also contribute to calorimetry readout, providing additional coherence.  

PIONEER anticipates use of the 16.1 Gbps Firefly implementation. This protocol serializes 64 bit words with a 64b/66b encoding, resulting in an effective transmission of 64 bit words at rates up to 244\,MHz.  Each channel supports both transmit and receive functions, which will allow for significant flexibility in the command and control system.

The Samtec Firefly system already supports PCI Express (PCIe) Gen 3 and Gen 4 integration, with commercial PCIe over firefly adapter cards already available for Gen 3.  As discussed further in Sec.~\ref{subsec:daq}, PIONEER proposes use of PCIe as the protocol underlying the DAQ system.

The main trigger APOLLO processor will receive trigger inputs, either via inputs connected to the FPGA standard IO pins, or from the firefly inputs and apply the appropriate logic algorithms to generate triggers and guide sparsification.  The processors  encode the resulting trigger and configuration information into the TCDS (or equivalent) datastream, which gets transmitted to all subdetectors.  The processor will also receive status and health information from the subdectector, communicating any alarm conditions to the DAQ system for appropriate action.  

The trigger processor will receive trigger inputs from the BEAM counters and from the tracker to form the PI, Track and PROMPT triggers, with configurable prescale factors applied as noted earlier.  The processor will also receive energy sums from subsections of the calorimeter (see Sec.~\ref{subsec:caloBE}) and complete the total energy sum used to form the CaloH trigger in coincidence with the (unprescaled) PI trigger.  The processor can flexibly accommodate the needs of the beta decay (Phase II) of the experiment, completing, for example, opening angle or mass estimates based on energy-weighted position sums from the calorimeter electronics.

The processor will also interact with the ATAR, level 3 trigger system and DAQ system to coordinate additional trigger information from the patterns in the ATAR as described above.

\subsubsection{Calorimeter backend electronics}
\label{subsec:caloBE}
\begin{figure}[tb]
\begin{center}
\includegraphics[width=0.9\textwidth]{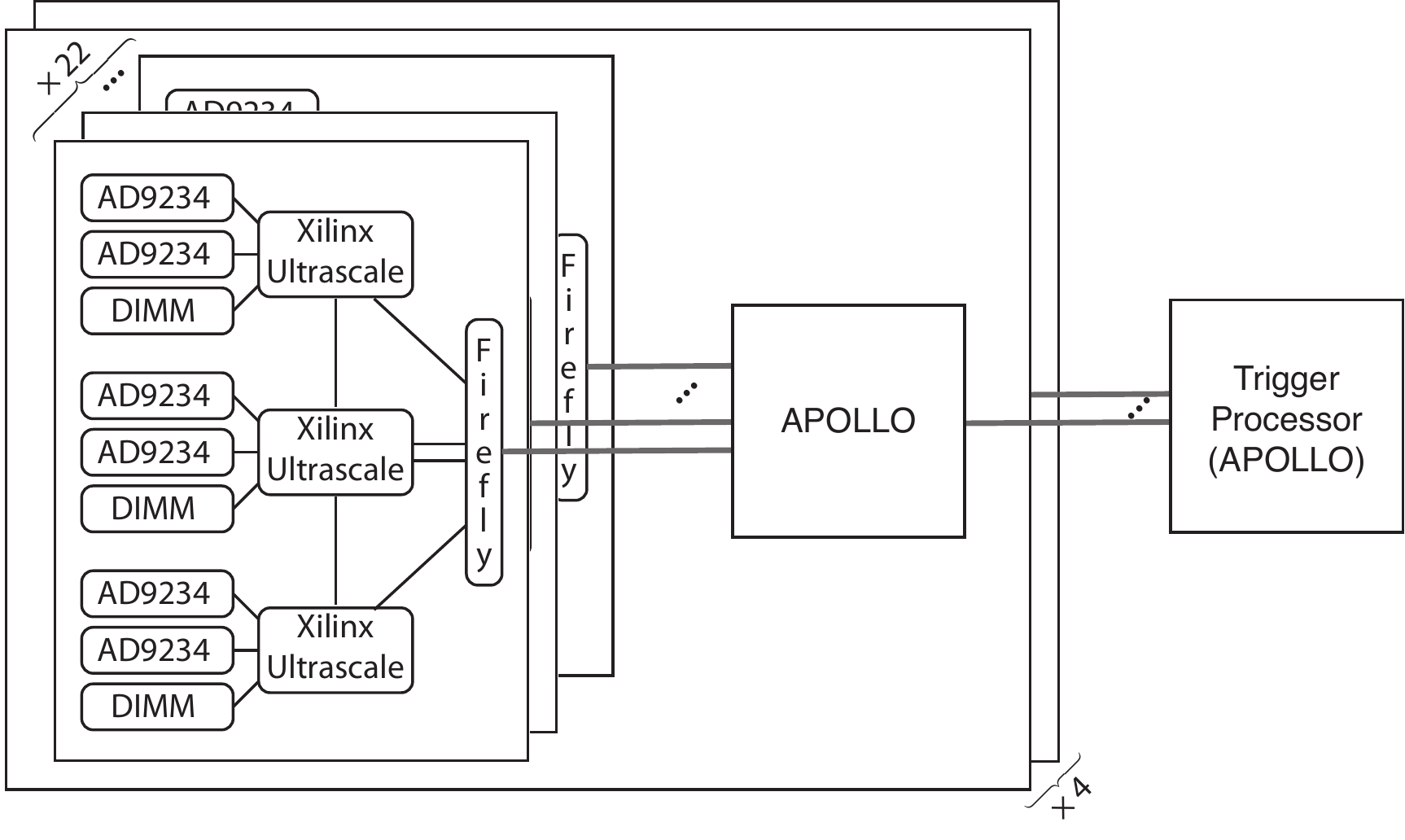}
\caption{Proposed calorimeter digitization and readout.  The 12 channel digitizer boards utlize the dual channel AD9234 12 bit, 1 GSPS ADCs, Xilinx Ultrascale FPGAs for contro, and Samtec Firefly\textsuperscript{TM} high speed communications to a CMS APOLLO board.  The APOLLO board can receive up to 22 boards in this configuration -- instrumenting quadrant of an order 1000 channel calorimeter.}
\label{fig:calodaq}
\end{center}
\end{figure}

Figure~\ref{fig:calodaq} illustrates the digitization and readout electronics proposed for the calorimeter.  The 12 channel digitization boards will utilize the Analog Devices AD 9234 dual channel, 12 bit 1 GSPS ADC chip, chosen for its low latency of 59\,ns from presentation of the signal at the front end to output of the digitized signal.  These chips communicate the digitized output via the JESD204 protocol, which will present the data for each ADC channel as 64 bit words containing four samples at 250\,MHz.  At these rates, the calorimeter information can be summed with a pipelined adder and potentially corrected with a running pedestal measurement as the first stage in a total energy measurement.  By clocking the ADC at the slightly lower rate of 976\,MHz, the ADC information can be synchronized to the firefly data transfer rate of 244\,MHz, simplifying synchronization of the system.

The ADCs will sample continuously with samples stored initially in a ring buffer on the FPGA.  Upon receipt of a trigger, a configurable time window will be stored in DIMM memory.  By deploying a DIMM with a 128 bit data path, simultaneous reading and writing can be accomplished via two 64 bit pathways.

A single FPGA will control one pair of ADCs (four calorimeter channels).  It can compare the energy sum from each channel against a channel activity threshold (or thresholds), as well as combine the four running energy sums as the first stage of a total energy sum for the high energy trigger.  That FPGA will also drive a single firefly channel.  The 64 bit word for the firefly communication provides much flexibility.  After reserving 48 bits for readout of the four channels upon receipt of a trigger, 16 bits can provide triggering information at the full 244\,MHz rate.  Possible configurations include a 12 bit sum of the four calorimeter channels along with a bit for each channel flagging a sum over threshold.  With a coarse 8 bit sum, 2 bits could be devoted to each individual channel.  For the beta decay phase, the trigger could pass energy-weighted position information for a mass calculation. Xilinx kintex ultrascale FPGAs provide sufficient numbers of high speed serial lines at a reasonable cost.

The digitizer boards will communicate with intermediate Apollo boards via the 16.1\,GHz firefly links, which come packaged with a minimum of four individual links.  Three of these links will provide the TCDS (or equivalent) clock and control signals and send the trigger and channel readout information to each of three sets of 4 ADC channels.  The fourth firefly line will provide PCIe or ethernet communication to allow board configuration and other slow control functions.

After taking into account communication links to the trigger processor and between the intermediate Apollo boards, each intermediate Apollo can accommodate 22 12-channel digitizer boards for a total of 264 channels. Four such boards would instrument a calorimeter with up to 1056 channels.  With a more complex, run-state dependent scheme for communications, 16 channels per board could be accommodated.

The intermediate APOLLO boards will collate the waveforms read from all channels for each trigger, and send that information directly to the DAQ system.  Each APOLLO can sustain a full readout rate of about 18.5 kHz for the 264 channels, allowing for significant headroom given the anticipated trigger rates (see Tab.~\ref{tab:triggers}) and use of sparsification for an order 1000 channel calorimeter.

The total calorimeter energy sum proceeds in several stages.  Each quad-channel FPGA adds those four signals as noted above.  Each intermediate Apollo will combine the those sums from the $22\times 3=66$ quad channels.  The trigger process will provide the final combination of the four intermediate sums needed for the complete calorimeter sum. 

The latency in the energy sum has three main contributions: the ADC latency, the FPGA summation time, and the serialization/deserialization overhead for the high speed links.  The Xilinx GTY transceivers have a high degree of configurability, resulting in a total serialize/deserialize latency for the readout chain above that ranges from 65\,ns to 175\,ns at the 16.1\,GHz serial rate.  Recent studies~\cite{OliveiraThesis} show that even the simpler, lower latency settings, can achieve reliable links.  With 100~ns as a reasonable estimate, 50\,ns as a limit for the energy summation at 244\,MHz, and the 50\,ns ADC latency, we estimate a total latency of 200\,ns.
\subsection{DAQ}
\label{subsec:daq}
\subsubsection{Data acquisition system}

The PIONEER data acquisition system must handle the read out, event assembly and data storage for the active target, positron tracker, electromagnetic calorimeter and other detector sub-systems of the experimental setup. It must provide a deadtime-free, distortion-free record of the datasets identified by the various physics and calibration triggers. It must facilitate the monitoring needed to guarantee the overall integrity of data taking and provide the metadata needed to document the experimental configuration during data taking. Finally, it must enable the configuration of the readout electronics and the associated trigger, clock and control system.

\begin{figure}[t!]
\includegraphics[width=0.99\textwidth]{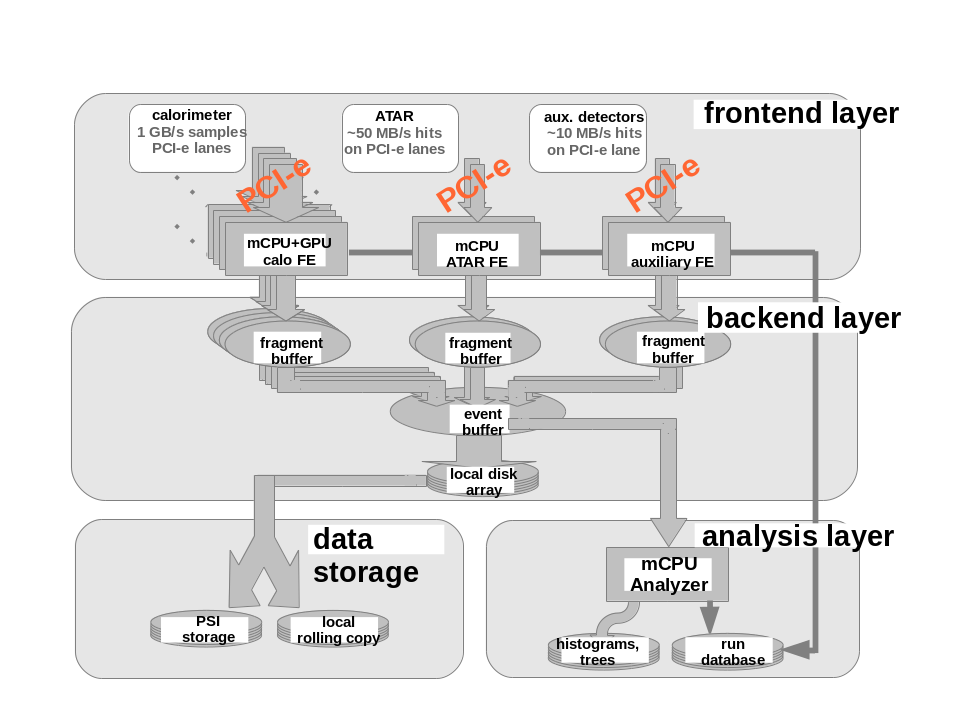}
\caption{Schematic of the data acquisition system showing the frontend layer for data readout and experiment configuration, the backend layer for event assembly and data storage, and the analysis layer for data quality monitoring. The number of frontends and the topology of the FPGA-to-frontend and frontend-to-backend networks will be based on the calorimeter, ATAR and FPGA technology choices.}
\label{f:daqlayout}
\end{figure}

The acquisition will be implemented as a modular, distributed system on a parallel, layered processor array consisting of networked, multi-core, commodity PC's running a scientific Linux operating system. The overall layout is depicted schematically  in Fig.\ \ref{f:daqlayout}.  It will comprise a frontend processor layer responsible for readout and processing of event fragments from the FPGA-based fast electronics instrumenting the various detector sub-systems, a backend layer responsible for event building and data storage,  a slow control layer responsible for configuration and readback, and a data quality layer responsible for monitoring of data integrity.

To maximize the bandwidth the frontends will utilize memory-mapped, Gen3 PCI Express communication with the Apollo readout boards and the downstream DAQ layers will utilize memory-mapped, Gen3 PCI Express host / target adaptors for data transfer and sharing system resources (disks, GPUs, {\it etc}). PCI-Express 3.0 delivers an effective data transfer rate of nearly 1 GB/s per serial lane. Commercial host-to-target optical interfaces are now available for the Apollo FPGA to frontend computer communication ({\it e.g.}, www.dolphinics.com/products/PXH842.html). Commercial peer-to-peer communication is also now available for direct PCI-Express data transfer between devices ({\it e.g.}, www.dolphinics.com/products/PXH810.html).   The setup will offer the capability of direct transfer of  data to / from FPGA and CPU / GPU memory for rapid realtime processing. 

The DAQ software will be based on the MIDAS data acquisition package \cite{midas}, CUDA GPU toolkit \cite{cuda}, ROOT data analysis package \cite{root}, and Linux PCI Express system and utility libraries. The MIDAS software consists of library functions for data flow between different processes on local / remote devices as well as infrastructure for data logging, experimental configuration and local /remote run control. It also incorporates an integrated alarm system and slow control system. The devices drivers for the configuration and the readout of the Apollo board FPGAs will be based on the PCI Express communication protocols / libraries.

As discussed in Sec.~\ref{Trigger} and Table~\ref{tab:triggers}, we anticipate the various physics triggers to include a prescaled beam trigger \tpi, a beam particle - high energy \calo\ trigger \tcaloh, a prescaled beam particle - low threshold \calo\ trigger \tprompt, and a beam particle - outgoing positron trigger \ttrack ~(as identified by either the ATAR or the tracker). The cumulative trigger rate in Table~\ref{tab:triggers} is estimated at approximately 8.8\,kHz. 

The ATAR data will comprise above-threshold islands of contiguously digitized ADC samples from the fired detector strips of the active stopping target. Based on a 1 GSPS sampling rate, 4 Bytes per sample, and 66 above-threshold strips per trigger, we anticipate an \atar\ data rate of about 70~MB/s for the aforementioned trigger types.

The CALO data  will comprise islands of contiguously digitized ADC samples from the SiPM / PMT readout of the calorimeter. Based on a 1 GSPS sampling rate, 2 Bytes per sample, 200 samples per island and 1000 islands / channels per trigger, we anticipate a \calo\ data rate of about 3.5 GB/s for the aforementioned trigger types.

Other detector systems will include the various beam detectors upstream of the ATAR and the positron tracking detector surrounding the ATAR. The data rates from beam detectors and tracking detector are expected to be a small fraction of the ATAR / CALO data rate.

Separate frontend processors running  MIDAS frontends will readout the ATAR, calorimeter, tracker and beam sub-systems via the  PCI-express bus using commericial transparent host / target PCI Express adapters to the Virtex FPGAs on the Apollo readout boards. The memory-mapped readout from the Virtex FPGAs chip will permit both high data transfer rates and direct data transition to system CPU and GPU memories. We plan to utilize IPBus over PCI Express for the configuration of the readout electronics and the trigger, clock and control system.

The data acquisition system will process in real-time the data from a roughly 3.5 GB/sec raw data rate to a roughly $\sim$300 MB/sec processed data rate for data storage on PSI's Petabyte archive. One option for the data processing is the lossless compression of the slow decay-time calorimeter signals via a combination of delta encoding and Golomb coding. The delta encoding will exploit the correlations between sequential ADC samples to shrink the word usage distribution while Golomb coding is suited to situations in which the occurrences of small values are significantly more likely than the occurrences of large values. Other possibilities are zero suppression of \calo\ islands and realtime fitting of \calo\ pulses. These algorithms are well suited to parallel processing using GPUs.

We are initiating an R\&D effort to demonstrate both the technology for the FPGA-to-CPU / GPU communication via optical PCI-express links and the performance of the data compression schemes. The R\&D setup will allow for code development, testing and debugging as well as the evaulation of the rate capabilities and the compression capabilities of the system. We plan to use commercial PCI-Express FPGA development boards as data simulators for the detector sub-systems in order to prototype and benchmark the DAQ.

The PIONEER DAQ group has developed and operated similar architectures of distributed data acquisition systems for the MuLan, MuCap and MuSun experiments at the Paul Scherrer Institute and the g-2 experiment at Fermi National Accelerator Lab.

\subsubsection{Data Policy}

The PIONEER collaboration will comply with the \emph{data policy for PSI research data} \cite{PSIdataPolicy} and will publish the experiment's data under PSI's custody after a suitable embargo period. We are also committed to publish our software under an open source license at a suitable time.


\section{Simulations} \label{Sims}

Each of the design elements discussed above is being actively studied 
using GEANT4-based \cite{GEANT4:2002zbu} simulations.
The simulation efforts include 
beamline and upstream detector simulations, simulation of the active target, and simulation of the calorimeter. 

The $\pi$E5 beamline at PSI is simulated using G4Beamline \cite{Roberts:2008zzc} discussed in Section \ref{sec:beam_studies}. The remainder of the simulation is done primarily using GEARS, 
an extension of GEANT4, which streamlines readout for rapidly iterating systems. The geometries for each of the experimental components are generated using a stand-alone Python script and geometry library, which takes as its input a \texttt{json} file of various parameters (e.g. diameters, number of elements, which detectors to implement, etc.) and exports a \texttt{GDML} file which is then read in by the Geant4 simulation. This file contains a full description of the physical geometry of the detectors, as well as their material properties (density, reflectivity, scintillation yield, etc.). 
Because of this workflow, it becomes trivial to implement scans over various parameters to perform systematic studies. 


In the following sections, we discuss the initial ATAR and calorimeter simulations.

\subsection{ATAR Simulations}

\begin{figure}
   \centering
   \includegraphics[width=0.35\textwidth]{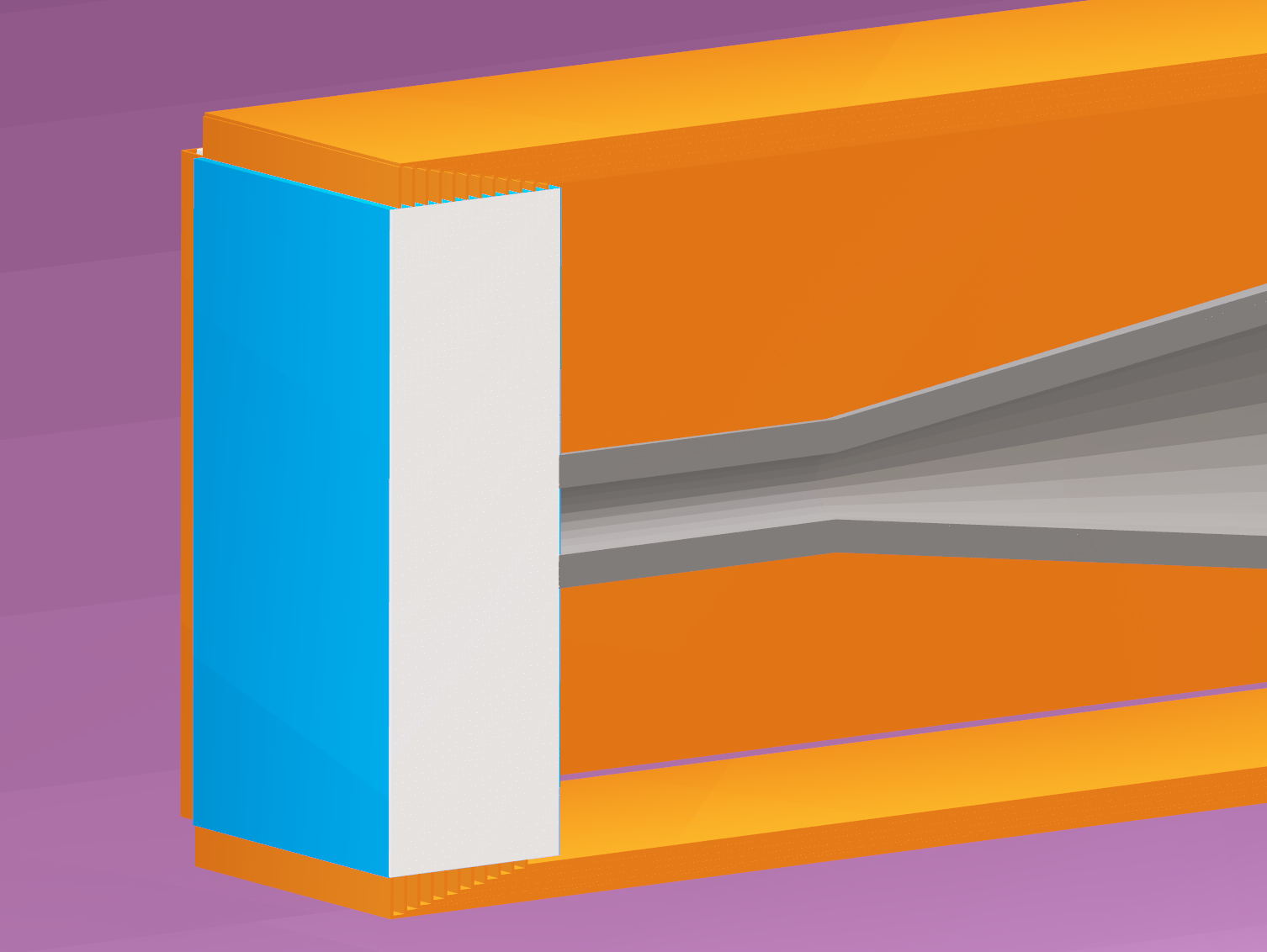}
   \includegraphics[width=0.48\textwidth]{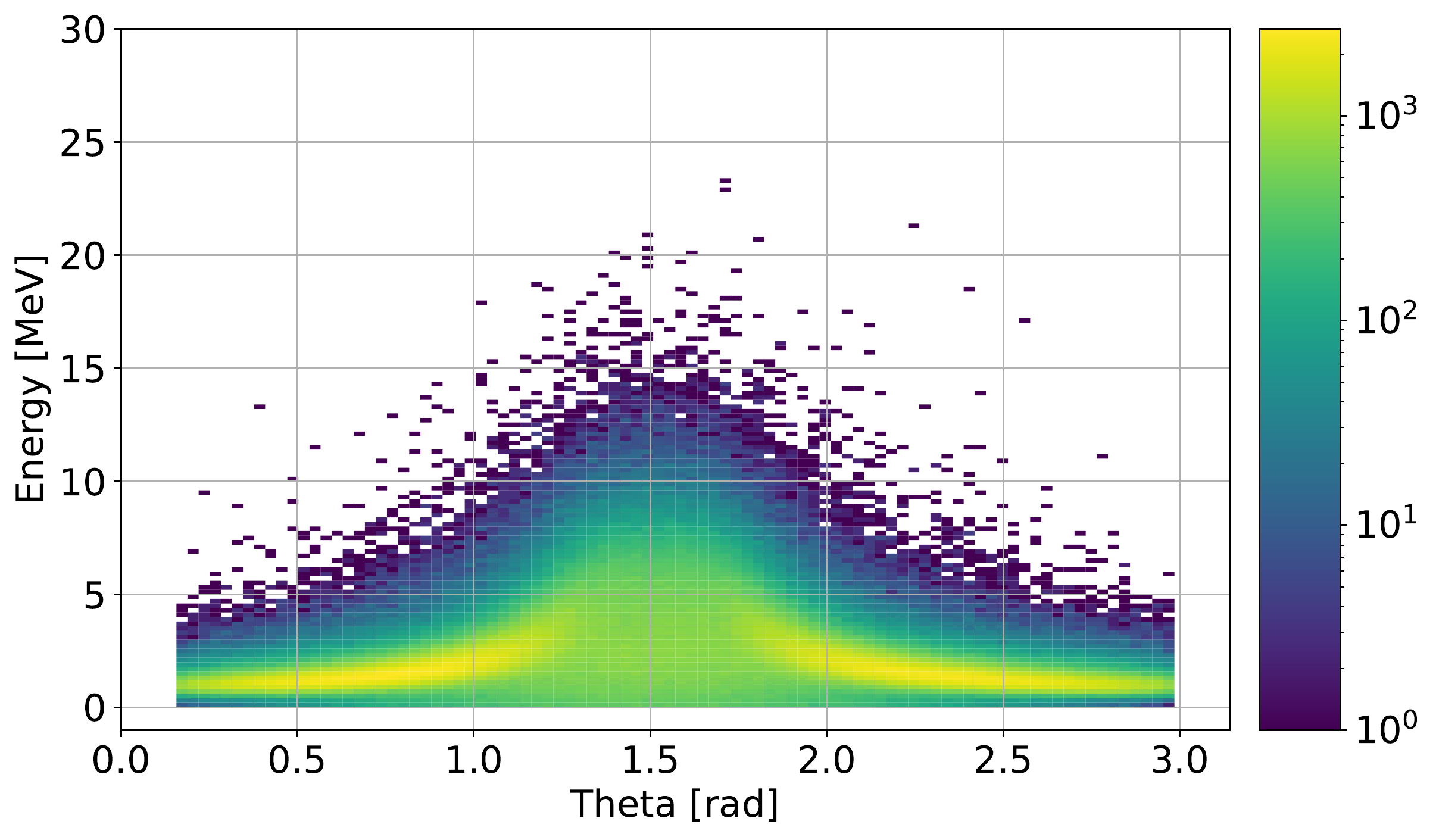}
   \caption{Left: An image of the simulated ATAR, showing the active planes (white) and readout strips (orange). Right: Total energy deposited in the ATAR by the decay positron as a function of the angle $\theta$ at which it enters the calorimeter. }
  \label{fig:atar_sim}
\end{figure}

Simulations of the active target are done with both GEANT4, which allows us to model pions and their decay products through the full chain of detectors, and 
TRIM\cite{ziegler1985stopping}, 
which allows us to track particle energy deposition precisely in a simulated detector readout. 
Figure \ref{fig:atar_sim} (left) shows the ATAR concept  and (right) a simulation of energy deposit by exiting positrons as a function of polar angle of emission.

The active target is implemented in the simulation as alternating layers of cross-hatched silicon strips. The number of strips per layer, their thicknesses, and the number of layers is fully configurable. The nominal configuration consists of 48 \unit[120]{$\mu$m} thick planes of 100 strips each for a total size of $\unit[20 \times 20 \times 5.76]{mm^3}$. Optional `dead' zones can be implemented in each pixel, to simulate the effect of trenches between pixels within a plane or a dead backing layer. Each plane of the ATAR has a fully simulated readout cable, consisting of Kapton and Silicon,  simulating the ultimate flex cable readout. These readouts introduce an asymmetry in the energy reconstructed by the calorimeter, and as such have the potential to introduce a systematic bias to our analysis. The design of the readouts will be informed by this simulation in order to minimize such an effect.

\subsubsection{ Background  Suppression for the Measurement of the \pie\ tail fraction}

\label{tail_suppression}

With an anticipated tail fraction $f= 0.5\%$ of \pie\ events, $\mu-e$ decays must be suppressed to a level that allows measurement of $\Delta f/f$  with accuracy 1--2\%. This can be accomplished by using ATAR information to identify stopped pions (and reject incoming muons) by energy loss, and range-energy relations, using a narrow time window (e.g. 3--35\,ns) following the pion stop, rejecting events with observation of the \tmu\ decay muon, and suppressing events with pion decay-in-flight (\pdif), and muon decay-in-flight (\mdif) following pion decay-at-rest (\pdar) by tracking and energy loss measurements. 

Assuming a 10\% energy resolution (comparable to that of the PIENU target) $\pi\to\mu\to e$ background in the tail region from the dominant pion decay at rest will be suppressed to a negligible level as found in PIENU. Radiative decays $\pi\to\mu \nu \gamma$ followed by $\mu\to e$ decays which may leave $<4.1$ MeV in the target will be suppressed by observation of the gamma and  detection of the muon pulse and residual energy. In addition, muons that originate from \pdar\ in the ATAR can decay in flight in the 12\,ps it takes  before stopping ($\mu\,\mathrm{DIF}$), leaving a positron with the same time distribution as $\pi^{+}\rightarrow e^{+}\nu$ events.  $\mu\,\mathrm{DIF}$ events, although the probability of occurrence is  small ($6 \times 10^{-6}$), would represent a significant background comparable to the low energy tail fraction since it has the same time distribution as \pie~ events. Thanks to the high granularity of the ATAR, it is possible to observe the energy deposition at the pion decay origin   and  to measure the local  energy loss along the positron track. Typically, a \tmu\ muon travels  0.8\,mm in the ATAR and leaves high dE/dx hits in 4 strips on average. An initial simulation of $\mu\,\mathrm{DIF}$ events, indicated a suppression factor of 20 from local dE/dx measurements and we expect another order of magnitude suppression from positron tracking, pulse detection, and total energy measurements. This would  put the $\mu\,\mathrm{DIF}$ at about $10\%$ of the tail fraction $f$. (We are also studying the possibility of reducing the LGAD strip thicknesses in the stopping region to enhance rejection of this background and/or using a separate thin layer ATAR setup for studying this effect.) Furthermore, we expect that experimental studies of observable  $\mu\,\mathrm{DIF}$ will enable us to know the background component accurately so it can be subtracted.

As illustrated in Fig.~\ref{fig:EventTypes}, another potentially significant  low energy background in the tail region comes from $\pi\,\mathrm{DIF}$ in the ATAR followed by muon  decay at rest. This component dominated the background suppressed spectrum in PIENU.
Simulations indicate that 0.1\% of pions decay in flight in the ATAR before stopping and initial studies indicate that the $\pi\,\mathrm{DIF}$ events can be suppressed by a factor of 5000 using ATAR tracking information which recognizes kinks in the topology and measures dE/dx along the track. 
Along with suppression of muon decays by selecting a narrow  time window e.g.  3--35\,ns after the pion stop, we estimate that the \pdif\ contribution to the uncertainty in the tail correction $f$ will be negligible.

There is an ongoing effort to apply machine learning tools to boost the sensitivity to $\pi^{+}\rightarrow e^{+}\nu$ events in the tail region and suppress the backgrounds. It has already been demonstrated with the in-ATAR $\pi\,\mathrm{DIF}$ events that gradient boosting decision trees (BDT) are able to  outperform the manual cut-based methods.  
Although the BDT model shows excellent classification performance, only manually-constructed physics-inspired high-level information has been used as input features for the model training. As deep neural networks, especially Convolutional Neural Networks (CNN), have shown extraordinary performance in various image processing and classification applications, it could be advantageous to make use of deep neural networks’ ability to automatically learn high-level features directly from the complete pixel-level energy deposit information.

\subsection{Calorimeter Simulations}
\label{sec:calo_simu}




\subsubsection{Pileup}

\begin{figure}[!tbp]
  \centering
  \subfloat[]{\includegraphics[width=0.22\textwidth]{
  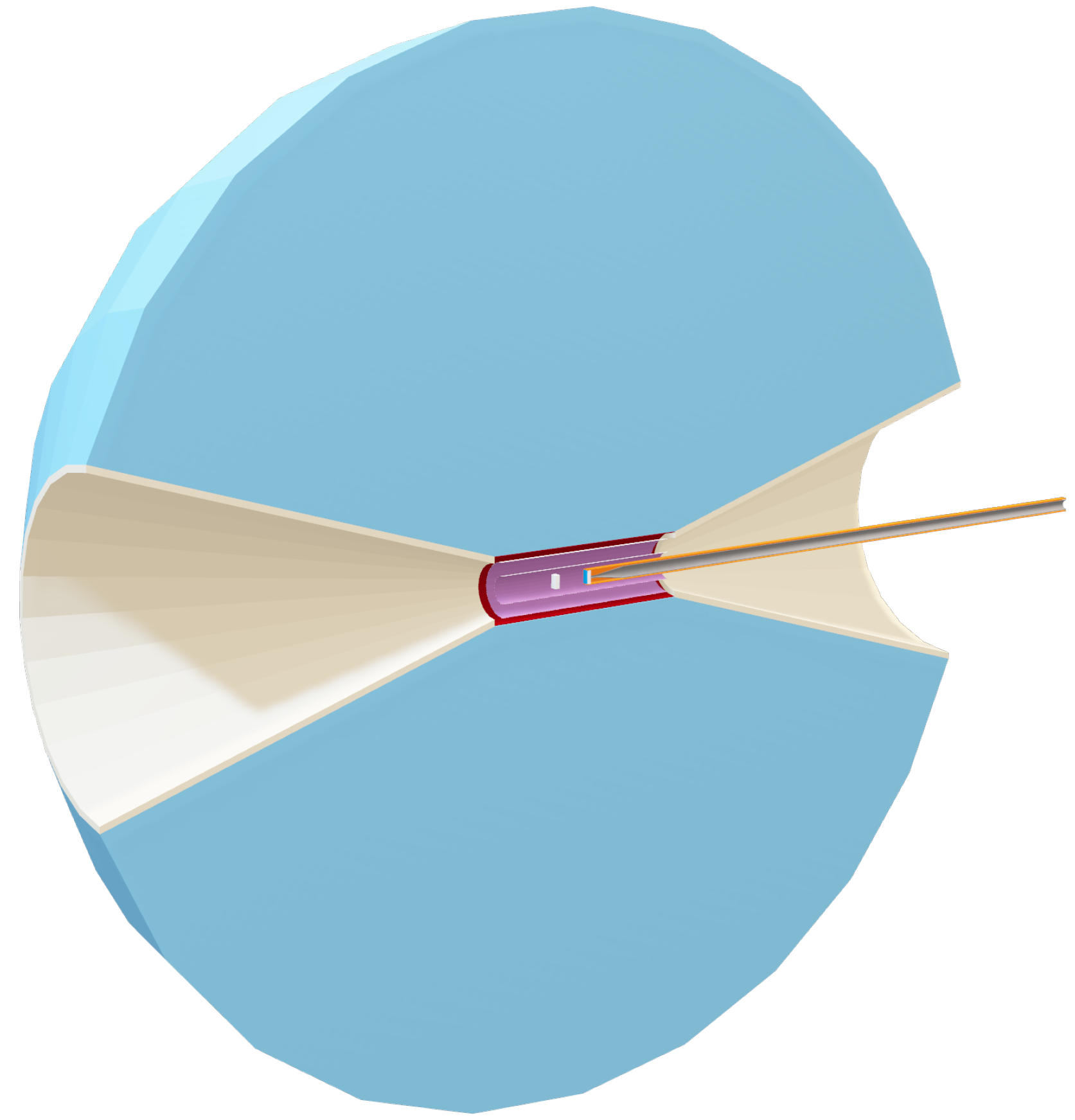}\label{fig:geometry_MC}}
  \hfill
  \subfloat[]{\includegraphics[width=0.7\textwidth]{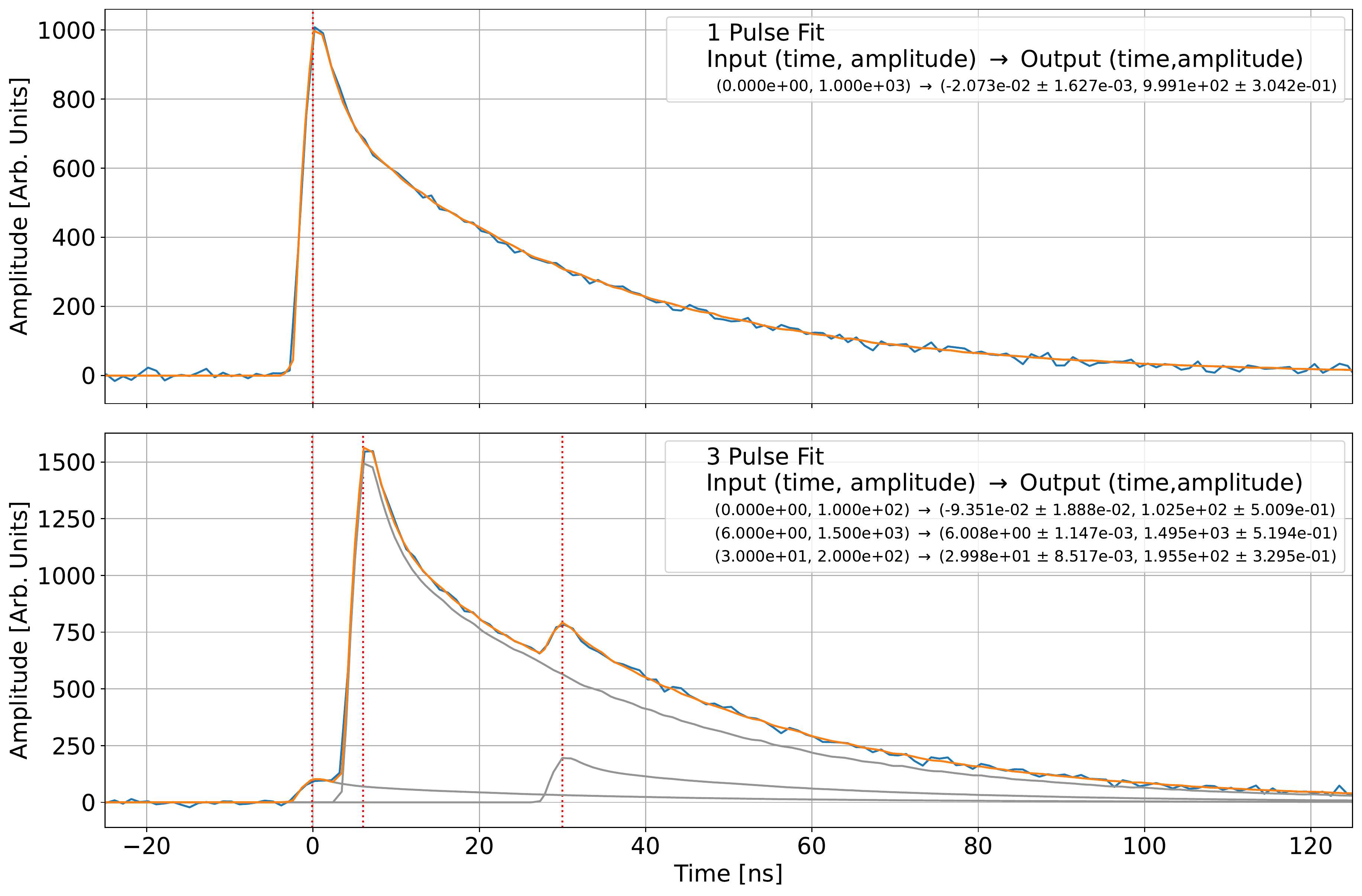}\label{fig:PSF}}
  \caption{(a) Rendering of the simplified geometry used in the first Monte-Carlo simulation of the PIONEER LXe calorimeter (b) Example of simulated pulse shapes of a single $\pi\rightarrow \mu \rightarrow e$ event (top) and 3 events happening closely in time (bottom) and recorded in the LXe calorimeter. Pulse shape fitting based on a template allows accurate identification and energy reconstruction of multiple pulses down to $\pm$\unit[5]{ns} separation.} 
\end{figure}

Simulation studies of the LXe calorimeter response have been performed using the GEANT4 package \cite{GEANT4:2002zbu} with optical photon tracking. A simple geometry (see Fig.~\ref{fig:geometry_MC}) that is representative of the current design was implemented. $\pi\rightarrow \mu\rightarrow e$ events were generated in the target at the center of the LXe sphere and optical photons originating from LXe scintillation induced by the showers generated by the positrons were tracked until the outside sensitive surface of the LXe sphere. The  simulated 
pulses obtained are similar to the ones reported by  MEG.
A pulse fitting algorithm was employed to evaluate the possibility of separating events that overlap in time i.e.~pulse  pileup. Pulse separation down to \unit[5]{ns} was achieved (see illustration in Fig.~\ref{fig:PSF}) across a wide range of amplitudes. This gives a first indication (without digitization) of the performance of the detector and subsequent data analysis with respect to dealing with  pileup.

Further studies are anticipated to introduce and optimize optical qualities of the surfaces, optimize photo-sensing detectors' coverage,
and improve the simulation of the originating scintillation photons. Much  of the input for these simulations will be  provided by test measurements. 
Position resolution will  also augment  pile-up handling capabilities. The position resolution capabilities of the detector and its importance for achieving the targeted rate will be modeled  and studies  are envisaged in the apparatus which will also be used to test the photo-sensors in LXe.  
Depending on the outcome of the pileup studies, segmentation of the LXe volume may be considered.

\subsubsection{Additional Calorimeter Studies}

At the current design stage and given the targeted level of precision, further simulations assessing the performance of the calorimeter will be carried out in conjunction with development of design parameters for the beam, ATAR, and  tracking detectors. 
One important aspect of the calorimeter design that is being informed by the simulations is understanding how energy from a decay positron is `lost' before it reaches the calorimeter. While energy lost in the ATAR is measured,  losses occurring in inactive material  e.g. the ATAR cabling, and calorimeter entrance windows, depend on the angle of emission.    Simulations are being used to study these effects on the positron energy  resolution.

Another important study involves modeling of photonuclear interactions. As decay positrons interact with atomic nuclei in the simulation, they will occasionally cause a nucleus to enter into an excited state with single or multiple neutron emission. When these neutrons escape the calorimeter without depositing their energy the observed energy is shifted down by multiples of the neutron binding energy (see Fig.~\ref{fig:Signal-Tail}). Photonuclear processes were studied in the PIENU experiment to determine the impact on the NaI(Tl) calorimeter lineshape.  The results were used as input to the calorimeter simulations. Modeling and prototype studies will be pursued for evaluating these effects in LXe. 


In summary, the robustness of the calorimeter performance for the determination of the low energy tail, the resolution,  and the pileup rejection will be evaluated by studying  the following aspects:

\begin{itemize}
    \item impact on resolution and reconstruction of energy loss in inactive materials,
    \item optical properties of the inner surfaces of the LXe detector (reflectivity, light absorption, etc),
    \item granularity, layout (inner and outer surfaces) and specification of photo-sensors (dark current, PDE, etc.), and
    \item characterizations of scintillation light in LXe;  e.g., variations in the absorption length.
\end{itemize} 
Understanding how these conditions affect the performance of the calorimeter will determine the critical parameters required for the next phase of the experimental design.

\section{Sensitivity}
\subsection {$\pi\to e \nu$}
The first phase of \nexp ~ ($R_{e/\mu}$) will employ a beam with pion stopping rate in the ATAR of approximately $3\times 10^5$/s
  at momentum of $55-70$~MeV/$c$ with $\frac{\Delta p}{p}\leq 2\%$
in  a spot size  $\leq2$\,cm diameter. Muon and positron contaminations will be reduced to $<10\%$ with the use of a separator as discussed above. These requirements are compatible with the beams available at the $\pi$E5 (and $\pi$E1) beamline. 

For the basis of estimating the running time to reach the proposed sensitivity, we assume that the beam will be available during 5 months per year. We also assume an overall event acceptance efficiency of 30\%, which is based on the fiducial volume, the timing window cuts, and reconstruction factors.  We assume a data-taking (operations) efficiency of 50\% based on the product of PSI beam delivery and experimental data-collection uptime, along with an allocation for non-production systematic uncertainty tests.  These factors result in $2\times 10^8$ $\pi^+\to e^+ \nu$ events for a 3-year run satisfying the statistics goal. 

\begin{table}[htbp]
\centering
\begin{tabular}{lrr}
\toprule
& PIENU 2015 & PIONEER Estimate\\
Error Source & \% & \% \\
\hline
Statistics & 0.19 & 0.007 \\
Tail Correction & 0.12 & <0.01 \\
$t_0$ Correction & 0.05 & <0.01 \\
Muon DIF & 0.05 & 0.005 \\
Parameter Fitting & 0.05 & <0.01 \\
Selection Cuts & 0.04 & <0.01 \\
Acceptance Correction & 0.03 & 0.003 \\
{\bf Total Uncertainty} & {\bf 0.24} & {\bf $\leq$ 0.01} \\
\bottomrule
\end{tabular}
\caption{$Br(\pi\to e\nu)$ precision for PIENU 2015\cite{PiENu:2015seu} (left) and estimated precision for PIONEER (right).}
\label{pienu_precision}
\end{table}


Systematic uncertainties for PIONEER have been estimated based on the experience of PIENU~\cite{PiENu:2015seu} and are shown in Table~\ref{pienu_precision}. The main systematic uncertainty for  PIENU was the uncertainty in the  tail correction for $\pi\to e\nu$ events below 52\,MeV. In \nexp ~the tail will be reduced from 3\% to $0.5\%$ due to the increased thickness of the calorimeter (25$X_0$ compared to $\leq$19$X_0$) and the more uniform acceptance due to the larger solid angle. The ATAR will allow suppression of $\pi DIF$ and $\mu DIF$ backgrounds enabling more precise measurement of the tail. Uncertainties in the other small corrections, e.g. the pion stop time ($t_0$) Correction, Selection Cuts, and Acceptance Correction, are estimated to be reduced due to the improvements such as in the calorimeter and ATAR timing resolutions. An additional uncertainty arises from the pion lifetime, presently known to 0.02\% precision~\cite{ParticleDataGroup:2020ssz}; the \nexp~ group intends to make additional measurements to reduce this uncertainty to $<0.01\%$.
\subsection {Exotics}
\subsubsection{Massive neutrino searches $\pi^+ \to \ell ^+ \nu _H$}

Searches for peaks in the positron energy spectum due to  $\pi^+ \to \ell ^+ \nu _H$ decays were performed in the PIENU experiment \cite{PIENU:2017wbj,PIENU:2019usb} sensitive to masses $65<m_H<135$ MeV  but no significant signal above statistical uncertainty was found. The measurement of $R_{e/\mu}$\cite{PiENu:2015seu} provides limits for $m_H<65$ MeV.
To estimate the expected sensitivities for PIONEER with $100\times$ the statistics (Phase I), reduced backgrounds, and improved detectors, toy MC simulations were performed.

\begin{figure}[htbp]
\centering
\includegraphics[scale=0.33]{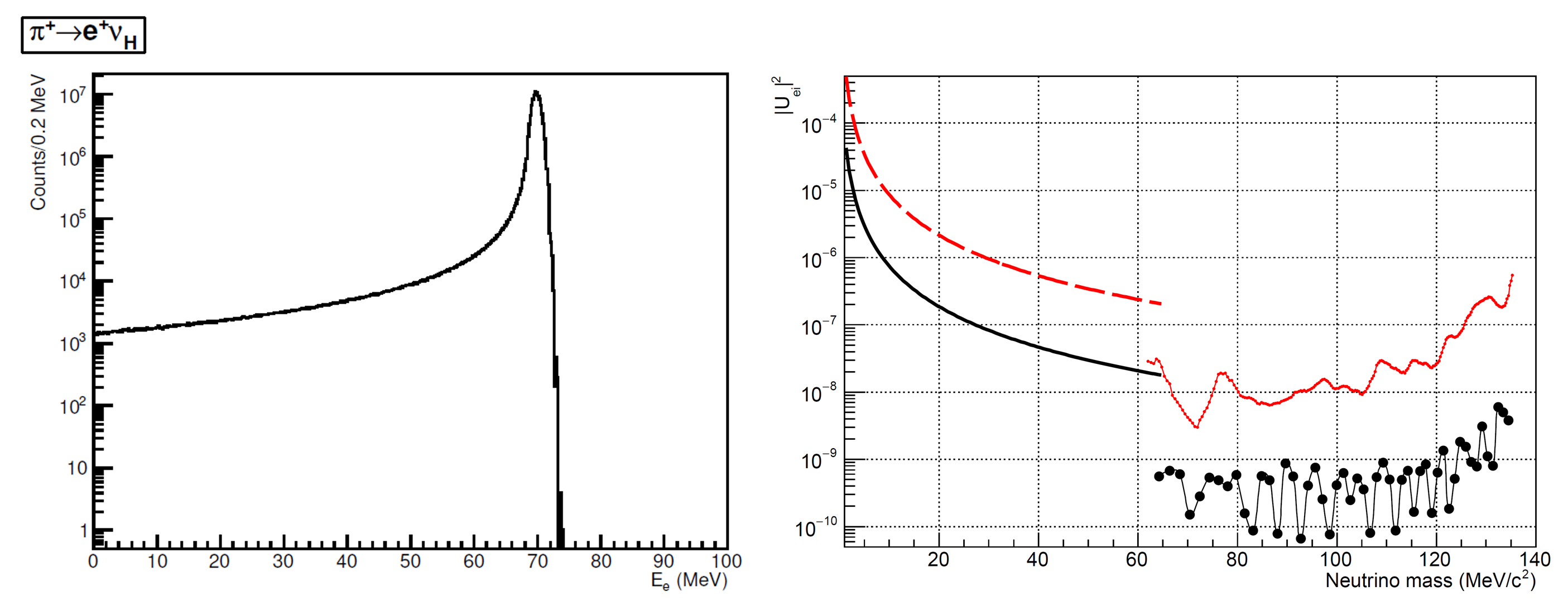}
\caption{Left: simulated $\pi^+ \to e^+ \nu_e$ energy spectrum. Right:Limits on  $|U_{e i}|^2$ (90\% C.L.) from PIENU (red)\cite{PiENu:2015seu,PIENU:2017wbj}) and expected  from  PIONEER (black). 
The  lower region limits ($m_H$<65 MeV) come from the branching ratio measurement and the  higher ($m_H$>65 MeV) region from the peak search.}
\label{pienuH}
\end{figure}

For $\pi^+ \to e ^+ \nu _H$ decays, the peak search sensitivity was limited by residual $\pi^+ \to \mu^+ \to e^+$ background from pion and muon decay-in-flight ($\pi$DIF and $\mu$DIF). 
The low energy calorimeter response tail and statistics of the $\pi^+ \to e^+ \nu_e$ decay also limits the sensitivity. 
Using an active target and a larger electromagnetic calorimeter, the background $\pi^+ \to \mu^+ \to e^+$ will  be significantly suppressed compared to PIENU, and a significantly smaller low energy $\pi^+ \to e^+ \nu_e$ tail is anticipated. 
Figure \ref{pienuH} shows the result of a toy MC study for the expected sensitivity (90\% C.L. upper limits) in PIONEER, assuming $1 \times 10^8$ $\pi^+ \to e^+ \nu_e$ events, 1\% tail fraction below 52\,MeV, no $\pi^+ \to \mu^+ \to e^+$ events, and negligible acceptance corrections due to the larger detector acceptance. 
Compared with PIENU (red  in Fig.~\ref{pienuH}), the expected sensitivity in PIONEER (black) would be improved by one order of magnitude. 

\begin{figure}[htbp]
\centering
\includegraphics[scale=0.34]{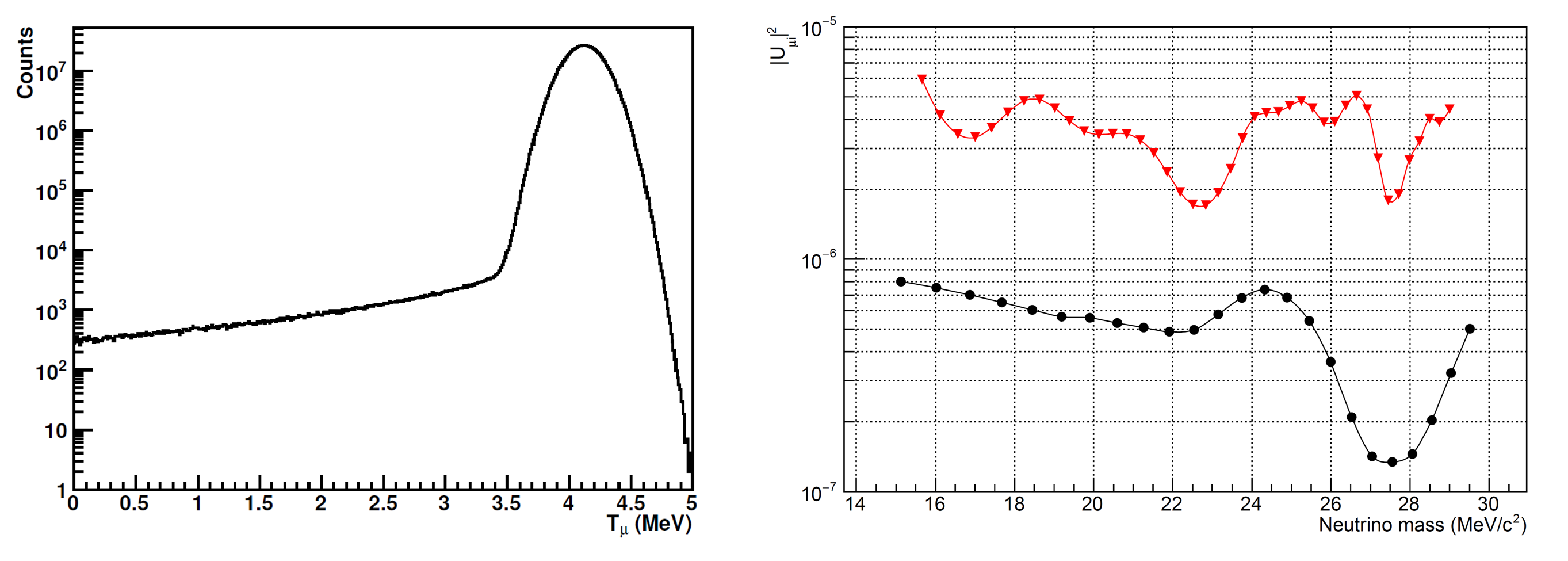}
\caption{Left: simulated $\pi^+ \to \mu^+ \nu_\mu$ kinetic energy spectrum. Right: The PIENU result (red triangles \cite{PIENU:2019usb}) and expected sensitivity with PIONEER (black circles) for the mixing matrix element $|U_{\mu i}|^2$ (90\% C.L. limit).}
\label{Result}
\end{figure}

 The sensitivity for $\pi^+ \to \mu^+ \nu_H$ decay will also be improved by the larger PIONEER statistics . 
The dominant background is mainly due to the radiative pion decay $\pi^+ \to \mu^+ \nu_{\mu} \gamma$ with branching fraction  $2\times10^{-4}$ \cite{PIMUNUG} (for $E_\gamma< 1$\,MeV). 
A toy MC simulation was performed with $1 \times 10^9~{\pi}^+\to\mu^+\nu_{\mu}$ decays ($\times 100$ larger statistics than PIENU) including $\pi^+ \to \mu^+ \nu_{\mu} \gamma$ background considering the proper branching fraction, and assuming the same detector resolution as in PIENU. 
Figure \ref{Result} shows the results of the simulation and the PIENU experiment \cite{PIENU:2019usb}. 

\subsubsection{Two body muon decay $\mu^+ \to e^+ X_H$}

Massive or massless weakly interacting neutral bosons $X$ such as axions \cite{Axion1, Axion2, Axion3, Axion4} and Majorons \cite{Majoron, Majoron2, Majoron3} have been suggested to extend the SM including models with dark matter candidates, baryogenesis, and solutions to the strong $CP$ problem. 
Wilczek suggested a model \cite{Wilczek} which may lead to charged lepton flavor violation (CLFV) where the boson X can be emitted in flavor changing interactions. 

When decay products from a massive boson $X_H$ are not detected due to a long lifetime, flavor violating two-body muon decays involving a massive boson $\mu^+ \to e^+ X_H$ can be sought by searching for extra peaks in the Michel spectrum. 
This search was performed with PIENU, resulting in the limit to the branching ratio $\Gamma (\mu^+ \to e^+ X_H)/\Gamma( \mu ^+ \to e^+ \nu \bar{\nu})$ at the level of $10^{-5}$ in mass range $47.8-95.1$ MeV/$c^2$ \cite{PIENU:2020loi}. 
The  statistics will improve by two orders of magnitude with respect to PIENU. 

\begin{figure}[htbp]
\centering
\includegraphics[scale=0.5]{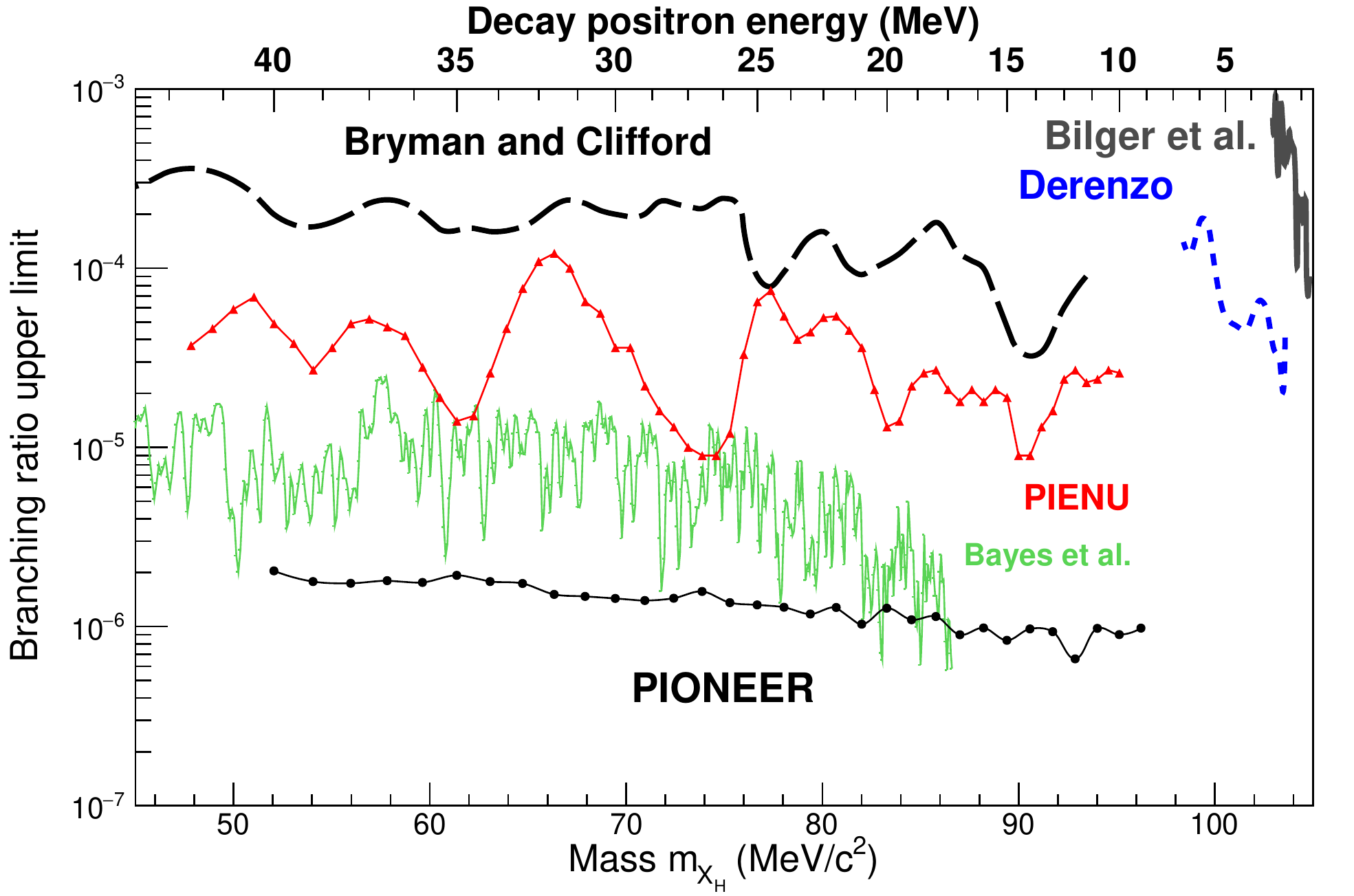}
\caption{90\% C.L. limit on the $\mu^+ \to e^+ X_H$ branching ratio anticipated for PIONEER (black circles) and other past experiments (see Refs. \cite{Exp1, Exp2, Exp3, Exp4} for more details).}
\label{FinalResult}
\end{figure}

To estimate the expected sensitivity, $2 \times 10^{10}$ muon decays were simulated.
Figure \ref{FinalResult} shows the 90\% C.L. upper limits of the branching ratio from different experiments and the expected sensitivity for PIONEER. 

\subsubsection{Other decays}

Three and four body pion decay modes can  be analyzed with the same method as for the $\pi^+ \to \ell ^+ \nu_H$ searches. However, the signal shapes are in these cases are represented by continuous lepton energy spectra. 
The expected sensitivities will also be improved by one order of magnitude. The current limits set by the PIENU experiment are described below.


Three body pion decays $\pi^+ \to \ell ^+ \nu X$, where $X$ is a massive or massless weakly interacting particle, were searched for in the PIENU experiment \cite{PIENU:2021clt}. 
The decay $\pi^+ \to e^+ \nu X$ was sought; no signal beyond the statistical uncertainty was observed, and 90\% C.L. upper limits were set on the branching ratio $\Gamma (\pi^+ \to e^+ \nu X)/\Gamma (\pi^+ \to \mu^+ \nu)$ with $10^{-7}-10^{-8}$ level in the mass range of $0<m_X<120$ MeV/$c^2$. 

The $\pi^+ \to \mu^+ \nu X$ decay was also searched for.
A 90\% C.L. upper limit was derived on the branching ratio $\Gamma (\pi^+ \to \mu^+ \nu X)/\Gamma (\pi^+ \to \mu^+ \nu)$ at the $10^{-5}-10^{-6}$ level in the mass region from 0 to 33.9 MeV/$c^2$.


The rare pion decays $\pi ^+ \to \ell^+ \nu_{\ell} \nu \bar{\nu}$ are highly suppressed. Thus, the experimental search for these processes could reveal small non-SM effects such as neutrino-neutrino interactions \cite{nunu} and six-fermion interactions \cite{6f,6f2}, which might compete with the SM processes at first order. 
The rare pion decays, considering three models (SM, neutrino-neutrino interaction, and six-fermion) were also searched for in PIENU \cite{PIENU:2020las}, and a first result for $\Gamma(\pi^+ \to \mu^+ \nu_{\mu} \nu \bar{\nu})/\Gamma(\pi^+ \to \mu^+ \nu_{\mu})<8.6 \times 10^{-6}$ and an improved measurement $\Gamma(\pi^+ \to e^+ \nu_{e} \nu \bar{\nu})/\Gamma(\pi^+ \to \mu^+ \nu_{\mu})<1.6 \times 10^{-7}$ were obtained. 

\subsection {Pion Beta Decay}

For the $\pi^+ \to \pi^0 e^+ \nu$ experiment the  positive pion stop rate would have to be higher, $\geq 10^7$/s, possibly with  a larger momentum bite $\frac{\Delta p}{p}\approx 3\%$ and likely  using higher pion momentum. These beam conditions are compatible with the $\pi$E5 beam line. This would result in $7\times 10^5$ $\pi^+ \to \pi^0 e^+ \nu$ events collected for 4 years of (5 months/yr) operation assuming similar efficiency factors as discussed for the $\pi\to e \nu$ measurement.\footnote{In the Phase I measurement of \pie~ \nexp\ will collect a sizeable sample of pion beta decay events, which will be helpful to inform the Phase~II(III) design.} This would be sufficient to achieve the required statistical precision to improve the pion beta decay branching ratio measurement precision  by a factor of 3 (Phase II). Systematic effects are expected to be reduced to the 0.06\% level ($10\times$ lower than for the previous PiBeta experiment) due to the combined improvements to the calorimetry (principally, the time and energy resolutions) and the ATAR which may facilitate the observation of the positron in $\pi^+ \to \pi^0 e^+ \nu$ decay in coincidence with the $\pi^0$ detection.

Running at higher rates may be possible leading to  a further precision improvement of 3 (Phase III) and will depend on the ability of the spectrometer to deal with higher rates of  pile-up of accidental events. In this regard, we are studying the possibility to optically segmenting the LXe volume.

\section{ Planning for Realization of PIONEER}
\subsection {Collaboration}

The PIONEER collaboration consists of participants from  PIENU, PEN/PiBeta, and MEG/MEGII as well as international experts in rare kaon decays, low-energy stopped muon experiments, the  Muon $g-2$ experimental campaign, high energy collider physics, neutrino physics, and other areas. The collaboration is still developing and welcomes new members.

We intend to  draft  a collaboration constitution and   institute an appropriate organizational structure. We have  good models, for example, from Muon g-2, that can be tailored to our smaller, but equally diverse and international, team.  We expect that there will be an elected Executive Board with a Chair that guides the key technical decisions as an advisory body to the elected Spokesperson(s) and leaders of the various important technical areas (Beam, ATAR, Calo, DAQ, etc.).

\subsection {R\&D }
In 2022-23, we anticipate performing detector R\&D in several areas including the following:
\begin{itemize}
\item Beam studies. We will carry out beam studies in $\pi E5$ (and possibly $\pi E1$) to establish the required beam conditions. A beam request for 2022 tests will be submitted.
    \item ATAR (see next section).
    \item Cylindrical positron tracker. Designs with standard 300$\mu m$ thick Si strips and with LGADs are being considered. We expect to construct and test  prototypes of various geometries.
    \item LXe prototype. The  objectives of this R\&D work include   determination of the properties of photo-sensors  and optical properties of materials for use in the  LXe calorimeter. We also want to  benchmark the photon transport simulations.  We are considering the development of a medium scale calorimeter prototype that would enable measurements of properties like energy resolution and photonuclear effects for validation of simulations.
    \item LXe calorimeter optical segmentation. Small prototypes will be used initially and UV compatible materials will be evaluated. Some of these studies may  be done using a LXe cryostat at McGill university containing $\sim$\unit[2]{l} of LXe developed  for SiPM tests for nEXO.  An assembly hosting SiPMs, reflective material and a retractable radioactive source will be prepared at TRIUMF and brought to McGill for measurements. 
    \item SiPMs. SiPM degradation at high rates will be studied. We will test available photosensors using small LXe prototypes in association with the McGill setup mentioned above.
    \item Crystal alternatives to LXe. Arrays of LYSO crystals with varying levels of doping will be evaluated from various manufacturers. See the Appendix for more details.
    \item DAQ. Rate testing of FPGA-to-CPU/GPU and CPU-to-CPU communication via optical PCI-express links will be done along with performance testing of data compression algorithms for \calo\ data.
    \item Trigger prototyping. We will use a prototype APOLLO Command Module that the Cornell CMS group will share with PIONEER, and build a 4-channel prototype of the digitizer board for evaluation and communications development.
\end{itemize}

\subsection{ATAR R\&D}
A brief summary of ATAR R\&D  follows.  A more detailed plan is presented in Appendix~\ref{sec:ATAR_timeline_app}. 
\begin{itemize}
    \item After initial sensor characterization and design optimization a PIONEER specific prototype production should happen by the end of 2023. The characterization includes studies on LGAD energy resolution and gain suppression mechanism.
    \item Building of a first ATAR demonstrator (ATAR0) with a few planes of available sensor prototypes (BNL strip LGADs, 2.5\,cm with 500\,$\mu$m  pitch). An electronics board with suitable characteristics needs to be designed and produced. The prototype would be then tested in a pion/muon beamline either at TRIUMF or PSI. This  prototype may be produced by the end of 2023.
    \item Identification of a suitable chips for the analog amplification and digitization by 2024. The effect of a short flex between sensor and chip will be studied within 2022.
    \item The support mechanics and thermal load needs to be studied well with mock-up prototypes and silicon heaters. These details needs to be fully understood by 2025.
    \item Full production of sensors and readout ASIC, once identified, should take less than a year given the modest area of the ATAR. Therefore final production and subsequent assembly can start in 2025.
\end{itemize}






\clearpage
\appendix
\appendixpage
\addappheadtotoc

\section{ PIENU and PEN}
{\bf Measurements of $R_{e/\mu}$ and associated exotic searches.}
\label{Appendix}

The PIENU experiment has provided the most precise measurement of the branching ratio $R_{e/\mu}=(1.2344\pm0.0023_{stat}\pm0.0019_{sys})\times10^{-4}$\cite{PiENu:2015seu}; a further factor two  improvement in precision is anticipated. The $\pi\to e\nu$ branching ratio provides the best test of electron–muon universality in charged current weak interactions resulting in the ratio of weak interaction strengths $\frac{g_\mu}{g_e}=1.0010\pm0.0009$\cite{Bryman:2021teu}.   The PEN experiment at PSI is aiming at comparable precision to PIENU.

\begin{figure}[h!]
\centering
\includegraphics[scale=0.485]{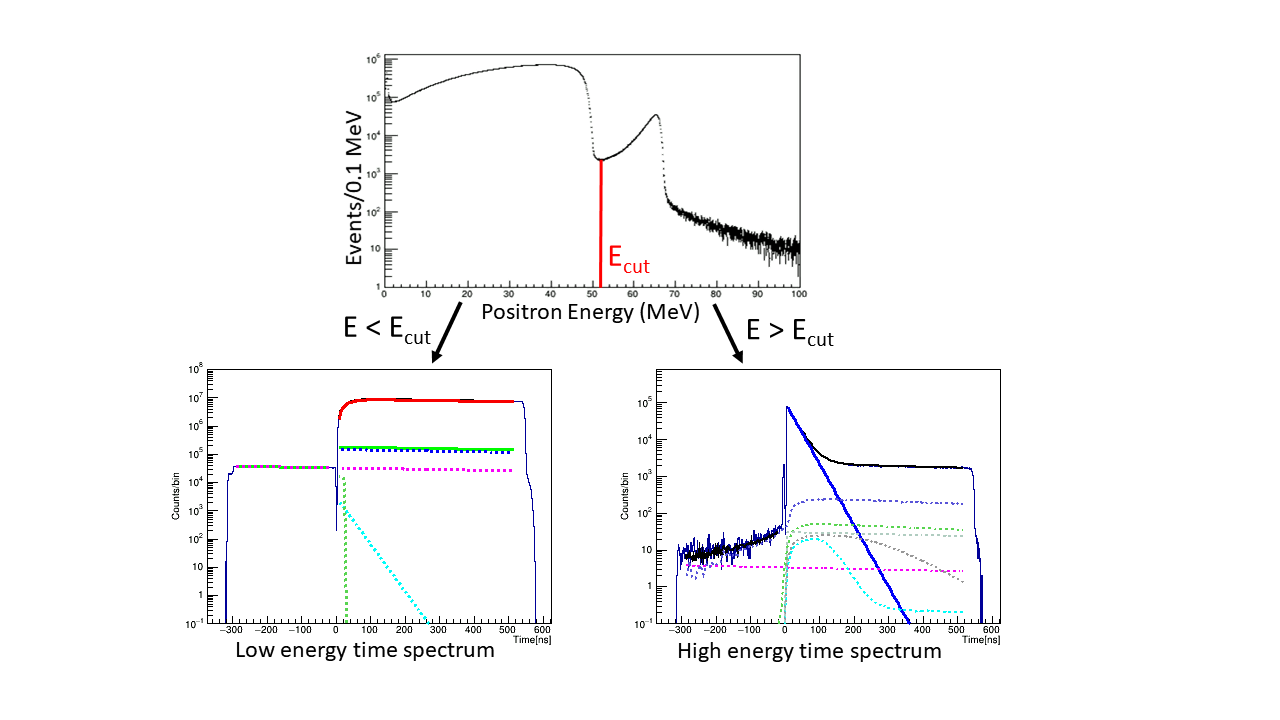}
\caption{The upper panel shows the positron energy spectrum with the red line indicating $E_{cut}$. The lower panels show the time distributions for events below and above $E_{cut}$. The black histograms are data, the red curve is the $\pi^+ \to \mu^+ \to e^+$ signal, and the blue line is the $\pi^+ \to e^+ \nu$ signal. The other histograms in various colors are the background terms related to pile-up, muon DIF, and other effects discussed in Ref.~\cite{PiENu:2015seu}.}
\label{fig:analysis}
\end{figure}

PIENU obtained the branching ratio by first separating events into high- and low-energy regions at an energy cut value ($E_{cut}$) as illustrated in Fig. \ref{fig:analysis}. The time spectra were fit in each region with the $\pi^+ \rightarrow e^+ \nu$ and $\pi^+ \rightarrow \mu^+ \rightarrow e^+$ shapes, plus backgrounds originating from different sources including pion decays in flight, contamination from old muon decays etc. The raw branching ratio $R^{raw}_{e/\mu}$ was the ratio of the $\pi^+ \rightarrow e^+ \nu$ amplitude to the $\pi^+ \rightarrow \mu^+ \rightarrow e^+$ amplitude. Corrections such as the tail correction for low energy events below the Michel spectrum,  were subsequently applied to obtain the final value.

High precision pion decay experiments also provide a plethora of constraints on exotic phenomena including heavy neutrinos and dark sector processes.
Extensions of the Standard Model postulate the
existence of additional (sterile) neutrinos \cite{Boyarsky:2009ix, Bryman:2019bjg}. These additional states may 
contribute to the solution of outstanding puzzles like
the nature of dark matter, early cosmological processes like small scale structure formation \cite{Bertoni:2014mva}, and Mesogenesis \cite{Elahi:2021jia}.  
Massive neutrino states $\nu_H$ are sought in the two-body pion decays
$\pi^+\rightarrow e^+\nu_H$ \cite{PIENU:2017wbj} and $\pi^+\rightarrow \mu^+\nu_H$ \cite{PIENU:2019usb}. Exploiting
large data sets of pion decays and the resulting decay muons,  exotic two-body muon decays like $\mu^+\rightarrow e^+X$ can
be sought \cite{PIENU:2020loi}, where X is a massive neutral boson (e.g. an axion or a Majoron).
Similarly, exotic particles have been searched for in three body decays like $\pi^+\rightarrow l^+ \nu X$ ($l=e^+,\mu^+$) \cite{PIENU:2021clt}.
The PIENU experiment also obtained upper limits for the rare decays $\pi^+\rightarrow e^+\nu_e\nu\bar{\nu}$ and
$\pi^+\rightarrow \mu^+\nu_{\mu}\nu\bar{\nu}$ at the $10^{-7}-10^{-6}$ level \cite{PIENU:2020las}.


\nexp~  with two orders of magnitude more statistics has the potential to improve the existing limits by at least an order of magnitude.
Since the searches are based on fits to the energy spectra of the visible final state particles, an improved experiment
can bring significant additional advantages in lowering the limits and in reducing the systematic errors. For example, 
the $\pi^+\rightarrow e^+\nu$ low energy tail represents the main background for the
$\pi\rightarrow e^+\nu_H$, $\pi^+\rightarrow e^+\nu X$, and $\pi^+\rightarrow e^+\nu_e\nu\bar{\nu}$ searches: 
more precise knowledge of the tail and its further reduction will significantly improve the upper limits beyond the statistics.
The search for rare and exotic decays involving  muons, like $\pi^+\rightarrow \mu^+\nu_H$, $\pi^+\rightarrow \mu^+\nu X$, and $\pi^+\rightarrow \mu^+\nu_{\mu}\nu\bar{\nu}$ will
benefit from an improved stopping target and faster electronics, which will allow better separation of  muons from pions and thus further improve the sensitivity.

\begin{figure}[!tbp]
  \centering
  \subfloat[]{\includegraphics[width=0.5\textwidth]{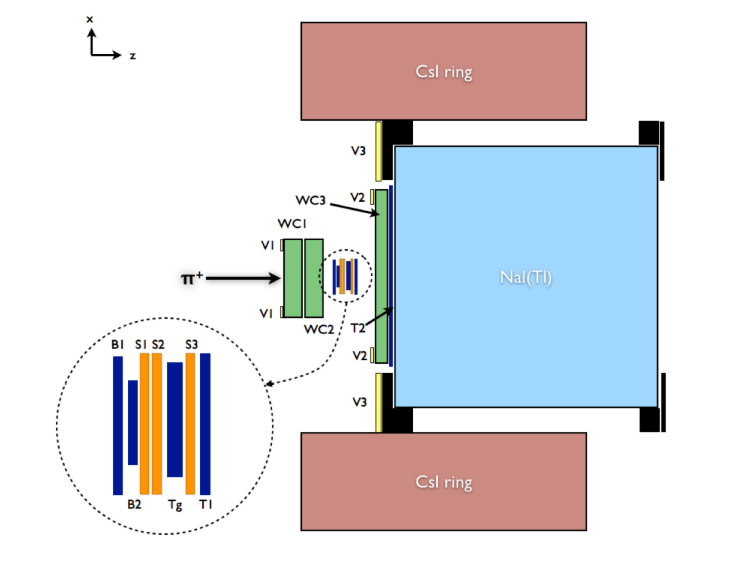}\label{fig:detector_PIENU}}
  \hfill
  \subfloat[]{\includegraphics[width=0.5\textwidth]{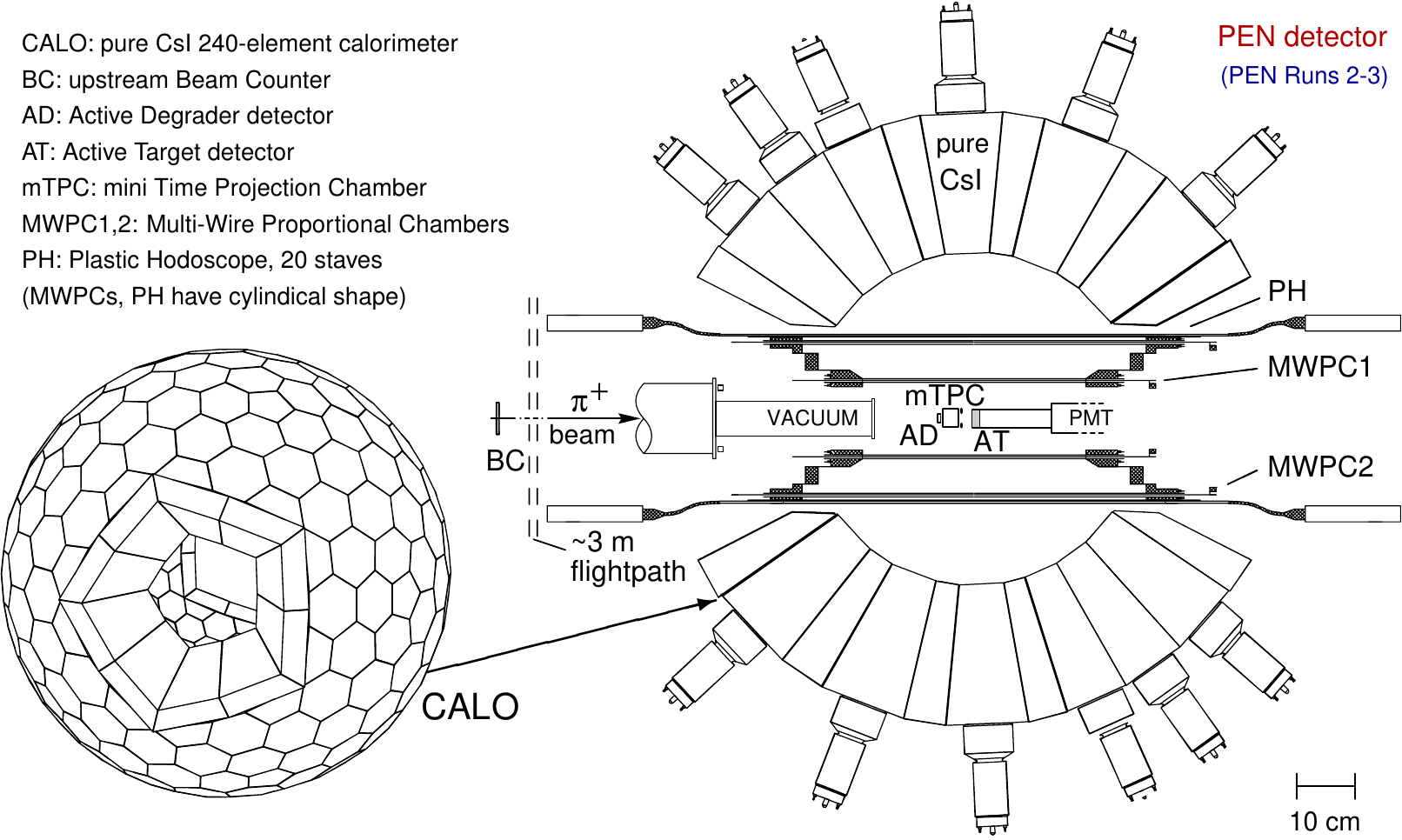}\label{fig:detector_PEN}}
  \caption{(a) Schematic view of the PIENU detector. Plastic scintillators are shown in dark blue, wire chambers in green, silicon strip trackers in orange and the calorimeter in light blue and red. (b) Schematic cross section of the PEN detector, with a view of the CsI crystal calorimeter.}
\end{figure}

The PEN/PiBeta and PIENU experiments relied on inorganic scintillator calorimetry.
PIENU used a high-resolution ($\sigma=1\%$) crystal calorimeter consisting of a single crystal NaI(Tl) detector  surrounded by an array of 97 pure CsI crystals for shower leakage containment. The PIENU detector is shown in Fig.~\ref{fig:detector_PIENU} and described in \cite{PiENu:2015pkq}. The large NaI(Tl) crystal was 19$X_0$ lengths thick and 19$X_0$ in diameter. The high energy resolution and long radiation-lengths of the Na(Tl) crystal  were essential for reducing the low energy tail. However, the slow decay constant of NaI(Tl) limited pile-up detection and rejection. 
The acceptance of the PIENU detector was relatively small $<\unit[20]{\%}$ which resulted in an important  source of systematic uncertainty. The PEN experiment, on the contrary, adopted a high solid angle geometry (see Fig.~\ref{fig:detector_PEN}). Its key components were a highly segmented (240 elements) spherical pure CsI crystal calorimeter covering $\sim \unit[3\pi$]{sr} of solid angle around the pion stopping target. The key limitations were related to the imperfect separation of
$\pi\rightarrow \mu \rightarrow e$ and $\pi\rightarrow e\nu$ decays.  The primary culprit was the 12$X_0$ thickness of the CsI calorimeter, which produced a substantial low energy tail for \unit[70]{MeV} positrons (and photons)
extending well under the $\pi\rightarrow \mu \rightarrow e$ spectrum. 

The PIONEER approach  using a high  resolution, uniform response LXe calorimeter with fast timing  and high solid angle combines the assets of both experiments. 

\section{PiBeta}

{\bf Pion beta decay measurements}

The branching ratio for pion beta decay was most accurately measured by the PiBeta experiment\footnote{The PiBeta and PEN experiments shared much of the same apparatus.} at PSI \cite{Pocanic1,Frlez:2003vg,Pocanic:2003pf,Frlez:2003pe,Bychkov:2008ws} to be
$\frac{\Gamma(\pi^+ \to  \pi^0 e^+ \nu)}{\Gamma\textrm{(Total)}}= [1.036 \pm 0.004 \textrm{(stat)} \pm 0.004\textrm{(syst)} \pm 0.003(\pi\to e\nu)] \times 10^{-8}$, where the first uncertainty is statistical, the second systematic, and the third is the $\pi\to e\nu$ branching ratio uncertainty.
Pion beta decay potentially provides the theoretically cleanest determination of the magnitude of the CKM matrix element \vud. With current input one obtains $\vud = 0.9739(28)_{\textrm{exp}}(1)_{\textrm{th}}$, where the experimental uncertainty comes almost entirely from the  $\pi^+ \rightarrow \pi^0 e^+ \nu (\gamma)$ branching ratio (BRPB).
The theory uncertainty has been reduced from $({\delta}V_{ud})_{\textrm{th}} = 0.0005$ \cite{Sirlin:1977sv, Cirigliano:2002ng, Passera:2011ae} to $({\delta}V_{ud})_{\textrm{th}} = 0.0001$ via a lattice QCD calculation of the radiative corrections \cite{Feng:2020zdc}. The current precision of \unit[0.3]{\%} on \vud makes $\pi^+ \rightarrow \pi^0 e^+ \nu (\gamma)$ not presently relevant for the CKM unitarity tests because super-allowed nuclear beta decays provide a nominal precision of 
\unit[0.03]{\%}. 
In order to make $\pi^+ \rightarrow \pi^0 e^+ \nu (\gamma)$ important for CKM unitarity tests, two precision experimental stages can be identified:
(1) As advocated in Ref.~\cite{Czarnecki:2019mwq}, a three-fold improvement in BRPB precision compared to Ref.~\cite{Pocanic:2003pf} would allow for a 0.2\% determination of $\left|V_{us}/V_{ud}\right|$ improving on measurement of the following ratio being currently 
      $  R_V = \frac{\Gamma\left(\textrm{K} \rightarrow \pi l \nu (\gamma) \right)}{\Gamma\left(\pi^+ \rightarrow \pi^0 e^+ \nu (\gamma)\right)}=1.3367(25)$ ,
    independent of the Fermi constant, short-distance, and structure-dependent radiative corrections. This
    would match the precision of the current extraction of $\left|V_{us} / V_{ud}\right|$ from the axial channels~\cite{Marciano:2004uf} 
        $R_A = \frac{\Gamma\left(\textrm{K} \rightarrow \mu \nu (\gamma) \right)}{\Gamma\left(\pi \rightarrow \mu \nu (\gamma)\right)}=1.9884(115)(42)$,
  providing a new competitive constraint on the \vus--\vud\ plane and probing new physics that might affect vector and axial-vector channels in different ways.
    The theoretical case for this approach was recently strengthened by improved analysis of radiative corrections in $K \to \pi e \nu $ decays \cite{Seng:2021nar}.  
(2)  In the second phase, an order of magnitude improvement  in the
BRPB precision will be sought. This would provide the theoretically cleanest extraction of \vud at the \unit[0.02]{\%} level. 
\section{Alternative LYSO Calorimeter Considerations}
\begin{figure}[h!]
\centering
\includegraphics[width=\linewidth]{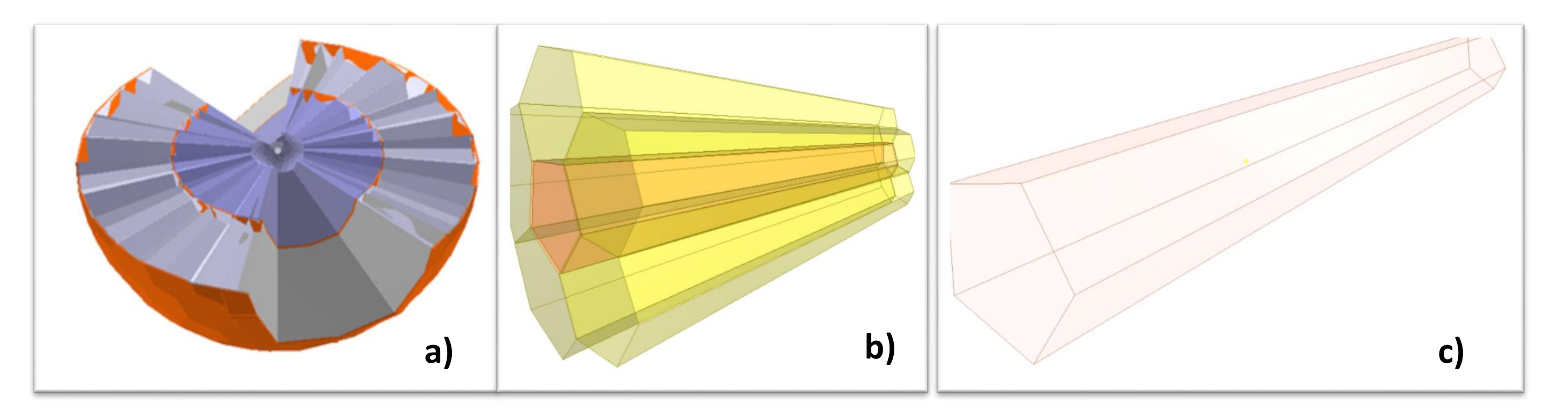}
\caption{Possible use of an inner array of tapered LYSO crystals within the open volume of the existing PEN CsI calorimeter.  a) Opened view showing in blue the array of LYSO crystals that matches one-to-one to the existing geometry of the PEN crystals shaded in gray.  b) An example array ideal for testing the concept.  c) An individual pentagonal crystal, 16$X_0$ in depth.  Each such crystal would be read out by a thin array of SiPMs.
}
\label{fig:LYSO-figs}
\end{figure}

A naturally segmented array of tapered LYSO crystals provides an attractive alternative to our proposed LXe-based calorimeter.  We are exploring a geometry that matches that of the PEN pure CsI detector. The PEN crystals were limited to 12$X_0$  depth, which is insufficient for an ideal $\pi \rightarrow e \nu$ measurement.  However, the segmentation and fast response allows for various trigger patterns and it is well-designed for the pion beta decay phase in the PiBeta configuration. This detector has an inner radius of 26\,cm, and an outer radius of 48\,cm.  CsI has a radiation length of 1.86\,cm. With the compact ATAR geometry we are proposing, a sufficient volume exists to insert an inner array of crystals between the ATAR and the existing PEN CsI array, see Fig.~\ref{fig:detector_PEN} and Fig.~\ref{fig:LYSO-figs}a.   On paper, LYSO crystals appear to be the ideal choice for such an array.  LYSO is a Cerium doped Lutetium based scintillator whose light output is comparable to doped NaI(Tl).  It has high density ($X_0$= 1.14\,cm, $R_{M}$=2.07 cm) and a light yield comparable to the highly luminous NaI(Tl), but with much faster light signals.  Its 420\,nm typical scintillation light has a 40\,ns single exponential decay time and the spectrum is well matched to conventional SiPM photosensors.  LYSO is both radiation hard and non-hygroscopic. To date, the main application for LYSO crystals has been in PET medical scanners, where small crystals are needed with high light output to resolve the 511\,keV gammas from positron annihilation.  The attractive properties of these crystals suggest that larger arrays could be made for use in particle physics applications and, indeed, several groups have explored this possibility with small arrays and somewhat limited success.

Experts in the field, such as Ren-Yuan Zhu from Caltech, have advised our collaboration on the pros and cons and experiences of other groups, see for example \cite{Zhu:2021sij}.  The growth of relatively long LYSO crystals is a fairly new and expensive R\&D effort, not easily justified without a demanding end-use application.  The Mu2e Collaboration built a $5 \times 5$ array of $3 \times 3 \times 20\,$cm$^3$ crystals and subjected the array to test beams at Frascati and Mainz~\cite{Atanov:2016blu}.  At 100\,MeV, the resolution was in the range of 3--4\%,  dominated by a 2.5\% constant term indicative of non-uniformity in light transmission and possibly crystal to crystal calibrations and surface preparation.  That performance met the specifications for Mu2e, but they did not push it further, choosing eventually to use undoped CsI based on an overall cost analysis.

Dr. Zhu has investigated crystals from various companies and published properties that will guide us going forward.  An example of a study of a 25$X_0$ crystal's longitudinal transmission and how to improve it is found in \cite{Mao:2012dr}.  It is imperative to improve on the uniformity of light production and transmission along the length of the crystal.  Experience has shown -- as it has for CsI -- that custom surface preparation on tapered crystals is also important and can be realized with careful lab bench work.  We aim to investigate the possibilities of using LYSO for PIONEER, but would not now claim this to be defensible until proper bench tests and manufacturer quotes are in hand.  Figures\,\ref{fig:LYSO-figs}b and c indicated prototype geometries that are designed to fit inside the PEN calorimeter.  We will issue requests to a variety of companies for quotes to see if these crystal shapes can be made to meet our specifications.
If tests are successful --- principally achieving the needed energy resolution below 3\% -- then a careful side-by-side comparison of costs and performance against our leading calorimeter candidate based on LXe can be made.

\section{ATAR technical details}
\label{sec:ATAR_appendix}


As introduced in Sec.~\ref{sec:ATAR}, the highly segmented active target (ATAR) is a key new feature of the proposed PIONEER experiment which will define the fiducial pion stop region, provide high resolution timing information, and furnish selective event triggers. Further technical details will be covered by this Appendix. 

\subsection{ATAR R\&D plan}
\label{sec:ATAR_timeline_app}
The ATAR is a small project in terms of production (roughly 0.01~m$^2$ of sensor area) but still requires significant R\&D for sensors, electronics, digitization and mechanics. An envisioned path forward for this project is the following:
\begin{itemize}
    \item Sensor characterization and design optimization is the top priority, LGAD prototypes from BNL and FBK are being studied. A PIONEER-specific prototype production is expected to happen within 2 years from now at BNL. Introduction to the LGAD technology is shown in Appendix~\ref{sec:LGAD_app} and studies on available high granularity LGAD prototypes are shown in Appendix~\ref{sec:GLGAD_app}. To steer the design effort, TCAD (Silvaco and Sentaurus) simulations are crucial (see Appendix~\ref{sec:ACLGAD_app}). 
    An important aspect to understand is the energy resolution of LGAD devices as well as the gain suppression mechanism~\cite{gainsuppr}. A test beam at the the ion beam line  at the University of Washington (CENPA)  will be used to study the response of LGADs to high ionizing particles. This will tentatively happen in  2022 with  available sensor prototypes and analog amplifier boards.
    \item Having the amplifier chip several cm away from the LGAD sensor is rather unconventional and the effect on the response needs to be understood; a first connection flex prototype was produced and will be thoroughly tested within 2023. Likely a second flex production with the lesson learned will happen in the same year.
    \item The building of a first ATAR demonstrator (ATAR0) is foreseen within the end of 2023 with available sensor prototypes. The project is described in Appendix~\ref{sec:ATAR0}.
    \item Identification of a suitable chip for the analog amplification; the ideal path would be to find an already existing chip (e.g. FAST2) and characterize it with LGAD prototypes. In parallel an effort to produce a new chip can be pursued through external companies (SBIR-like funding). A prototype readout chip needs to be ready by 2024.
    \item Identification of a digitizer chip: currently available digitizers are too expensive for the number of channels in the ATAR. Small companies might be available to develop a new chip with the needed performance by modifying existing designs. A suitable digitizer chip needs to be identified/developed by 2024.
    \item The support mechanics and thermal transport calculation needs to be studied well for the success of the ATAR. The support needs to introduce as little dead material as possible to avoid degradation of the positron energy. A discussion is ongoing between technicians from UCSC and UW for the design of the mechanic support and mounting procedure. Thermal load tests can be conducted with an ATAR mock-up made with Silicon heaters, thermal calculations are also foreseen. The radiation damage during the \pie~phase of the experiments is low enough that the heat dissipation of the sensors is not a concern.
\end{itemize}

\subsection{Sensor technology}
\label{sec:LGAD_app}

The chosen technology for the ATAR is based on  Low Gain Avalanche Detectors (LGAD) \cite{bib:LGAD}, thin silicon detectors with moderate internal gain. Due to the internal gain and thin bulk, LGADs have fast rise time and short full charge collection time.
The best estimate at present for the sensor thickness is around 120\,$\mu$m to avoid support structures for the sensor, which would introduce dead areas and inactive material within the beam.
Using fast electronics is expected to result in a pulse rise time of about 1\,ns for an LGAD of this thickness. Such a sensor would be able to separate two closely overlapping hits if they arrive more than 1.5\,ns apart. The time resolution on the rising edge should be less than 100\,ps for a minimum ionizing signal, down to much better time resolution for large $\pi/\mu$ signals.

Current standard LGADs are limited in terms of granularity to the mm scale.
To achieve a ~100\% active area, several technologies still at prototype level are being evaluated for PIONEER, such as AC-LGADs~\cite{Apresyan:2020ipp} (studies shown in Appendix~\ref{sec:ACLGAD_app}), TI-LGADs~\cite{9081916} (studies shown in Appendix~\ref{sec:TILGAD_app}) and DJ-LGAD~\cite{Ayyoub:2021dgk} (a prototype run is expected to be finished and tested by Q1 2022).

Preliminary data taken at the Stanford Light source (SSRL) \cite{GALLOWAY20195} show that LGADs can detect low energy X-rays with a reasonable energy resolution (8\% to 15\%) thanks to the internal gain.
The beamline at SSRL had a 2\,ns repetition rate and single pulses were completely separated with 50\,$\mu$m thick LGADs.

A dynamic range from MIP (positron) to several MeV (pion/muon) of deposited charge is expected in the ATAR. 
Since the event reconstruction relies on temporal pulse separation, the response to successive MiP and high charge deposition have to be studied \cite{Pulserep}.
Furthermore the effect of gain suppression for large charge deposition in LGADs has to be taken into account \cite{gainsuppr}.
To study these effects a test beam will be organized at the ion beam line of the University of Washington (CENPA) in 2022 to study the response of the aforementioned sensors to high ionizing events. 
Furthermore, laboratory tests will be conducted with an alpha source.


\subsection{High granularity LGAD ongoing R\&D}
\label{sec:GLGAD_app}

\subsubsection{AC-LGADs}
\label{sec:ACLGAD_app}

\begin{figure}[htbp]
\centering
\includegraphics[width=0.4\textwidth]{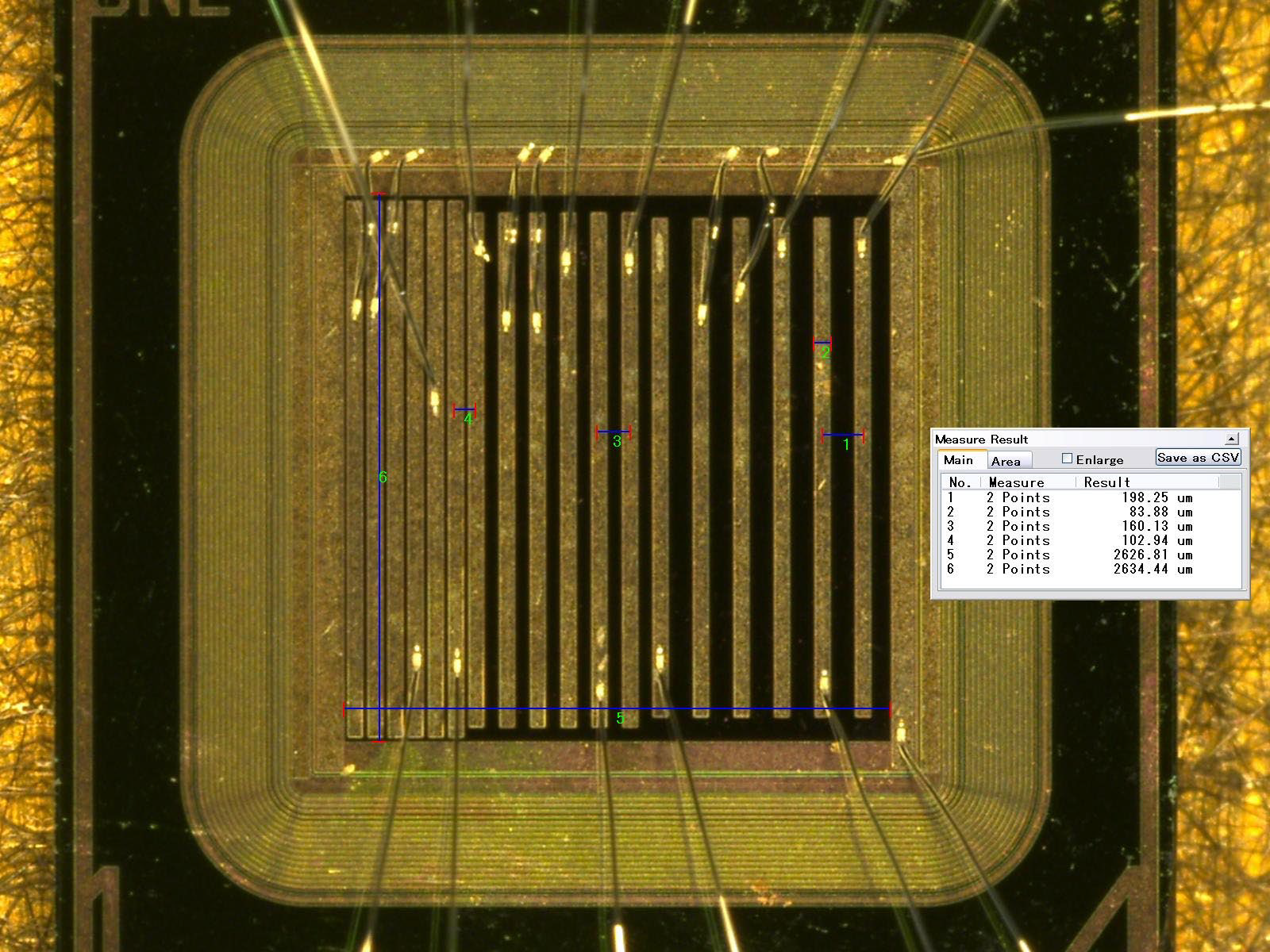}
\includegraphics[width=0.5\textwidth]{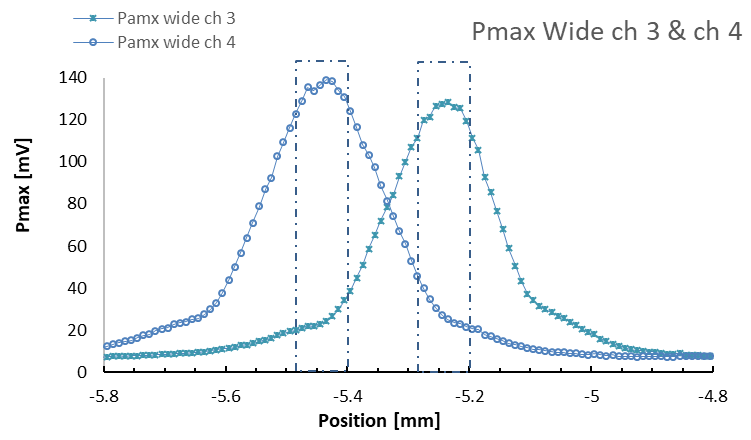}
\caption{Left: prototype BNL AC-LGAD strip sensor with 80\,$\mu$m wide strips and pitch of (left to right) 100, 150, 200\,$\mu$m. Right: sensor response ($P_{max}$) as a function of position (perpendicular to the strip) of two strips with 200~$\mu m$ of pitch~\cite{ACLGADpico}. The dashed lines highlight the position of the two strips in the plot. Data taken at the FNAL 120\,GeV proton test beam facility.}
\label{fig:FNAL_data}
\end{figure}

AC-LGADs overcome the granularity limitation of traditional LGADs and have been shown to provide spatial resolution of the order of tens of $\mu$m~\cite{Tornago:2020otn}. AC-LGAD design also allows to have a completely active sensor with no dead regions.
Studies were conducted on strip AC-LGAD prototypes from Brookhaven National Lab (BNL) (Fig.~\ref{fig:FNAL_data}, Left). 
The sensors have been tested with a laboratory IR laser TCT station~\cite{Particulars} and at a Fermilab (FNAL) test beam~\cite{ACLGADpico}.
The response of two strips of a 200\,$\mu$m pitch BNL AC-LGAD as a function of position can be seen in Fig.~\ref{fig:FNAL_data} (Right). The position resolution of this prototype sensor varies between 5-15~$\mu m$ in the direction perpendicular to the strip.
In the next few years several prototypes will be tested in laboratories and test beams to identify the right parameter configuration, then a PIONEER-specific production will be made at Brookhaven.

The envisioned metal size for the 200\,$\mu$m pitch strips, as well as other parameters of interest for the sensor such as the doping profile, needs to be confirmed after a testing R\&D campaign and TCAD simulations. 
These parameters have been studied with the TCAD Silvaco \cite{Silvaco} to have a good representation of the observed sensor performance. Simulations with TCAD software are important to compare with existing prototype data and to help in optimizing the design, currently the simulated sensor response show a reasonably good match with FNAL TB data. 
Studies made with simulations tools will provide crucial input for the future PIONEER sensor production.

\subsubsection{TI-LGADs}
\label{sec:TILGAD_app}
Trench Isolated (TI) LGADs are a novel silicon sensor technology that utilizes a deep narrow trench to electrically isolate neighboring pixels to prevent breakdown, as opposed to standard LGADs which use a junction termination extension to prevent breakdown at the pixel edges \cite{9081916}. By utilizing the deep trench isolation technology, the no-gain region is reduced to a few micrometers, thus achieving a higher fill factor than regular LGADs. 
Prototypes TI-LGADs sensors from Fondazione Bruno Kessler (FBK) \cite{9081916} were studied at UCSC.
TI-LGADs show the standard response of a conventional LGAD and exhibits a small amount of “cross-talk”. Furthermore the response of the sensor is constant along the strip.
The maximum values of the pulse shape for the neighbor strips correspond to ~3$\%$ of the maximum value of the red pulse showing there is good isolation between strips.
New TI-LGADs productions are envisioned at FBK and other vendors, the prototypes will be tested in laboratories and test beams to find a suitable alternative sensor technology for PIONEER.


\subsection{Electronics and readout chain}
\label{sec:ATAR_electronics_app}

To read out the ATAR sensors, two crucial electronic components need to be identified: an amplifier chip and a digitizer board.
The ASIC needs to be fast enough for the sensor in use; for the signal rise time in the 120\,$\mu$m-thick prototype sensors,  a bandwidth of 1\,GHz should be sufficient.
However, the high dynamic range (2000) requirement for the ATAR brings major complications to the readout.
Current fast readout chips usually have a dynamic range of $<$~1000, since they are targeted at MIPs-only detection in tracker sub-systems.
One possibility is to develop an amplifier chip with logarithmic response or dynamic gain switching as well as a high enough bandwidth, currently no such chip exists with the necessary characteristics.
Already available integrated chips, such as FAST~\cite{OLAVE2021164615} and FAST2, are being evaluated.
Some new ASIC technologies that are being developed at UCSC in collaboration with external companies can run with 2.5V maximum signal, this allows for an increased dynamic range.

Since the amplification chip has to be positioned away from the active region, the effect of placing a short (5\,cm) flex cable between the sensor and the amplification stage has to be studied.
To study this a prototype flex was produced and the effect on LGAD signals will be studied.

To successfully reconstruct the decay chains, the ATAR is expected to be fully digitized at each event.
To achieve this goal, a high bandwidth digitizer with sufficient bandwidth and sampling rate have to be identified. 
The same issue afflicting the amplifier, the high dynamic range, is also problematic for the digitization stage. 
A digitizer that would suit PIONEER's requirements needs to be identified, a ready commercial solution would be the best option but the cost per channel might be prohibitive. 
For this reason the collaboration is exploring the possibility to develop a new kind of digitizer specific to this application.

\subsection{ATAR0 prototype}
\label{sec:ATAR0}
A first ATAR demonstrator (ATAR0) is foreseen with available sensor prototypes: the current BNL AC-LGAD production has 2.5~cm long strips with a pitch of 500~$\mu$m, which is close to the final 200~$\mu$m 2x2~cm$^2$ ATAR design (however the sensor thickness is 50~$\mu$m instead of the final design 120~$\mu$m). Since a 50~$\mu$m thick sensor is fabricated on a support wafer of a few 100s $\mu$m, the devices would need to undergo an etching procedure to have full active volume, the thinning procedure can be executed at BNL. The BNL sensor group has experience with wafer lapping and chemical mechanical planarization (CMP). The sensor would be initially lapped to remove most of the thickness of the handling thick substrate, followed by CMP for final polish. The etching rate would be determined on dummy wafers and the process would be optimized to reach the required final thickness.
The prototype would have a few layers (5-10) with a reduced number of channels to detect the temporal development of a muon or pion decay. As the layers would need to be very close to each other, a suitable readout board needs to be developed. The board might be built with discrete components or using a chip such as FAST2.
The prototype would be then tested in a pion/muon beamline either at TRIUMF or a PSI. Hopefully such a prototype can be produced by the end of 2023.

\bibliography{References}

\end{document}